\documentclass[apjl]{emulateapj}
\usepackage{comment}
\usepackage{ifthen}
\usepackage[switch]{lineno}
\modulolinenumbers[5]
\newcommand{\forloop}[5][1]%
{%
\setcounter{#2}{#3}%
\ifthenelse{#4}%
	{%
	#5%
	\addtocounter{#2}{#1}%
	\forloop[#1]{#2}{\value{#2}}{#4}{#5}%
	}%
	{%
	}%
}% 

\newcommand{\ctbd}[1]{}

\newcommand{\lc}{light curve}
\newcommand{\lcs}{light curves}
\newcommand{\Lc}{Light curve}

\newcommand{\tel}{telescope}

\newcommand{\aper}{aperture}
\newcommand{\oot}{out-of-transit}

\newcommand{\ccdsize}[1]{\ensuremath{\rm #1\times\rm#1}}
\newcommand{\fovsize}[2]{\ensuremath{\rm #1 #2\times\rm#1 #2}}
\newcommand{\tsize}[1]{\mbox{\rm #1 m}}

\newcommand{\Ks}{\ensuremath{K_s}}
\newcommand{\masy}{\ensuremath{\rm mas\,yr^{-1}}}
\newcommand{\kms}{\ensuremath{\rm km\,s^{-1}}}
\newcommand{\ms}{\ensuremath{\rm m\,s^{-1}}}

\newcommand{\gcmc}{\ensuremath{\rm g\,cm^{-3}}}

\newcommand{\el}{\ensuremath{e^-}}

\newcommand{\pxs}{\ensuremath{\rm \arcsec pixel^{-1}}}

\newcommand{\eladu}{\ensuremath{\lbrack \el/ADU \rbrack}}

\newcommand{\teff}{\ensuremath{T_{\rm eff}}}
\newcommand{\logg}{\ensuremath{\log{g}}}
\newcommand{\vsini}{\ensuremath{v \sin{i}}}
\newcommand{\feh}{\ensuremath{\rm [Fe/H]}}

\newcommand{\rsun}{\ensuremath{R_\sun}}
\newcommand{\msun}{\ensuremath{M_\sun}}
\newcommand{\lsun}{\ensuremath{L_\sun}}

\newcommand{\rstar}{\ensuremath{R_\star}}
\newcommand{\mstar}{\ensuremath{M_\star}}
\newcommand{\lstar}{\ensuremath{L_\star}}

\newcommand{\teffstar}{\ensuremath{T_{\rm eff\star}}}
\newcommand{\rhostar}{\ensuremath{\rho_\star}}
\newcommand{\loggstar}{\ensuremath{\log{g_{\star}}}}

\newcommand{\rpl}{\ensuremath{R_{p}}}
\newcommand{\mpl}{\ensuremath{M_{p}}}

\newcommand{\rhopl}{\ensuremath{\rho_{p}}}

\newcommand{\arstar}{\ensuremath{a/\rstar}}
\newcommand{\zrstar}{\ensuremath{\zeta/\rstar}}

\newcommand{\rjup}{\ensuremath{R_{\rm J}}}
\newcommand{\mjup}{\ensuremath{M_{\rm J}}}

\newcommand{\reffigl}[1]{Figure~\ref{fig:#1}}
\newcommand{\refsecl}[1]{\mbox{Section \ref{sec:#1}}}

\newcommand{\reftabl}[1]{Table~\ref{tab:#1}}

\newcommand{\reftabls}[2]{Tables~\ref{tab:#1}-\ref{tab:#2}}

\newcommand{\loopand}{\ifnum\value{planetcounter}=2 and \else\fi}
\newcommand{\loopcomma}{\ifnum\value{planetcounter}<2 ,\else. \fi}
\newcommand{\loopcommanoperiod}{\ifnum\value{planetcounter}<2 ,\else \space\fi}
\newcommand{\loopcommanospace}{\ifnum\value{planetcounter}<2 ,\else \fi}

\newcommand{\hatcurhtrxxxxxA}{HATS579-019}                             % Original HTR name of target
\newcommand{\hatcurfieldxxxxxA}{\ensuremath{string}}                   % HTR field
\newcommand{\hatcurCCraxxxxxA}{\ensuremath{19^{\mathrm h}17^{\mathrm m}36.24{\mathrm s}}}                            % Right Ascension
\newcommand{\hatcurCCdecxxxxxA}{\ensuremath{-22{\arcdeg}23{\arcmin}23.7{\arcsec}}}                           % Declination
\newcommand{\hatcurCCmagxxxxxA}{14.018}                                % apparent V-band magnitude
\newcommand{\hatcurCCtwomassxxxxxA}{2MASS~19173618-2223236}            % 2MASS identifier
\newcommand{\hatcurCCgscxxxxxA}{GSC~6308-00430}                        % GSC(1.2) identifier
\newcommand{\hatcurCCtassmvxxxxxA}{\ensuremath{14.018\pm0.081}}        % APASS V-band magnitude
\newcommand{\hatcurCCtassmvshortxxxxxA}{\ensuremath{14.0}}             % APASS V-band magnitude
\newcommand{\hatcurCCtassmBxxxxxA}{\ensuremath{14.657\pm0.019}}        % APASS B-band magnitude
\newcommand{\hatcurCCtassmBshortxxxxxA}{\ensuremath{14.7}}             % APASS B-band magnitude
\newcommand{\hatcurCCtassmIxxxxxA}{\ensuremath{nff\pmnff}}             % TASS I-band magnitude
\newcommand{\hatcurCCtassmIshortxxxxxA}{\ensuremath{0.0}}              % TASS I-band magnitude
\newcommand{\hatcurCCtassmgxxxxxA}{\ensuremath{14.268\pm0.012}}        % APASS g-band magnitude
\newcommand{\hatcurCCtassmgshortxxxxxA}{\ensuremath{14.3}}             % APASS g-band magnitude
\newcommand{\hatcurCCtassmrxxxxxA}{\ensuremath{13.865\pm0.039}}        % APASS r-band magnitude
\newcommand{\hatcurCCtassmrshortxxxxxA}{\ensuremath{13.9}}             % APASS r-band magnitude
\newcommand{\hatcurCCtassmixxxxxA}{\ensuremath{13.506\pm0.234}}          % APASS i-band magnitude
\newcommand{\hatcurCCtassmishortxxxxxA}{\ensuremath{13.5}}             % APASS i-band magnitude
\newcommand{\hatcurCCtwomassJmagxxxxxA}{\ensuremath{12.647\pm0.023}}   % 2MASS ORIG MAG
\newcommand{\hatcurCCtwomassHmagxxxxxA}{\ensuremath{12.315\pm0.025}}   % 2MASS ORIG MAG
\newcommand{\hatcurCCtwomassKmagxxxxxA}{\ensuremath{12.243\pm0.025}}   % 2MASS ORIG MAG
\newcommand{\hatcurCCcitJmagxxxxxA}{\ensuremath{12.662\pm0.023}}       % 2MASS CIT MAG
\newcommand{\hatcurCCcitHmagxxxxxA}{\ensuremath{12.310\pm0.026}}       % 2MASS CIT MAG
\newcommand{\hatcurCCcitKmagxxxxxA}{\ensuremath{12.267\pm0.025}}       % 2MASS CIT MAG
\newcommand{\hatcurCCbbJmagxxxxxA}{\ensuremath{12.714\pm0.025}}        % 2MASS BB MAG
\newcommand{\hatcurCCbbHmagxxxxxA}{\ensuremath{12.331\pm0.026}}        % 2MASS BB MAG
\newcommand{\hatcurCCbbKmagxxxxxA}{\ensuremath{12.287\pm0.025}}        % 2MASS BB MAG
\newcommand{\hatcurCCesoJmagxxxxxA}{\ensuremath{12.716\pm0.027}}       % 2MASS ESO MAG
\newcommand{\hatcurCCesoHmagxxxxxA}{\ensuremath{12.326\pm0.027}}       % 2MASS ESO MAG
\newcommand{\hatcurCCesoKmagxxxxxA}{\ensuremath{12.286\pm0.026}}       % 2MASS ESO MAG
\newcommand{\hatcurCCesoJHmagxxxxxA}{\ensuremath{0.391\pm0.037}}       % 2MASS ESO JH COLOR
\newcommand{\hatcurCCesoJKmagxxxxxA}{\ensuremath{0.431\pm0.037}}       % 2MASS ESO JK COLOR
\newcommand{\hatcurCCesoHKmagxxxxxA}{\ensuremath{0.0400\pm0.0090}}     % 2MASS ESO HK COLOR
\newcommand{\hatcurLCdipxxxxxA}{\ensuremath{12.9}}                     % BLS detected dip (mmag)
\newcommand{\hatcurLCrprstarxxxxxA}{\ensuremath{0.1076\pm0.0028}}      % Rp/R*
\newcommand{\hatcurLCbsqxxxxxA}{\ensuremath{0.041_{-0.029}^{+0.060}}}  % impact parameter square
\newcommand{\hatcurLCimpxxxxxA}{\ensuremath{0.203_{-0.094}^{+0.115}}}  % impact parameter
\newcommand{\hatcurLCzetaxxxxxA}{\ensuremath{12.23\pm0.26}}            % zeta/R*
\newcommand{\hatcurLCdurxxxxxA}{\ensuremath{0.1819\pm0.0039}}          % transit duration (days)
\newcommand{\hatcurLCdurshortxxxxxA}{\ensuremath{0.1819}}              % transit duration (days)
\newcommand{\hatcurLCdurhrxxxxxA}{\ensuremath{4.367\pm0.094}}          % transit duration (hours)
\newcommand{\hatcurLCdurhrshortxxxxxA}{\ensuremath{4.367}}             % transit duration (hours)
\newcommand{\hatcurLCqxxxxxA}{\ensuremath{0.0503\pm0.0011}}            % fractional transit duration (days)
\newcommand{\hatcurLCqshortxxxxxA}{\ensuremath{0.050}}                 % fractional transit duration (days)
\newcommand{\hatcurLCingdurxxxxxA}{\ensuremath{0.0184\pm0.0012}}       % ingress/egress duration (days)
\newcommand{\hatcurLCPxxxxxA}{\ensuremath{3.6191613\pm0.0000099}}      % period (days)
\newcommand{\hatcurLCPprecxxxxxA}{\ensuremath{3.6191613}}              % period (days)
\newcommand{\hatcurLCPshortxxxxxA}{\ensuremath{3.6192}}                % period (days)
\newcommand{\hatcurLCTxxxxxA}{\ensuremath{2456574.9657\pm0.0013}}      % epoch (BJD)
\newcommand{\hatcurLCTAxxxxxA}{\ensuremath{2455272.0675\pm0.0034}}     % TA (BJD)
\newcommand{\hatcurLCTBxxxxxA}{\ensuremath{2456911.5477\pm0.0018}}     % TB (BJD)
\newcommand{\hatcurLChatnetmAxxxxxA}{\ensuremath{13.76551\pm0.00022}}  % HATNet OOT level
\newcommand{\hatcurLCiblendAxxxxxA}{\ensuremath{0.73\pm0.11}}          % HATNet iblend factor
\newcommand{\hatcurLChatnetmBxxxxxA}{\ensuremath{13.76555\pm0.00011}}  % HATNet OOT level
\newcommand{\hatcurLCiblendBxxxxxA}{\ensuremath{0.838\pm0.070}}        % HATNet iblend factor
\newcommand{\hatcurSMEiteffxxxxxA}{\ensuremath{6080\pm150}}            % Ini SME, stellar effective temperature
\newcommand{\hatcurSMEizfehxxxxxA}{\ensuremath{-0.380\pm0.080}}        % Ini SME, stellar metallicity
\newcommand{\hatcurSMEizfehshortxxxxxA}{\ensuremath{-0.38}}            % Ini SME, stellar metallicity
\newcommand{\hatcurSMEiloggxxxxxA}{\ensuremath{4.19\pm0.19}}           % Ini SME, stellar surface gravity
\newcommand{\hatcurSMEivsinxxxxxA}{\ensuremath{3.63\pm0.60}}           % Ini SME, stellar rotational velocity
\newcommand{\hatcurSMEivmacxxxxxA}{\ensuremath{0.0}}                   % Ini SME, stellar macroturbulence
\newcommand{\hatcurSMEivmicxxxxxA}{\ensuremath{0.0}}                   % Ini SME, stellar microturbulence
\newcommand{\hatcurSMEiiteffxxxxxA}{\ensuremath{6060\pm150}}           % Final SME, stellar effective temperature
\newcommand{\hatcurSMEiizfehxxxxxA}{\ensuremath{-0.390\pm0.060}}       % Final SME, stellar metallicity
\newcommand{\hatcurSMEiizfehshortxxxxxA}{\ensuremath{-0.39}}           % Final SME, stellar metallicity
\newcommand{\hatcurSMEiiloggxxxxxA}{\ensuremath{4.140\pm0.030}}        % Final SME, stellar surface gravity
\newcommand{\hatcurSMEiivsinxxxxxA}{\ensuremath{3.8\pm1.0}}            % Final SME, stellar rotational velocity
\newcommand{\hatcurLBizxxxxxA}{\ensuremath{0.1621}}                    % Limb darkening parameters, Gamma1, z-band
\newcommand{\hatcurLBiizxxxxxA}{\ensuremath{0.3369}}                   % Limb darkening parameters, Gamma2, z-band
\newcommand{\hatcurLBiixxxxxA}{\ensuremath{0.2115}}                    % Limb darkening parameters, Gamma1, i-band
\newcommand{\hatcurLBiiixxxxxA}{\ensuremath{0.3464}}                   % Limb darkening parameters, Gamma2, i-band
\newcommand{\hatcurLBiIxxxxxA}{\ensuremath{0.1947}}                    % Limb darkening parameters, Gamma1, I-band
\newcommand{\hatcurLBiiIxxxxxA}{\ensuremath{0.3436}}                   % Limb darkening parameters, Gamma2, I-band
\newcommand{\hatcurLBigxxxxxA}{\ensuremath{0.4393}}                    % Limb darkening parameters, Gamma1, g-band
\newcommand{\hatcurLBiigxxxxxA}{\ensuremath{0.3122}}                   % Limb darkening parameters, Gamma2, g-band
\newcommand{\hatcurLBirxxxxxA}{\ensuremath{0.2778}}                    % Limb darkening parameters, Gamma1, r-band
\newcommand{\hatcurLBiirxxxxxA}{\ensuremath{0.3619}}                   % Limb darkening parameters, Gamma2, r-band
\newcommand{\hatcurLBiRxxxxxA}{\ensuremath{0.2590}}                    % Limb darkening parameters, Gamma1, R-band
\newcommand{\hatcurLBiiRxxxxxA}{\ensuremath{0.3590}}                   % Limb darkening parameters, Gamma2, R-band
\newcommand{\hatcurLBiKxxxxxA}{\ensuremath{-0.0081}}                   % Limb darkening parameters, Gamma1, K-band
\newcommand{\hatcurLBiiKxxxxxA}{\ensuremath{0.3305}}                   % Limb darkening parameters, Gamma2, K-band
\newcommand{\hatcurLBikepxxxxxA}{\ensuremath{0.1000}}                  % Limb darkening parameters, Gamma1, Kep-band
\newcommand{\hatcurLBiikepxxxxxA}{\ensuremath{0.1000}}                 % Limb darkening parameters, Gamma2, Kep-band
\newcommand{\hatcurISOmxxxxxA}{\ensuremath{1.000\pm0.060}}             % stellar mass
\newcommand{\hatcurISOmshortxxxxxA}{\ensuremath{1.00}}                 % stellar mass
\newcommand{\hatcurISOmlongxxxxxA}{\ensuremath{1.000\pm0.060}}         % stellar mass
\newcommand{\hatcurISOrxxxxxA}{\ensuremath{1.444\pm0.057}}             % stellar radius
\newcommand{\hatcurISOrshortxxxxxA}{\ensuremath{1.44}}                 % stellar radius
\newcommand{\hatcurISOrlongxxxxxA}{\ensuremath{1.444\pm0.057}}         % stellar radius
\newcommand{\hatcurISOrhoxxxxxA}{\ensuremath{0.471_{-0.052}^{+0.037}}} % stellar density (cgs)
\newcommand{\hatcurISOrholongxxxxxA}{\ensuremath{0.471_{-0.052}^{+0.037}}} % stellar density (cgs)
\newcommand{\hatcurISOloggxxxxxA}{\ensuremath{4.118\pm0.026}}          % stellar surface gravity from isochrones
\newcommand{\hatcurISOlumxxxxxA}{\ensuremath{2.53\pm0.39}}             % stellar luminosity
\newcommand{\hatcurISOlumshortxxxxxA}{\ensuremath{2.53}}               % stellar luminosity
\newcommand{\hatcurISOmvxxxxxA}{\ensuremath{3.81\pm0.18}}              % stellar absolute magnitude
\newcommand{\hatcurISOvixxxxxA}{\ensuremath{0.612\pm0.041}}            % stellar V-I index
\newcommand{\hatcurISOagexxxxxA}{\ensuremath{7.7_{-1.6}^{+2.2}}}       % stellar age
\newcommand{\hatcurISOsigmaxxxxxA}{\ensuremath{0.00050\pm0.00010}}     % system mass-correction sigma parameter
\newcommand{\hatcurISOMJxxxxxA}{\ensuremath{2.79\pm0.12}}              % stellar absolute J magnitude
\newcommand{\hatcurISOMHxxxxxA}{\ensuremath{2.49\pm0.10}}              % stellar absolute H magnitude
\newcommand{\hatcurISOMKxxxxxA}{\ensuremath{2.44\pm0.10}}              % stellar absolute K magnitude
\newcommand{\hatcurISOJKxxxxxA}{\ensuremath{0.360\pm0.030}}            % J-K color index from isochrones.
\newcommand{\hatcurISOspecxxxxxA}{G0}                                  % stellar spectral type
\newcommand{\hatcurRVKxxxxxA}{\ensuremath{112\pm15}}                   % RV semi-amplitude [m/s]
\newcommand{\hatcurRVrkxxxxxA}{\ensuremath{0\pm0}}                     % sqrt(e)*cos(omega)
\newcommand{\hatcurRVrhxxxxxA}{\ensuremath{0\pm0}}                     % sqrt(e)*sin(omega)
\newcommand{\hatcurRVkxxxxxA}{\ensuremath{0\pm0}}                      % e*cos(omega)
\newcommand{\hatcurRVhxxxxxA}{\ensuremath{0\pm0}}                      % e*sin(omega)
\newcommand{\hatcurRVtronexxxxxA}{\ensuremath{0\pm0}}                  % RV linear trend tr1 factor
\newcommand{\hatcurRVtrtwoxxxxxA}{\ensuremath{0\pm0}}                  % RV linear trend tr2 factor
\newcommand{\hatcurRVgammaAxxxxxA}{\ensuremath{-58412\pm30}}           % RV gamma velocity, relative scale
\newcommand{\hatcurRVjitterAxxxxxA}{\ensuremath{0\pm12}}               % RV jitter (m/s)
\newcommand{\hatcurRVjitterAtwosiglimxxxxxA}{\ensuremath{<26.3}}       % RV jitter (m/s)
\newcommand{\hatcurRVfitrmsAxxxxxA}{\ensuremath{0.0}}                  % RVfitrms
\newcommand{\hatcurRVgammaBxxxxxA}{\ensuremath{-58324\pm12}}           % RV gamma velocity, relative scale
\newcommand{\hatcurRVjitterBxxxxxA}{\ensuremath{0\pm10}}               % RV jitter (m/s)
\newcommand{\hatcurRVjitterBtwosiglimxxxxxA}{\ensuremath{<22.1}}       % RV jitter (m/s)
\newcommand{\hatcurRVfitrmsBxxxxxA}{\ensuremath{0.0}}                  % RVfitrms
\newcommand{\hatcurRVeccenxxxxxA}{\ensuremath{0\pm0}}                  % eccentricity
\newcommand{\hatcurRVeccentwosiglimxxxxxA}{\ensuremath{<0.000}}        % eccentricity
\newcommand{\hatcurRVomegaxxxxxA}{\ensuremath{0\pm0}}                  % argument of pericenter
\newcommand{\hatcurPPixxxxxA}{\ensuremath{88.31\pm0.86}}               % orbital inclination
\newcommand{\hatcurPPgxxxxxA}{\ensuremath{9.1_{-1.3}^{+1.7}}}          % planetary surface gravity (m/s^2)
\newcommand{\hatcurPPloggxxxxxA}{\ensuremath{2.959\pm0.071}}           % planetary surface gravity (log cgs)
\newcommand{\hatcurPParxxxxxA}{\ensuremath{6.88_{-0.27}^{+0.18}}}      % relative orbital radius (a/R*)
\newcommand{\hatcurPParelxxxxxA}{\ensuremath{0.04614\pm0.00093}}       % semimajor axis (AU)
\newcommand{\hatcurPPrhoxxxxxA}{\ensuremath{0.299_{-0.050}^{+0.071}}}  % planetary density (cgs)
\newcommand{\hatcurPPmxxxxxA}{\ensuremath{0.85\pm0.12}}                % planetary mass (M_jup)
\newcommand{\hatcurPPmshortxxxxxA}{\ensuremath{0.85}}                  % planetary mass (M_jup)
\newcommand{\hatcurPPmlongxxxxxA}{\ensuremath{0.85\pm0.12}}            % planetary mass (M_jup)
\newcommand{\hatcurPPmexxxxxA}{\ensuremath{269\pm39}}                  % planetary mass (M_earth)
\newcommand{\hatcurPPmeshortxxxxxA}{\ensuremath{268.8}}                % planetary mass (M_earth)
\newcommand{\hatcurPPmelongxxxxxA}{\ensuremath{269\pm39}}              % planetary mass (M_earth)
\newcommand{\hatcurPPrxxxxxA}{\ensuremath{1.510\pm0.078}}              % planetary radius (R_jup)
\newcommand{\hatcurPPrshortxxxxxA}{\ensuremath{1.51}}                  % planetary radius (R_jup)
\newcommand{\hatcurPPrlongxxxxxA}{\ensuremath{1.510\pm0.078}}          % planetary radius (R_jup)
\newcommand{\hatcurPPrexxxxxA}{\ensuremath{16.92\pm0.88}}              % planetary radius (R_earth)
\newcommand{\hatcurPPreshortxxxxxA}{\ensuremath{16.9}}                 % planetary radius (R_earth)
\newcommand{\hatcurPPrelongxxxxxA}{\ensuremath{16.92\pm0.88}}          % planetary radius (R_earth)
\newcommand{\hatcurPPmrcorrxxxxxA}{\ensuremath{0.15}}                  % mass/radius correlation
\newcommand{\hatcurPPteffxxxxxA}{\ensuremath{1637\pm48}}               % planetary temperature (K)
\newcommand{\hatcurPPthetaxxxxxA}{\ensuremath{0.0514\pm0.0076}}        % Safranov number
\newcommand{\hatcurPPfluxperixxxxxA}{\ensuremath{1.62\pm0.19}}         % flux @ periastron (CGS)
\newcommand{\hatcurPPfluxperidimxxxxxA}{\ensuremath{9}}                % flux @ periastron (CGS) units.
\newcommand{\hatcurPPfluxapxxxxxA}{\ensuremath{1.62\pm0.19}}           % flux @ apastron (CGS)
\newcommand{\hatcurPPfluxapdimxxxxxA}{\ensuremath{9}}                  % flux @ apastron (CGS) units.
\newcommand{\hatcurPPfluxavgxxxxxA}{\ensuremath{1.62\pm0.19}}          % flux on average (CGS)
\newcommand{\hatcurPPfluxavgdimxxxxxA}{\ensuremath{9}}                 % flux average (CGS) units.
\newcommand{\hatcurPPfluxavglogxxxxxA}{\ensuremath{9.211\pm0.050}}     % log10 flux on average (CGS)
\newcommand{\hatcurXsecphasexxxxxA}{\ensuremath{0\pm0}}                % Phase of secondary eclipse
\newcommand{\hatcurXsecondaryxxxxxA}{\ensuremath{2456576.7752\pm0.0013}} % Secondary eclipse epoch
\newcommand{\hatcurXsecdurxxxxxA}{\ensuremath{0.1819\pm0.0039}}        % sec eclipse duration (days)
\newcommand{\hatcurXsecingdurxxxxxA}{\ensuremath{0.0184\pm0.0012}}     % sec I/E duration (days)
\newcommand{\hatcurPPphiconjxxxxxA}{\ensuremath{0\pm0}}                % phase diff between conjunction and periastron
\newcommand{\hatcurPPperixxxxxA}{\ensuremath{2456574.0609\pm0.0013}}   % time of periastron passage.
\newcommand{\hatcurPPaequivxxxxxA}{\ensuremath{0.0290\pm0.0017}}       % equivalent semi-major axis
\newcommand{\hatcurPPtcircxxxxxA}{\ensuremath{85_{-20}^{+27}}}         % circularization timescale
\newcommand{\hatcurPPtinfallxxxxxA}{\ensuremath{610\pm130}}            % infall timescale
\newcommand{\hatcurXdistxxxxxA}{\ensuremath{933\pm45}}                 % distance (pc), no reddenning correction
\newcommand{\hatcurXAvxxxxxA}{\ensuremath{0.42\pm0.15}}                % Av (mag)
\newcommand{\hatcurXdistredxxxxxA}{\ensuremath{906\pm41}}              % distance with Av correction (pc)
\newcommand{\hatcurXEBVxxxxxA}{\ensuremath{0.135\pm0.048}}             % E(B-V) (mag)
\newcommand{\hatcurXmvisoredxxxxxA}{\ensuremath{14.022\pm0.074}}       % Expected m_v with reddening (mag)
\newcommand{\hatcurXmiisoredxxxxxA}{\ensuremath{13.190\pm0.031}}       % Expected m_i with reddening (mag)
\newcommand{\hatcurXmjisoredxxxxxA}{\ensuremath{12.700\pm0.018}}       % Expected m_j with reddening (mag)
\newcommand{\hatcurXmhisoredxxxxxA}{\ensuremath{12.356\pm0.016}}       % Expected m_h with reddening (mag)
\newcommand{\hatcurXmkisoredxxxxxA}{\ensuremath{12.273\pm0.018}}       % Expected m_k with reddening (mag)
\newcommand{\hatcurXviisoredxxxxxA}{\ensuremath{0.832\pm0.052}}        % Expected V-I with reddening (mag)
\newcommand{\hatcurXvkisoredxxxxxA}{\ensuremath{1.750\pm0.082}}        % Expected V-K with reddening (mag)
\newcommand{\hatcurXjhisoredxxxxxA}{\ensuremath{0.344\pm0.016}}        % Expected J-H with reddening (mag)
\newcommand{\hatcurXjkisoredxxxxxA}{\ensuremath{0.427\pm0.018}}        % Expected J-K with reddening (mag)
\newcommand{\hatcurCCpmraxxxxxA}{\ensuremath{4.5\pm1.9}}               % proper motion, in RA
\newcommand{\hatcurCCpmdecxxxxxA}{\ensuremath{-21.5\pm1.9}}            % proper motion, in DEC
\newcommand{\hatcurCCpmxxxxxA}{\ensuremath{22.0\pm2.7}}                % proper motion

\newcommand{\hatcurhtrxxxxxB}{HATS579-023}                             % Original HTR name of target
\newcommand{\hatcurfieldxxxxxB}{\ensuremath{string}}                   % HTR field
\newcommand{\hatcurCCraxxxxxB}{\ensuremath{19^{\mathrm h}16^{\mathrm m}48.72{\mathrm s}}}                            % Right Ascension
\newcommand{\hatcurCCdecxxxxxB}{\ensuremath{-19{\arcdeg}21{\arcmin}21.2{\arcsec}}}                           % Declination
\newcommand{\hatcurCCmagxxxxxB}{12.767}                                % apparent V-band magnitude
\newcommand{\hatcurCCtwomassxxxxxB}{2MASS~19164857-1921212}            % 2MASS identifier
\newcommand{\hatcurCCgscxxxxxB}{GSC~6304-00396}                        % GSC(1.2) identifier
\newcommand{\hatcurCCtassmvxxxxxB}{\ensuremath{12.756\pm0.021}}        % APASS V-band magnitude
\newcommand{\hatcurCCtassmvshortxxxxxB}{\ensuremath{12.8}}             % APASS V-band magnitude
\newcommand{\hatcurCCtassmBxxxxxB}{\ensuremath{13.258\pm0.043}}        % APASS B-band magnitude
\newcommand{\hatcurCCtassmBshortxxxxxB}{\ensuremath{13.3}}             % APASS B-band magnitude
\newcommand{\hatcurCCtassmIxxxxxB}{\ensuremath{100\pm100}}             % TASS I-band magnitude
\newcommand{\hatcurCCtassmIshortxxxxxB}{\ensuremath{100.0}}            % TASS I-band magnitude
\newcommand{\hatcurCCtassmgxxxxxB}{\ensuremath{12.972\pm0.005}}        % APASS g-band magnitude
\newcommand{\hatcurCCtassmgshortxxxxxB}{\ensuremath{13.0}}             % APASS g-band magnitude
\newcommand{\hatcurCCtassmrxxxxxB}{\ensuremath{12.654\pm0.015}}        % APASS r-band magnitude
\newcommand{\hatcurCCtassmrshortxxxxxB}{\ensuremath{12.7}}             % APASS r-band magnitude
\newcommand{\hatcurCCtassmixxxxxB}{\ensuremath{12.595\pm0.021}}        % APASS i-band magnitude
\newcommand{\hatcurCCtassmishortxxxxxB}{\ensuremath{12.6}}             % APASS i-band magnitude
\newcommand{\hatcurCCtwomassJmagxxxxxB}{\ensuremath{11.725\pm0.023}}   % 2MASS ORIG MAG
\newcommand{\hatcurCCtwomassHmagxxxxxB}{\ensuremath{11.507\pm0.024}}   % 2MASS ORIG MAG
\newcommand{\hatcurCCtwomassKmagxxxxxB}{\ensuremath{11.391\pm0.023}}   % 2MASS ORIG MAG
\newcommand{\hatcurCCcitJmagxxxxxB}{\ensuremath{11.744\pm0.024}}       % 2MASS CIT MAG
\newcommand{\hatcurCCcitHmagxxxxxB}{\ensuremath{11.501\pm0.024}}       % 2MASS CIT MAG
\newcommand{\hatcurCCcitKmagxxxxxB}{\ensuremath{11.415\pm0.023}}       % 2MASS CIT MAG
\newcommand{\hatcurCCbbJmagxxxxxB}{\ensuremath{11.790\pm0.024}}        % 2MASS BB MAG
\newcommand{\hatcurCCbbHmagxxxxxB}{\ensuremath{11.523\pm0.025}}        % 2MASS BB MAG
\newcommand{\hatcurCCbbKmagxxxxxB}{\ensuremath{11.435\pm0.023}}        % 2MASS BB MAG
\newcommand{\hatcurCCesoJmagxxxxxB}{\ensuremath{11.792\pm0.025}}       % 2MASS ESO MAG
\newcommand{\hatcurCCesoHmagxxxxxB}{\ensuremath{11.520\pm0.030}}       % 2MASS ESO MAG
\newcommand{\hatcurCCesoKmagxxxxxB}{\ensuremath{11.434\pm0.024}}       % 2MASS ESO MAG
\newcommand{\hatcurCCesoJHmagxxxxxB}{\ensuremath{0.272\pm0.038}}       % 2MASS ESO JH COLOR
\newcommand{\hatcurCCesoJKmagxxxxxB}{\ensuremath{0.3580\pm0.0090}}     % 2MASS ESO JK COLOR
\newcommand{\hatcurCCesoHKmagxxxxxB}{\ensuremath{0.086\pm0.038}}       % 2MASS ESO HK COLOR
\newcommand{\hatcurLCdipxxxxxB}{\ensuremath{5.1}}                      % BLS detected dip (mmag)
\newcommand{\hatcurLCrprstarxxxxxB}{\ensuremath{0.0630\pm0.0022}}      % Rp/R*
\newcommand{\hatcurLCbsqxxxxxB}{\ensuremath{0.347_{-0.132}^{+0.099}}}  % impact parameter square
\newcommand{\hatcurLCimpxxxxxB}{\ensuremath{0.589_{-0.126}^{+0.079}}}  % impact parameter
\newcommand{\hatcurLCzetaxxxxxB}{\ensuremath{11.54\pm0.12}}            % zeta/R*
\newcommand{\hatcurLCdurxxxxxB}{\ensuremath{0.1899\pm0.0031}}          % transit duration (days)
\newcommand{\hatcurLCdurshortxxxxxB}{\ensuremath{0.1899}}              % transit duration (days)
\newcommand{\hatcurLCdurhrxxxxxB}{\ensuremath{4.558\pm0.075}}          % transit duration (hours)
\newcommand{\hatcurLCdurhrshortxxxxxB}{\ensuremath{4.558}}             % transit duration (hours)
\newcommand{\hatcurLCqxxxxxB}{\ensuremath{0.06040\pm0.00099}}          % fractional transit duration (days)
\newcommand{\hatcurLCqshortxxxxxB}{\ensuremath{0.060}}                 % fractional transit duration (days)
\newcommand{\hatcurLCingdurxxxxxB}{\ensuremath{0.0167\pm0.0033}}       % ingress/egress duration (days)
\newcommand{\hatcurLCPxxxxxB}{\ensuremath{3.142833\pm0.000011}}        % period (days)
\newcommand{\hatcurLCPprecxxxxxB}{\ensuremath{3.1428330}}              % period (days)
\newcommand{\hatcurLCPshortxxxxxB}{\ensuremath{3.1428}}                % period (days)
\newcommand{\hatcurLCTxxxxxB}{\ensuremath{2456798.9556\pm0.0012}}      % epoch (BJD)
\newcommand{\hatcurLCTAxxxxxB}{\ensuremath{2455271.5387\pm0.0047}}     % TA (BJD)
\newcommand{\hatcurLCTBxxxxxB}{\ensuremath{2457169.8097\pm0.0023}}     % TB (BJD)
\newcommand{\hatcurLChatnetmxxxxxB}{\ensuremath{12.687820\pm0.000077}} % HATNet OOT level
\newcommand{\hatcurLCiblendxxxxxB}{\ensuremath{0.745\pm0.090}}         % HATNet iblend factor
\newcommand{\hatcurSMEiteffxxxxxB}{\ensuremath{6315\pm67}}             % Ini SME, stellar effective temperature
\newcommand{\hatcurSMEizfehxxxxxB}{\ensuremath{-0.290\pm0.020}}        % Ini SME, stellar metallicity
\newcommand{\hatcurSMEizfehshortxxxxxB}{\ensuremath{-0.29}}            % Ini SME, stellar metallicity
\newcommand{\hatcurSMEiloggxxxxxB}{\ensuremath{3.64\pm0.12}}           % Ini SME, stellar surface gravity
\newcommand{\hatcurSMEivsinxxxxxB}{\ensuremath{5.89\pm0.50}}           % Ini SME, stellar rotational velocity
\newcommand{\hatcurSMEivmacxxxxxB}{\ensuremath{0.0}}                   % Ini SME, stellar macroturbulence
\newcommand{\hatcurSMEivmicxxxxxB}{\ensuremath{0.0}}                   % Ini SME, stellar microturbulence
\newcommand{\hatcurSMEiiteffxxxxxB}{\ensuremath{6408\pm75}}            % Final SME, stellar effective temperature
\newcommand{\hatcurSMEiizfehxxxxxB}{\ensuremath{-0.100\pm0.040}}       % Final SME, stellar metallicity
\newcommand{\hatcurSMEiizfehshortxxxxxB}{\ensuremath{-0.1}}            % Final SME, stellar metallicity
\newcommand{\hatcurSMEiiloggxxxxxB}{\ensuremath{4\pm0}}                % Final SME, stellar surface gravity
\newcommand{\hatcurSMEiivsinxxxxxB}{\ensuremath{6.57\pm0.50}}          % Final SME, stellar rotational velocity
\newcommand{\hatcurLBizxxxxxB}{\ensuremath{0.1276}}                    % Limb darkening parameters, Gamma1, z-band
\newcommand{\hatcurLBiizxxxxxB}{\ensuremath{0.3601}}                   % Limb darkening parameters, Gamma2, z-band
\newcommand{\hatcurLBiixxxxxB}{\ensuremath{0.1762}}                    % Limb darkening parameters, Gamma1, i-band
\newcommand{\hatcurLBiiixxxxxB}{\ensuremath{0.3706}}                   % Limb darkening parameters, Gamma2, i-band
\newcommand{\hatcurLBiIxxxxxB}{\ensuremath{0.1580}}                    % Limb darkening parameters, Gamma1, I-band
\newcommand{\hatcurLBiiIxxxxxB}{\ensuremath{0.3685}}                   % Limb darkening parameters, Gamma2, I-band
\newcommand{\hatcurLBigxxxxxB}{\ensuremath{0.4153}}                    % Limb darkening parameters, Gamma1, g-band
\newcommand{\hatcurLBiigxxxxxB}{\ensuremath{0.3276}}                   % Limb darkening parameters, Gamma2, g-band
\newcommand{\hatcurLBirxxxxxB}{\ensuremath{0.2509}}                    % Limb darkening parameters, Gamma1, r-band
\newcommand{\hatcurLBiirxxxxxB}{\ensuremath{0.3789}}                   % Limb darkening parameters, Gamma2, r-band
\newcommand{\hatcurLBiRxxxxxB}{\ensuremath{0.2295}}                    % Limb darkening parameters, Gamma1, R-band
\newcommand{\hatcurLBiiRxxxxxB}{\ensuremath{0.3783}}                   % Limb darkening parameters, Gamma2, R-band
\newcommand{\hatcurLBikepxxxxxB}{\ensuremath{0.1000}}                  % Limb darkening parameters, Gamma1, Kep-band
\newcommand{\hatcurLBiikepxxxxxB}{\ensuremath{0.1000}}                 % Limb darkening parameters, Gamma2, Kep-band
\newcommand{\hatcurISOmxxxxxB}{\ensuremath{1.489\pm0.071}}             % stellar mass
\newcommand{\hatcurISOmshortxxxxxB}{\ensuremath{1.49}}                 % stellar mass
\newcommand{\hatcurISOmlongxxxxxB}{\ensuremath{1.489\pm0.071}}         % stellar mass
\newcommand{\hatcurISOrxxxxxB}{\ensuremath{2.21\pm0.21}}               % stellar radius
\newcommand{\hatcurISOrshortxxxxxB}{\ensuremath{2.21}}                 % stellar radius
\newcommand{\hatcurISOrlongxxxxxB}{\ensuremath{2.21\pm0.21}}           % stellar radius
\newcommand{\hatcurISOrhoxxxxxB}{\ensuremath{0.196_{-0.044}^{+0.057}}} % stellar density (cgs)
\newcommand{\hatcurISOrholongxxxxxB}{\ensuremath{0.196_{-0.044}^{+0.057}}} % stellar density (cgs)
\newcommand{\hatcurISOloggxxxxxB}{\ensuremath{3.923\pm0.065}}          % stellar surface gravity from isochrones
\newcommand{\hatcurISOlumxxxxxB}{\ensuremath{7.3\pm1.5}}               % stellar luminosity
\newcommand{\hatcurISOlumshortxxxxxB}{\ensuremath{7.35}}               % stellar luminosity
\newcommand{\hatcurISOmvxxxxxB}{\ensuremath{2.58\pm0.22}}              % stellar absolute magnitude
\newcommand{\hatcurISOvixxxxxB}{\ensuremath{0.519\pm0.019}}            % stellar V-I index
\newcommand{\hatcurISOagexxxxxB}{\ensuremath{2.36\pm0.31}}             % stellar age
\newcommand{\hatcurISOsigmaxxxxxB}{\ensuremath{0.00060\pm0.00013}}     % system mass-correction sigma parameter
\newcommand{\hatcurISOMJxxxxxB}{\ensuremath{1.75\pm0.21}}              % stellar absolute J magnitude
\newcommand{\hatcurISOMHxxxxxB}{\ensuremath{1.51\pm0.21}}              % stellar absolute H magnitude
\newcommand{\hatcurISOMKxxxxxB}{\ensuremath{1.47\pm0.21}}              % stellar absolute K magnitude
\newcommand{\hatcurISOJKxxxxxB}{\ensuremath{0.280\pm0.020}}            % J-K color index from isochrones.
\newcommand{\hatcurISOspecxxxxxB}{F}                                   % stellar spectral type
\newcommand{\hatcurRVKxxxxxB}{\ensuremath{250.6\pm9.4}}                % RV semi-amplitude [m/s]
\newcommand{\hatcurRVrkxxxxxB}{\ensuremath{0\pm0}}                     % sqrt(e)*cos(omega)
\newcommand{\hatcurRVrhxxxxxB}{\ensuremath{0\pm0}}                     % sqrt(e)*sin(omega)
\newcommand{\hatcurRVkxxxxxB}{\ensuremath{0\pm0}}                      % e*cos(omega)
\newcommand{\hatcurRVhxxxxxB}{\ensuremath{0\pm0}}                      % e*sin(omega)
\newcommand{\hatcurRVtronexxxxxB}{\ensuremath{0\pm0}}                  % RV linear trend tr1 factor
\newcommand{\hatcurRVtrtwoxxxxxB}{\ensuremath{0\pm0}}                  % RV linear trend tr2 factor
\newcommand{\hatcurRVgammaAxxxxxB}{\ensuremath{-21661\pm12}}           % RV gamma velocity, relative scale
\newcommand{\hatcurRVjitterAxxxxxB}{\ensuremath{31\pm16}}              % RV jitter (m/s)
\newcommand{\hatcurRVjitterAtwosiglimxxxxxB}{\ensuremath{<56.4}}       % RV jitter (m/s)
\newcommand{\hatcurRVfitrmsAxxxxxB}{\ensuremath{0.0}}                  % RVfitrms
\newcommand{\hatcurRVgammaBxxxxxB}{\ensuremath{71.3\pm6.0}}            % RV gamma velocity, relative scale
\newcommand{\hatcurRVjitterBxxxxxB}{\ensuremath{0.0\pm7.6}}            % RV jitter (m/s)
\newcommand{\hatcurRVjitterBtwosiglimxxxxxB}{\ensuremath{<20.0}}       % RV jitter (m/s)
\newcommand{\hatcurRVfitrmsBxxxxxB}{\ensuremath{0.0}}                  % RVfitrms
\newcommand{\hatcurRVeccenxxxxxB}{\ensuremath{0\pm0}}                  % eccentricity
\newcommand{\hatcurRVeccentwosiglimxxxxxB}{\ensuremath{<0.000}}        % eccentricity
\newcommand{\hatcurRVomegaxxxxxB}{\ensuremath{0\pm0}}                  % argument of pericenter
\newcommand{\hatcurPPixxxxxB}{\ensuremath{82.7\pm1.9}}                 % orbital inclination
\newcommand{\hatcurPPgxxxxxB}{\ensuremath{32.1_{-5.7}^{+8.5}}}         % planetary surface gravity (m/s^2)
\newcommand{\hatcurPPloggxxxxxB}{\ensuremath{3.506\pm0.095}}           % planetary surface gravity (log cgs)
\newcommand{\hatcurPParxxxxxB}{\ensuremath{4.67\pm0.38}}               % relative orbital radius (a/R*)
\newcommand{\hatcurPParelxxxxxB}{\ensuremath{0.04795\pm0.00077}}       % semimajor axis (AU)
\newcommand{\hatcurPPrhoxxxxxB}{\ensuremath{1.19_{-0.32}^{+0.54}}}     % planetary density (cgs)
\newcommand{\hatcurPPmxxxxxB}{\ensuremath{2.38\pm0.11}}                % planetary mass (M_jup)
\newcommand{\hatcurPPmshortxxxxxB}{\ensuremath{2.38}}                  % planetary mass (M_jup)
\newcommand{\hatcurPPmlongxxxxxB}{\ensuremath{2.38\pm0.11}}            % planetary mass (M_jup)
\newcommand{\hatcurPPmexxxxxB}{\ensuremath{757\pm36}}                  % planetary mass (M_earth)
\newcommand{\hatcurPPmeshortxxxxxB}{\ensuremath{757.4}}                % planetary mass (M_earth)
\newcommand{\hatcurPPmelongxxxxxB}{\ensuremath{757\pm36}}              % planetary mass (M_earth)
\newcommand{\hatcurPPrxxxxxB}{\ensuremath{1.35\pm0.17}}                % planetary radius (R_jup)
\newcommand{\hatcurPPrshortxxxxxB}{\ensuremath{1.35}}                  % planetary radius (R_jup)
\newcommand{\hatcurPPrlongxxxxxB}{\ensuremath{1.35\pm0.17}}            % planetary radius (R_jup)
\newcommand{\hatcurPPrexxxxxB}{\ensuremath{15.2\pm1.9}}                % planetary radius (R_earth)
\newcommand{\hatcurPPreshortxxxxxB}{\ensuremath{15.2}}                 % planetary radius (R_earth)
\newcommand{\hatcurPPrelongxxxxxB}{\ensuremath{15.2\pm1.9}}            % planetary radius (R_earth)
\newcommand{\hatcurPPmrcorrxxxxxB}{\ensuremath{0.57}}                  % mass/radius correlation
\newcommand{\hatcurPPteffxxxxxB}{\ensuremath{2097\pm89}}               % planetary temperature (K)
\newcommand{\hatcurPPthetaxxxxxB}{\ensuremath{0.112_{-0.012}^{+0.018}}} % Safranov number
\newcommand{\hatcurPPfluxperixxxxxB}{\ensuremath{4.36\pm0.76}}         % flux @ periastron (CGS)
\newcommand{\hatcurPPfluxperidimxxxxxB}{\ensuremath{9}}                % flux @ periastron (CGS) units.
\newcommand{\hatcurPPfluxapxxxxxB}{\ensuremath{4.36\pm0.76}}           % flux @ apastron (CGS)
\newcommand{\hatcurPPfluxapdimxxxxxB}{\ensuremath{9}}                  % flux @ apastron (CGS) units.
\newcommand{\hatcurPPfluxavgxxxxxB}{\ensuremath{4.36\pm0.76}}          % flux on average (CGS)
\newcommand{\hatcurPPfluxavgdimxxxxxB}{\ensuremath{9}}                 % flux average (CGS) units.
\newcommand{\hatcurPPfluxavglogxxxxxB}{\ensuremath{9.640\pm0.074}}     % log10 flux on average (CGS)
\newcommand{\hatcurXsecphasexxxxxB}{\ensuremath{0\pm0}}                % Phase of secondary eclipse
\newcommand{\hatcurXsecondaryxxxxxB}{\ensuremath{2456800.5270\pm0.0012}} % Secondary eclipse epoch
\newcommand{\hatcurXsecdurxxxxxB}{\ensuremath{0.1899\pm0.0031}}        % sec eclipse duration (days)
\newcommand{\hatcurXsecingdurxxxxxB}{\ensuremath{0.0167\pm0.0033}}     % sec I/E duration (days)
\newcommand{\hatcurPPphiconjxxxxxB}{\ensuremath{0\pm0}}                % phase diff between conjunction and periastron
\newcommand{\hatcurPPperixxxxxB}{\ensuremath{2456798.1699\pm0.0012}}   % time of periastron passage.
\newcommand{\hatcurPPaequivxxxxxB}{\ensuremath{0.0177\pm0.0015}}       % equivalent semi-major axis
\newcommand{\hatcurPPtcircxxxxxB}{\ensuremath{290_{-120}^{+240}}}      % circularization timescale
\newcommand{\hatcurPPtinfallxxxxxB}{\ensuremath{41_{-14}^{+21}}}       % infall timescale
\newcommand{\hatcurXdistxxxxxB}{\ensuremath{983\pm95}}                 % distance (pc), no reddenning correction
\newcommand{\hatcurXAvxxxxxB}{\ensuremath{0.228\pm0.068}}              % Av (mag)
\newcommand{\hatcurXdistredxxxxxB}{\ensuremath{981\pm94}}              % distance with Av correction (pc)
\newcommand{\hatcurXEBVxxxxxB}{\ensuremath{0.073\pm0.022}}             % E(B-V) (mag)
\newcommand{\hatcurXmvisoredxxxxxB}{\ensuremath{12.770\pm0.020}}       % Expected m_v with reddening (mag)
\newcommand{\hatcurXmiisoredxxxxxB}{\ensuremath{12.133\pm0.017}}       % Expected m_i with reddening (mag)
\newcommand{\hatcurXmjisoredxxxxxB}{\ensuremath{11.769\pm0.014}}       % Expected m_j with reddening (mag)
\newcommand{\hatcurXmhisoredxxxxxB}{\ensuremath{11.515\pm0.016}}       % Expected m_h with reddening (mag)
\newcommand{\hatcurXmkisoredxxxxxB}{\ensuremath{11.454\pm0.017}}       % Expected m_k with reddening (mag)
\newcommand{\hatcurXviisoredxxxxxB}{\ensuremath{0.637\pm0.021}}        % Expected V-I with reddening (mag)
\newcommand{\hatcurXvkisoredxxxxxB}{\ensuremath{1.316\pm0.027}}        % Expected V-K with reddening (mag)
\newcommand{\hatcurXjhisoredxxxxxB}{\ensuremath{0.2540\pm0.0092}}      % Expected J-H with reddening (mag)
\newcommand{\hatcurXjkisoredxxxxxB}{\ensuremath{0.3150\pm0.0078}}      % Expected J-K with reddening (mag)
\newcommand{\hatcurCCpmraxxxxxB}{\ensuremath{6.2\pm4.0}}               % proper motion, in RA
\newcommand{\hatcurCCpmdecxxxxxB}{\ensuremath{5.2\pm2.7}}              % proper motion, in DEC
\newcommand{\hatcurCCpmxxxxxB}{\ensuremath{8.1\pm4.8}}                 % proper motion

\newcommand{\hatcurCCbbHmag}[1]{\ifnum#1=11 %
\hatcurCCbbHmagxxxxxA
\else
\ifnum#1=12 %
\hatcurCCbbHmagxxxxxB
\else
??????\fi
\fi
}
\newcommand{\hatcurCCbbJmag}[1]{\ifnum#1=11 %
\hatcurCCbbJmagxxxxxA
\else
\ifnum#1=12 %
\hatcurCCbbJmagxxxxxB
\else
??????\fi
\fi
}
\newcommand{\hatcurCCbbKmag}[1]{\ifnum#1=11 %
\hatcurCCbbKmagxxxxxA
\else
\ifnum#1=12 %
\hatcurCCbbKmagxxxxxB
\else
??????\fi
\fi
}
\newcommand{\hatcurCCcitHmag}[1]{\ifnum#1=11 %
\hatcurCCcitHmagxxxxxA
\else
\ifnum#1=12 %
\hatcurCCcitHmagxxxxxB
\else
??????\fi
\fi
}
\newcommand{\hatcurCCcitJmag}[1]{\ifnum#1=11 %
\hatcurCCcitJmagxxxxxA
\else
\ifnum#1=12 %
\hatcurCCcitJmagxxxxxB
\else
??????\fi
\fi
}
\newcommand{\hatcurCCcitKmag}[1]{\ifnum#1=11 %
\hatcurCCcitKmagxxxxxA
\else
\ifnum#1=12 %
\hatcurCCcitKmagxxxxxB
\else
??????\fi
\fi
}
\newcommand{\hatcurCCdec}[1]{\ifnum#1=11 %
\hatcurCCdecxxxxxA
\else
\ifnum#1=12 %
\hatcurCCdecxxxxxB
\else
??????\fi
\fi
}
\newcommand{\hatcurCCesoHKmag}[1]{\ifnum#1=11 %
\hatcurCCesoHKmagxxxxxA
\else
\ifnum#1=12 %
\hatcurCCesoHKmagxxxxxB
\else
??????\fi
\fi
}
\newcommand{\hatcurCCesoHmag}[1]{\ifnum#1=11 %
\hatcurCCesoHmagxxxxxA
\else
\ifnum#1=12 %
\hatcurCCesoHmagxxxxxB
\else
??????\fi
\fi
}
\newcommand{\hatcurCCesoJHmag}[1]{\ifnum#1=11 %
\hatcurCCesoJHmagxxxxxA
\else
\ifnum#1=12 %
\hatcurCCesoJHmagxxxxxB
\else
??????\fi
\fi
}
\newcommand{\hatcurCCesoJKmag}[1]{\ifnum#1=11 %
\hatcurCCesoJKmagxxxxxA
\else
\ifnum#1=12 %
\hatcurCCesoJKmagxxxxxB
\else
??????\fi
\fi
}
\newcommand{\hatcurCCesoJmag}[1]{\ifnum#1=11 %
\hatcurCCesoJmagxxxxxA
\else
\ifnum#1=12 %
\hatcurCCesoJmagxxxxxB
\else
??????\fi
\fi
}
\newcommand{\hatcurCCesoKmag}[1]{\ifnum#1=11 %
\hatcurCCesoKmagxxxxxA
\else
\ifnum#1=12 %
\hatcurCCesoKmagxxxxxB
\else
??????\fi
\fi
}
\newcommand{\hatcurCCgsc}[1]{\ifnum#1=11 %
\hatcurCCgscxxxxxA
\else
\ifnum#1=12 %
\hatcurCCgscxxxxxB
\else
??????\fi
\fi
}
\newcommand{\hatcurCCmag}[1]{\ifnum#1=11 %
\hatcurCCmagxxxxxA
\else
\ifnum#1=12 %
\hatcurCCmagxxxxxB
\else
??????\fi
\fi
}
\newcommand{\hatcurCCpm}[1]{\ifnum#1=11 %
\hatcurCCpmxxxxxA
\else
\ifnum#1=12 %
\hatcurCCpmxxxxxB
\else
??????\fi
\fi
}
\newcommand{\hatcurCCpmdec}[1]{\ifnum#1=11 %
\hatcurCCpmdecxxxxxA
\else
\ifnum#1=12 %
\hatcurCCpmdecxxxxxB
\else
??????\fi
\fi
}
\newcommand{\hatcurCCpmra}[1]{\ifnum#1=11 %
\hatcurCCpmraxxxxxA
\else
\ifnum#1=12 %
\hatcurCCpmraxxxxxB
\else
??????\fi
\fi
}
\newcommand{\hatcurCCra}[1]{\ifnum#1=11 %
\hatcurCCraxxxxxA
\else
\ifnum#1=12 %
\hatcurCCraxxxxxB
\else
??????\fi
\fi
}
\newcommand{\hatcurCCtassmB}[1]{\ifnum#1=11 %
\hatcurCCtassmBxxxxxA
\else
\ifnum#1=12 %
\hatcurCCtassmBxxxxxB
\else
??????\fi
\fi
}
\newcommand{\hatcurCCtassmBshort}[1]{\ifnum#1=11 %
\hatcurCCtassmBshortxxxxxA
\else
\ifnum#1=12 %
\hatcurCCtassmBshortxxxxxB
\else
??????\fi
\fi
}
\newcommand{\hatcurCCtassmg}[1]{\ifnum#1=11 %
\hatcurCCtassmgxxxxxA
\else
\ifnum#1=12 %
\hatcurCCtassmgxxxxxB
\else
??????\fi
\fi
}
\newcommand{\hatcurCCtassmgshort}[1]{\ifnum#1=11 %
\hatcurCCtassmgshortxxxxxA
\else
\ifnum#1=12 %
\hatcurCCtassmgshortxxxxxB
\else
??????\fi
\fi
}
\newcommand{\hatcurCCtassmi}[1]{\ifnum#1=11 %
\hatcurCCtassmixxxxxA
\else
\ifnum#1=12 %
\hatcurCCtassmixxxxxB
\else
??????\fi
\fi
}
\newcommand{\hatcurCCtassmI}[1]{\ifnum#1=11 %
\hatcurCCtassmIxxxxxA
\else
\ifnum#1=12 %
\hatcurCCtassmIxxxxxB
\else
??????\fi
\fi
}
\newcommand{\hatcurCCtassmishort}[1]{\ifnum#1=11 %
\hatcurCCtassmishortxxxxxA
\else
\ifnum#1=12 %
\hatcurCCtassmishortxxxxxB
\else
??????\fi
\fi
}
\newcommand{\hatcurCCtassmIshort}[1]{\ifnum#1=11 %
\hatcurCCtassmIshortxxxxxA
\else
\ifnum#1=12 %
\hatcurCCtassmIshortxxxxxB
\else
??????\fi
\fi
}
\newcommand{\hatcurCCtassmr}[1]{\ifnum#1=11 %
\hatcurCCtassmrxxxxxA
\else
\ifnum#1=12 %
\hatcurCCtassmrxxxxxB
\else
??????\fi
\fi
}
\newcommand{\hatcurCCtassmrshort}[1]{\ifnum#1=11 %
\hatcurCCtassmrshortxxxxxA
\else
\ifnum#1=12 %
\hatcurCCtassmrshortxxxxxB
\else
??????\fi
\fi
}
\newcommand{\hatcurCCtassmv}[1]{\ifnum#1=11 %
\hatcurCCtassmvxxxxxA
\else
\ifnum#1=12 %
\hatcurCCtassmvxxxxxB
\else
??????\fi
\fi
}
\newcommand{\hatcurCCtassmvshort}[1]{\ifnum#1=11 %
\hatcurCCtassmvshortxxxxxA
\else
\ifnum#1=12 %
\hatcurCCtassmvshortxxxxxB
\else
??????\fi
\fi
}
\newcommand{\hatcurCCtwomass}[1]{\ifnum#1=11 %
\hatcurCCtwomassxxxxxA
\else
\ifnum#1=12 %
\hatcurCCtwomassxxxxxB
\else
??????\fi
\fi
}
\newcommand{\hatcurCCtwomassHmag}[1]{\ifnum#1=11 %
\hatcurCCtwomassHmagxxxxxA
\else
\ifnum#1=12 %
\hatcurCCtwomassHmagxxxxxB
\else
??????\fi
\fi
}
\newcommand{\hatcurCCtwomassJmag}[1]{\ifnum#1=11 %
\hatcurCCtwomassJmagxxxxxA
\else
\ifnum#1=12 %
\hatcurCCtwomassJmagxxxxxB
\else
??????\fi
\fi
}
\newcommand{\hatcurCCtwomassKmag}[1]{\ifnum#1=11 %
\hatcurCCtwomassKmagxxxxxA
\else
\ifnum#1=12 %
\hatcurCCtwomassKmagxxxxxB
\else
??????\fi
\fi
}
\newcommand{\hatcurfield}[1]{\ifnum#1=11 %
\hatcurfieldxxxxxA
\else
\ifnum#1=12 %
\hatcurfieldxxxxxB
\else
??????\fi
\fi
}
\newcommand{\hatcurhtr}[1]{\ifnum#1=11 %
\hatcurhtrxxxxxA
\else
\ifnum#1=12 %
\hatcurhtrxxxxxB
\else
??????\fi
\fi
}
\newcommand{\hatcurISOage}[1]{\ifnum#1=11 %
\hatcurISOagexxxxxA
\else
\ifnum#1=12 %
\hatcurISOagexxxxxB
\else
??????\fi
\fi
}
\newcommand{\hatcurISOJK}[1]{\ifnum#1=11 %
\hatcurISOJKxxxxxA
\else
\ifnum#1=12 %
\hatcurISOJKxxxxxB
\else
??????\fi
\fi
}
\newcommand{\hatcurISOlogg}[1]{\ifnum#1=11 %
\hatcurISOloggxxxxxA
\else
\ifnum#1=12 %
\hatcurISOloggxxxxxB
\else
??????\fi
\fi
}
\newcommand{\hatcurISOlum}[1]{\ifnum#1=11 %
\hatcurISOlumxxxxxA
\else
\ifnum#1=12 %
\hatcurISOlumxxxxxB
\else
??????\fi
\fi
}
\newcommand{\hatcurISOlumshort}[1]{\ifnum#1=11 %
\hatcurISOlumshortxxxxxA
\else
\ifnum#1=12 %
\hatcurISOlumshortxxxxxB
\else
??????\fi
\fi
}
\newcommand{\hatcurISOm}[1]{\ifnum#1=11 %
\hatcurISOmxxxxxA
\else
\ifnum#1=12 %
\hatcurISOmxxxxxB
\else
??????\fi
\fi
}
\newcommand{\hatcurISOMH}[1]{\ifnum#1=11 %
\hatcurISOMHxxxxxA
\else
\ifnum#1=12 %
\hatcurISOMHxxxxxB
\else
??????\fi
\fi
}
\newcommand{\hatcurISOMJ}[1]{\ifnum#1=11 %
\hatcurISOMJxxxxxA
\else
\ifnum#1=12 %
\hatcurISOMJxxxxxB
\else
??????\fi
\fi
}
\newcommand{\hatcurISOMK}[1]{\ifnum#1=11 %
\hatcurISOMKxxxxxA
\else
\ifnum#1=12 %
\hatcurISOMKxxxxxB
\else
??????\fi
\fi
}
\newcommand{\hatcurISOmlong}[1]{\ifnum#1=11 %
\hatcurISOmlongxxxxxA
\else
\ifnum#1=12 %
\hatcurISOmlongxxxxxB
\else
??????\fi
\fi
}
\newcommand{\hatcurISOmshort}[1]{\ifnum#1=11 %
\hatcurISOmshortxxxxxA
\else
\ifnum#1=12 %
\hatcurISOmshortxxxxxB
\else
??????\fi
\fi
}
\newcommand{\hatcurISOmv}[1]{\ifnum#1=11 %
\hatcurISOmvxxxxxA
\else
\ifnum#1=12 %
\hatcurISOmvxxxxxB
\else
??????\fi
\fi
}
\newcommand{\hatcurISOr}[1]{\ifnum#1=11 %
\hatcurISOrxxxxxA
\else
\ifnum#1=12 %
\hatcurISOrxxxxxB
\else
??????\fi
\fi
}
\newcommand{\hatcurISOrho}[1]{\ifnum#1=11 %
\hatcurISOrhoxxxxxA
\else
\ifnum#1=12 %
\hatcurISOrhoxxxxxB
\else
??????\fi
\fi
}
\newcommand{\hatcurISOrholong}[1]{\ifnum#1=11 %
\hatcurISOrholongxxxxxA
\else
\ifnum#1=12 %
\hatcurISOrholongxxxxxB
\else
??????\fi
\fi
}
\newcommand{\hatcurISOrlong}[1]{\ifnum#1=11 %
\hatcurISOrlongxxxxxA
\else
\ifnum#1=12 %
\hatcurISOrlongxxxxxB
\else
??????\fi
\fi
}
\newcommand{\hatcurISOrshort}[1]{\ifnum#1=11 %
\hatcurISOrshortxxxxxA
\else
\ifnum#1=12 %
\hatcurISOrshortxxxxxB
\else
??????\fi
\fi
}
\newcommand{\hatcurISOsigma}[1]{\ifnum#1=11 %
\hatcurISOsigmaxxxxxA
\else
\ifnum#1=12 %
\hatcurISOsigmaxxxxxB
\else
??????\fi
\fi
}
\newcommand{\hatcurISOspec}[1]{\ifnum#1=11 %
\hatcurISOspecxxxxxA
\else
\ifnum#1=12 %
\hatcurISOspecxxxxxB
\else
??????\fi
\fi
}
\newcommand{\hatcurISOvi}[1]{\ifnum#1=11 %
\hatcurISOvixxxxxA
\else
\ifnum#1=12 %
\hatcurISOvixxxxxB
\else
??????\fi
\fi
}
\newcommand{\hatcurLBig}[1]{\ifnum#1=11 %
\hatcurLBigxxxxxA
\else
\ifnum#1=12 %
\hatcurLBigxxxxxB
\else
??????\fi
\fi
}
\newcommand{\hatcurLBii}[1]{\ifnum#1=11 %
\hatcurLBiixxxxxA
\else
\ifnum#1=12 %
\hatcurLBiixxxxxB
\else
??????\fi
\fi
}
\newcommand{\hatcurLBiI}[1]{\ifnum#1=11 %
\hatcurLBiIxxxxxA
\else
\ifnum#1=12 %
\hatcurLBiIxxxxxB
\else
??????\fi
\fi
}
\newcommand{\hatcurLBiig}[1]{\ifnum#1=11 %
\hatcurLBiigxxxxxA
\else
\ifnum#1=12 %
\hatcurLBiigxxxxxB
\else
??????\fi
\fi
}
\newcommand{\hatcurLBiii}[1]{\ifnum#1=11 %
\hatcurLBiiixxxxxA
\else
\ifnum#1=12 %
\hatcurLBiiixxxxxB
\else
??????\fi
\fi
}
\newcommand{\hatcurLBiiI}[1]{\ifnum#1=11 %
\hatcurLBiiIxxxxxA
\else
\ifnum#1=12 %
\hatcurLBiiIxxxxxB
\else
??????\fi
\fi
}
\newcommand{\hatcurLBiiK}[1]{\ifnum#1=11 %
\hatcurLBiiKxxxxxA
\else
??????\fi
}
\newcommand{\hatcurLBiikep}[1]{\ifnum#1=11 %
\hatcurLBiikepxxxxxA
\else
\ifnum#1=12 %
\hatcurLBiikepxxxxxB
\else
??????\fi
\fi
}
\newcommand{\hatcurLBiir}[1]{\ifnum#1=11 %
\hatcurLBiirxxxxxA
\else
\ifnum#1=12 %
\hatcurLBiirxxxxxB
\else
??????\fi
\fi
}
\newcommand{\hatcurLBiiR}[1]{\ifnum#1=11 %
\hatcurLBiiRxxxxxA
\else
\ifnum#1=12 %
\hatcurLBiiRxxxxxB
\else
??????\fi
\fi
}
\newcommand{\hatcurLBiiz}[1]{\ifnum#1=11 %
\hatcurLBiizxxxxxA
\else
\ifnum#1=12 %
\hatcurLBiizxxxxxB
\else
??????\fi
\fi
}
\newcommand{\hatcurLBiK}[1]{\ifnum#1=11 %
\hatcurLBiKxxxxxA
\else
??????\fi
}
\newcommand{\hatcurLBikep}[1]{\ifnum#1=11 %
\hatcurLBikepxxxxxA
\else
\ifnum#1=12 %
\hatcurLBikepxxxxxB
\else
??????\fi
\fi
}
\newcommand{\hatcurLBir}[1]{\ifnum#1=11 %
\hatcurLBirxxxxxA
\else
\ifnum#1=12 %
\hatcurLBirxxxxxB
\else
??????\fi
\fi
}
\newcommand{\hatcurLBiR}[1]{\ifnum#1=11 %
\hatcurLBiRxxxxxA
\else
\ifnum#1=12 %
\hatcurLBiRxxxxxB
\else
??????\fi
\fi
}
\newcommand{\hatcurLBiz}[1]{\ifnum#1=11 %
\hatcurLBizxxxxxA
\else
\ifnum#1=12 %
\hatcurLBizxxxxxB
\else
??????\fi
\fi
}
\newcommand{\hatcurLCbsq}[1]{\ifnum#1=11 %
\hatcurLCbsqxxxxxA
\else
\ifnum#1=12 %
\hatcurLCbsqxxxxxB
\else
??????\fi
\fi
}
\newcommand{\hatcurLCdip}[1]{\ifnum#1=11 %
\hatcurLCdipxxxxxA
\else
\ifnum#1=12 %
\hatcurLCdipxxxxxB
\else
??????\fi
\fi
}
\newcommand{\hatcurLCdur}[1]{\ifnum#1=11 %
\hatcurLCdurxxxxxA
\else
\ifnum#1=12 %
\hatcurLCdurxxxxxB
\else
??????\fi
\fi
}
\newcommand{\hatcurLCdurhr}[1]{\ifnum#1=11 %
\hatcurLCdurhrxxxxxA
\else
\ifnum#1=12 %
\hatcurLCdurhrxxxxxB
\else
??????\fi
\fi
}
\newcommand{\hatcurLCdurhrshort}[1]{\ifnum#1=11 %
\hatcurLCdurhrshortxxxxxA
\else
\ifnum#1=12 %
\hatcurLCdurhrshortxxxxxB
\else
??????\fi
\fi
}
\newcommand{\hatcurLCdurshort}[1]{\ifnum#1=11 %
\hatcurLCdurshortxxxxxA
\else
\ifnum#1=12 %
\hatcurLCdurshortxxxxxB
\else
??????\fi
\fi
}
\newcommand{\hatcurLChatnetm}[1]{\ifnum#1=12 %
\hatcurLChatnetmxxxxxB
\else
??????\fi
}
\newcommand{\hatcurLChatnetmA}[1]{\ifnum#1=11 %
\hatcurLChatnetmAxxxxxA
\else
??????\fi
}
\newcommand{\hatcurLChatnetmB}[1]{\ifnum#1=11 %
\hatcurLChatnetmBxxxxxA
\else
??????\fi
}
\newcommand{\hatcurLCiblend}[1]{\ifnum#1=12 %
\hatcurLCiblendxxxxxB
\else
??????\fi
}
\newcommand{\hatcurLCiblendA}[1]{\ifnum#1=11 %
\hatcurLCiblendAxxxxxA
\else
??????\fi
}
\newcommand{\hatcurLCiblendB}[1]{\ifnum#1=11 %
\hatcurLCiblendBxxxxxA
\else
??????\fi
}
\newcommand{\hatcurLCimp}[1]{\ifnum#1=11 %
\hatcurLCimpxxxxxA
\else
\ifnum#1=12 %
\hatcurLCimpxxxxxB
\else
??????\fi
\fi
}
\newcommand{\hatcurLCingdur}[1]{\ifnum#1=11 %
\hatcurLCingdurxxxxxA
\else
\ifnum#1=12 %
\hatcurLCingdurxxxxxB
\else
??????\fi
\fi
}
\newcommand{\hatcurLCP}[1]{\ifnum#1=11 %
\hatcurLCPxxxxxA
\else
\ifnum#1=12 %
\hatcurLCPxxxxxB
\else
??????\fi
\fi
}
\newcommand{\hatcurLCPprec}[1]{\ifnum#1=11 %
\hatcurLCPprecxxxxxA
\else
\ifnum#1=12 %
\hatcurLCPprecxxxxxB
\else
??????\fi
\fi
}
\newcommand{\hatcurLCPshort}[1]{\ifnum#1=11 %
\hatcurLCPshortxxxxxA
\else
\ifnum#1=12 %
\hatcurLCPshortxxxxxB
\else
??????\fi
\fi
}
\newcommand{\hatcurLCq}[1]{\ifnum#1=11 %
\hatcurLCqxxxxxA
\else
\ifnum#1=12 %
\hatcurLCqxxxxxB
\else
??????\fi
\fi
}
\newcommand{\hatcurLCqshort}[1]{\ifnum#1=11 %
\hatcurLCqshortxxxxxA
\else
\ifnum#1=12 %
\hatcurLCqshortxxxxxB
\else
??????\fi
\fi
}
\newcommand{\hatcurLCrprstar}[1]{\ifnum#1=11 %
\hatcurLCrprstarxxxxxA
\else
\ifnum#1=12 %
\hatcurLCrprstarxxxxxB
\else
??????\fi
\fi
}
\newcommand{\hatcurLCT}[1]{\ifnum#1=11 %
\hatcurLCTxxxxxA
\else
\ifnum#1=12 %
\hatcurLCTxxxxxB
\else
??????\fi
\fi
}
\newcommand{\hatcurLCTA}[1]{\ifnum#1=11 %
\hatcurLCTAxxxxxA
\else
\ifnum#1=12 %
\hatcurLCTAxxxxxB
\else
??????\fi
\fi
}
\newcommand{\hatcurLCTB}[1]{\ifnum#1=11 %
\hatcurLCTBxxxxxA
\else
\ifnum#1=12 %
\hatcurLCTBxxxxxB
\else
??????\fi
\fi
}
\newcommand{\hatcurLCzeta}[1]{\ifnum#1=11 %
\hatcurLCzetaxxxxxA
\else
\ifnum#1=12 %
\hatcurLCzetaxxxxxB
\else
??????\fi
\fi
}
\newcommand{\hatcurPPaequiv}[1]{\ifnum#1=11 %
\hatcurPPaequivxxxxxA
\else
\ifnum#1=12 %
\hatcurPPaequivxxxxxB
\else
??????\fi
\fi
}
\newcommand{\hatcurPPar}[1]{\ifnum#1=11 %
\hatcurPParxxxxxA
\else
\ifnum#1=12 %
\hatcurPParxxxxxB
\else
??????\fi
\fi
}
\newcommand{\hatcurPParel}[1]{\ifnum#1=11 %
\hatcurPParelxxxxxA
\else
\ifnum#1=12 %
\hatcurPParelxxxxxB
\else
??????\fi
\fi
}
\newcommand{\hatcurPPfluxap}[1]{\ifnum#1=11 %
\hatcurPPfluxapxxxxxA
\else
\ifnum#1=12 %
\hatcurPPfluxapxxxxxB
\else
??????\fi
\fi
}
\newcommand{\hatcurPPfluxapdim}[1]{\ifnum#1=11 %
\hatcurPPfluxapdimxxxxxA
\else
\ifnum#1=12 %
\hatcurPPfluxapdimxxxxxB
\else
??????\fi
\fi
}
\newcommand{\hatcurPPfluxavg}[1]{\ifnum#1=11 %
\hatcurPPfluxavgxxxxxA
\else
\ifnum#1=12 %
\hatcurPPfluxavgxxxxxB
\else
??????\fi
\fi
}
\newcommand{\hatcurPPfluxavgdim}[1]{\ifnum#1=11 %
\hatcurPPfluxavgdimxxxxxA
\else
\ifnum#1=12 %
\hatcurPPfluxavgdimxxxxxB
\else
??????\fi
\fi
}
\newcommand{\hatcurPPfluxavglog}[1]{\ifnum#1=11 %
\hatcurPPfluxavglogxxxxxA
\else
\ifnum#1=12 %
\hatcurPPfluxavglogxxxxxB
\else
??????\fi
\fi
}
\newcommand{\hatcurPPfluxperi}[1]{\ifnum#1=11 %
\hatcurPPfluxperixxxxxA
\else
\ifnum#1=12 %
\hatcurPPfluxperixxxxxB
\else
??????\fi
\fi
}
\newcommand{\hatcurPPfluxperidim}[1]{\ifnum#1=11 %
\hatcurPPfluxperidimxxxxxA
\else
\ifnum#1=12 %
\hatcurPPfluxperidimxxxxxB
\else
??????\fi
\fi
}
\newcommand{\hatcurPPg}[1]{\ifnum#1=11 %
\hatcurPPgxxxxxA
\else
\ifnum#1=12 %
\hatcurPPgxxxxxB
\else
??????\fi
\fi
}
\newcommand{\hatcurPPi}[1]{\ifnum#1=11 %
\hatcurPPixxxxxA
\else
\ifnum#1=12 %
\hatcurPPixxxxxB
\else
??????\fi
\fi
}
\newcommand{\hatcurPPlogg}[1]{\ifnum#1=11 %
\hatcurPPloggxxxxxA
\else
\ifnum#1=12 %
\hatcurPPloggxxxxxB
\else
??????\fi
\fi
}
\newcommand{\hatcurPPm}[1]{\ifnum#1=11 %
\hatcurPPmxxxxxA
\else
\ifnum#1=12 %
\hatcurPPmxxxxxB
\else
??????\fi
\fi
}
\newcommand{\hatcurPPme}[1]{\ifnum#1=11 %
\hatcurPPmexxxxxA
\else
\ifnum#1=12 %
\hatcurPPmexxxxxB
\else
??????\fi
\fi
}
\newcommand{\hatcurPPmelong}[1]{\ifnum#1=11 %
\hatcurPPmelongxxxxxA
\else
\ifnum#1=12 %
\hatcurPPmelongxxxxxB
\else
??????\fi
\fi
}
\newcommand{\hatcurPPmeshort}[1]{\ifnum#1=11 %
\hatcurPPmeshortxxxxxA
\else
\ifnum#1=12 %
\hatcurPPmeshortxxxxxB
\else
??????\fi
\fi
}
\newcommand{\hatcurPPmlong}[1]{\ifnum#1=11 %
\hatcurPPmlongxxxxxA
\else
\ifnum#1=12 %
\hatcurPPmlongxxxxxB
\else
??????\fi
\fi
}
\newcommand{\hatcurPPmrcorr}[1]{\ifnum#1=11 %
\hatcurPPmrcorrxxxxxA
\else
\ifnum#1=12 %
\hatcurPPmrcorrxxxxxB
\else
??????\fi
\fi
}
\newcommand{\hatcurPPmshort}[1]{\ifnum#1=11 %
\hatcurPPmshortxxxxxA
\else
\ifnum#1=12 %
\hatcurPPmshortxxxxxB
\else
??????\fi
\fi
}
\newcommand{\hatcurPPperi}[1]{\ifnum#1=11 %
\hatcurPPperixxxxxA
\else
\ifnum#1=12 %
\hatcurPPperixxxxxB
\else
??????\fi
\fi
}
\newcommand{\hatcurPPphiconj}[1]{\ifnum#1=11 %
\hatcurPPphiconjxxxxxA
\else
\ifnum#1=12 %
\hatcurPPphiconjxxxxxB
\else
??????\fi
\fi
}
\newcommand{\hatcurPPr}[1]{\ifnum#1=11 %
\hatcurPPrxxxxxA
\else
\ifnum#1=12 %
\hatcurPPrxxxxxB
\else
??????\fi
\fi
}
\newcommand{\hatcurPPre}[1]{\ifnum#1=11 %
\hatcurPPrexxxxxA
\else
\ifnum#1=12 %
\hatcurPPrexxxxxB
\else
??????\fi
\fi
}
\newcommand{\hatcurPPrelong}[1]{\ifnum#1=11 %
\hatcurPPrelongxxxxxA
\else
\ifnum#1=12 %
\hatcurPPrelongxxxxxB
\else
??????\fi
\fi
}
\newcommand{\hatcurPPreshort}[1]{\ifnum#1=11 %
\hatcurPPreshortxxxxxA
\else
\ifnum#1=12 %
\hatcurPPreshortxxxxxB
\else
??????\fi
\fi
}
\newcommand{\hatcurPPrho}[1]{\ifnum#1=11 %
\hatcurPPrhoxxxxxA
\else
\ifnum#1=12 %
\hatcurPPrhoxxxxxB
\else
??????\fi
\fi
}
\newcommand{\hatcurPPrlong}[1]{\ifnum#1=11 %
\hatcurPPrlongxxxxxA
\else
\ifnum#1=12 %
\hatcurPPrlongxxxxxB
\else
??????\fi
\fi
}
\newcommand{\hatcurPPrshort}[1]{\ifnum#1=11 %
\hatcurPPrshortxxxxxA
\else
\ifnum#1=12 %
\hatcurPPrshortxxxxxB
\else
??????\fi
\fi
}
\newcommand{\hatcurPPtcirc}[1]{\ifnum#1=11 %
\hatcurPPtcircxxxxxA
\else
\ifnum#1=12 %
\hatcurPPtcircxxxxxB
\else
??????\fi
\fi
}
\newcommand{\hatcurPPteff}[1]{\ifnum#1=11 %
\hatcurPPteffxxxxxA
\else
\ifnum#1=12 %
\hatcurPPteffxxxxxB
\else
??????\fi
\fi
}
\newcommand{\hatcurPPtheta}[1]{\ifnum#1=11 %
\hatcurPPthetaxxxxxA
\else
\ifnum#1=12 %
\hatcurPPthetaxxxxxB
\else
??????\fi
\fi
}
\newcommand{\hatcurPPtinfall}[1]{\ifnum#1=11 %
\hatcurPPtinfallxxxxxA
\else
\ifnum#1=12 %
\hatcurPPtinfallxxxxxB
\else
??????\fi
\fi
}
\newcommand{\hatcurRVeccen}[1]{\ifnum#1=11 %
\hatcurRVeccenxxxxxA
\else
\ifnum#1=12 %
\hatcurRVeccenxxxxxB
\else
??????\fi
\fi
}
\newcommand{\hatcurRVeccentwosiglim}[1]{\ifnum#1=11 %
\hatcurRVeccentwosiglimxxxxxA
\else
\ifnum#1=12 %
\hatcurRVeccentwosiglimxxxxxB
\else
??????\fi
\fi
}
\newcommand{\hatcurRVfitrmsA}[1]{\ifnum#1=11 %
\hatcurRVfitrmsAxxxxxA
\else
\ifnum#1=12 %
\hatcurRVfitrmsAxxxxxB
\else
??????\fi
\fi
}
\newcommand{\hatcurRVfitrmsB}[1]{\ifnum#1=11 %
\hatcurRVfitrmsBxxxxxA
\else
\ifnum#1=12 %
\hatcurRVfitrmsBxxxxxB
\else
??????\fi
\fi
}
\newcommand{\hatcurRVgammaA}[1]{\ifnum#1=11 %
\hatcurRVgammaAxxxxxA
\else
\ifnum#1=12 %
\hatcurRVgammaAxxxxxB
\else
??????\fi
\fi
}
\newcommand{\hatcurRVgammaB}[1]{\ifnum#1=11 %
\hatcurRVgammaBxxxxxA
\else
\ifnum#1=12 %
\hatcurRVgammaBxxxxxB
\else
??????\fi
\fi
}
\newcommand{\hatcurRVh}[1]{\ifnum#1=11 %
\hatcurRVhxxxxxA
\else
\ifnum#1=12 %
\hatcurRVhxxxxxB
\else
??????\fi
\fi
}
\newcommand{\hatcurRVjitterA}[1]{\ifnum#1=11 %
\hatcurRVjitterAxxxxxA
\else
\ifnum#1=12 %
\hatcurRVjitterAxxxxxB
\else
??????\fi
\fi
}
\newcommand{\hatcurRVjitterAtwosiglim}[1]{\ifnum#1=11 %
\hatcurRVjitterAtwosiglimxxxxxA
\else
\ifnum#1=12 %
\hatcurRVjitterAtwosiglimxxxxxB
\else
??????\fi
\fi
}
\newcommand{\hatcurRVjitterB}[1]{\ifnum#1=11 %
\hatcurRVjitterBxxxxxA
\else
\ifnum#1=12 %
\hatcurRVjitterBxxxxxB
\else
??????\fi
\fi
}
\newcommand{\hatcurRVjitterBtwosiglim}[1]{\ifnum#1=11 %
\hatcurRVjitterBtwosiglimxxxxxA
\else
\ifnum#1=12 %
\hatcurRVjitterBtwosiglimxxxxxB
\else
??????\fi
\fi
}
\newcommand{\hatcurRVk}[1]{\ifnum#1=11 %
\hatcurRVkxxxxxA
\else
\ifnum#1=12 %
\hatcurRVkxxxxxB
\else
??????\fi
\fi
}
\newcommand{\hatcurRVK}[1]{\ifnum#1=11 %
\hatcurRVKxxxxxA
\else
\ifnum#1=12 %
\hatcurRVKxxxxxB
\else
??????\fi
\fi
}
\newcommand{\hatcurRVomega}[1]{\ifnum#1=11 %
\hatcurRVomegaxxxxxA
\else
\ifnum#1=12 %
\hatcurRVomegaxxxxxB
\else
??????\fi
\fi
}
\newcommand{\hatcurRVrh}[1]{\ifnum#1=11 %
\hatcurRVrhxxxxxA
\else
\ifnum#1=12 %
\hatcurRVrhxxxxxB
\else
??????\fi
\fi
}
\newcommand{\hatcurRVrk}[1]{\ifnum#1=11 %
\hatcurRVrkxxxxxA
\else
\ifnum#1=12 %
\hatcurRVrkxxxxxB
\else
??????\fi
\fi
}
\newcommand{\hatcurRVtrone}[1]{\ifnum#1=11 %
\hatcurRVtronexxxxxA
\else
\ifnum#1=12 %
\hatcurRVtronexxxxxB
\else
??????\fi
\fi
}
\newcommand{\hatcurRVtrtwo}[1]{\ifnum#1=11 %
\hatcurRVtrtwoxxxxxA
\else
\ifnum#1=12 %
\hatcurRVtrtwoxxxxxB
\else
??????\fi
\fi
}
\newcommand{\hatcurSMEiilogg}[1]{\ifnum#1=11 %
\hatcurSMEiiloggxxxxxA
\else
\ifnum#1=12 %
\hatcurSMEiiloggxxxxxB
\else
??????\fi
\fi
}
\newcommand{\hatcurSMEiiteff}[1]{\ifnum#1=11 %
\hatcurSMEiiteffxxxxxA
\else
\ifnum#1=12 %
\hatcurSMEiiteffxxxxxB
\else
??????\fi
\fi
}
\newcommand{\hatcurSMEiivsin}[1]{\ifnum#1=11 %
\hatcurSMEiivsinxxxxxA
\else
\ifnum#1=12 %
\hatcurSMEiivsinxxxxxB
\else
??????\fi
\fi
}
\newcommand{\hatcurSMEiizfeh}[1]{\ifnum#1=11 %
\hatcurSMEiizfehxxxxxA
\else
\ifnum#1=12 %
\hatcurSMEiizfehxxxxxB
\else
??????\fi
\fi
}
\newcommand{\hatcurSMEiizfehshort}[1]{\ifnum#1=11 %
\hatcurSMEiizfehshortxxxxxA
\else
\ifnum#1=12 %
\hatcurSMEiizfehshortxxxxxB
\else
??????\fi
\fi
}
\newcommand{\hatcurSMEilogg}[1]{\ifnum#1=11 %
\hatcurSMEiloggxxxxxA
\else
\ifnum#1=12 %
\hatcurSMEiloggxxxxxB
\else
??????\fi
\fi
}
\newcommand{\hatcurSMEiteff}[1]{\ifnum#1=11 %
\hatcurSMEiteffxxxxxA
\else
\ifnum#1=12 %
\hatcurSMEiteffxxxxxB
\else
??????\fi
\fi
}
\newcommand{\hatcurSMEivmac}[1]{\ifnum#1=11 %
\hatcurSMEivmacxxxxxA
\else
\ifnum#1=12 %
\hatcurSMEivmacxxxxxB
\else
??????\fi
\fi
}
\newcommand{\hatcurSMEivmic}[1]{\ifnum#1=11 %
\hatcurSMEivmicxxxxxA
\else
\ifnum#1=12 %
\hatcurSMEivmicxxxxxB
\else
??????\fi
\fi
}
\newcommand{\hatcurSMEivsin}[1]{\ifnum#1=11 %
\hatcurSMEivsinxxxxxA
\else
\ifnum#1=12 %
\hatcurSMEivsinxxxxxB
\else
??????\fi
\fi
}
\newcommand{\hatcurSMEizfeh}[1]{\ifnum#1=11 %
\hatcurSMEizfehxxxxxA
\else
\ifnum#1=12 %
\hatcurSMEizfehxxxxxB
\else
??????\fi
\fi
}
\newcommand{\hatcurSMEizfehshort}[1]{\ifnum#1=11 %
\hatcurSMEizfehshortxxxxxA
\else
\ifnum#1=12 %
\hatcurSMEizfehshortxxxxxB
\else
??????\fi
\fi
}
\newcommand{\hatcurXAv}[1]{\ifnum#1=11 %
\hatcurXAvxxxxxA
\else
\ifnum#1=12 %
\hatcurXAvxxxxxB
\else
??????\fi
\fi
}
\newcommand{\hatcurXdist}[1]{\ifnum#1=11 %
\hatcurXdistxxxxxA
\else
\ifnum#1=12 %
\hatcurXdistxxxxxB
\else
??????\fi
\fi
}
\newcommand{\hatcurXdistred}[1]{\ifnum#1=11 %
\hatcurXdistredxxxxxA
\else
\ifnum#1=12 %
\hatcurXdistredxxxxxB
\else
??????\fi
\fi
}
\newcommand{\hatcurXEBV}[1]{\ifnum#1=11 %
\hatcurXEBVxxxxxA
\else
\ifnum#1=12 %
\hatcurXEBVxxxxxB
\else
??????\fi
\fi
}
\newcommand{\hatcurXjhisored}[1]{\ifnum#1=11 %
\hatcurXjhisoredxxxxxA
\else
\ifnum#1=12 %
\hatcurXjhisoredxxxxxB
\else
??????\fi
\fi
}
\newcommand{\hatcurXjkisored}[1]{\ifnum#1=11 %
\hatcurXjkisoredxxxxxA
\else
\ifnum#1=12 %
\hatcurXjkisoredxxxxxB
\else
??????\fi
\fi
}
\newcommand{\hatcurXmhisored}[1]{\ifnum#1=11 %
\hatcurXmhisoredxxxxxA
\else
\ifnum#1=12 %
\hatcurXmhisoredxxxxxB
\else
??????\fi
\fi
}
\newcommand{\hatcurXmiisored}[1]{\ifnum#1=11 %
\hatcurXmiisoredxxxxxA
\else
\ifnum#1=12 %
\hatcurXmiisoredxxxxxB
\else
??????\fi
\fi
}
\newcommand{\hatcurXmjisored}[1]{\ifnum#1=11 %
\hatcurXmjisoredxxxxxA
\else
\ifnum#1=12 %
\hatcurXmjisoredxxxxxB
\else
??????\fi
\fi
}
\newcommand{\hatcurXmkisored}[1]{\ifnum#1=11 %
\hatcurXmkisoredxxxxxA
\else
\ifnum#1=12 %
\hatcurXmkisoredxxxxxB
\else
??????\fi
\fi
}
\newcommand{\hatcurXmvisored}[1]{\ifnum#1=11 %
\hatcurXmvisoredxxxxxA
\else
\ifnum#1=12 %
\hatcurXmvisoredxxxxxB
\else
??????\fi
\fi
}
\newcommand{\hatcurXsecdur}[1]{\ifnum#1=11 %
\hatcurXsecdurxxxxxA
\else
\ifnum#1=12 %
\hatcurXsecdurxxxxxB
\else
??????\fi
\fi
}
\newcommand{\hatcurXsecingdur}[1]{\ifnum#1=11 %
\hatcurXsecingdurxxxxxA
\else
\ifnum#1=12 %
\hatcurXsecingdurxxxxxB
\else
??????\fi
\fi
}
\newcommand{\hatcurXsecondary}[1]{\ifnum#1=11 %
\hatcurXsecondaryxxxxxA
\else
\ifnum#1=12 %
\hatcurXsecondaryxxxxxB
\else
??????\fi
\fi
}
\newcommand{\hatcurXsecphase}[1]{\ifnum#1=11 %
\hatcurXsecphasexxxxxA
\else
\ifnum#1=12 %
\hatcurXsecphasexxxxxB
\else
??????\fi
\fi
}
\newcommand{\hatcurXviisored}[1]{\ifnum#1=11 %
\hatcurXviisoredxxxxxA
\else
\ifnum#1=12 %
\hatcurXviisoredxxxxxB
\else
??????\fi
\fi
}
\newcommand{\hatcurXvkisored}[1]{\ifnum#1=11 %
\hatcurXvkisoredxxxxxA
\else
\ifnum#1=12 %
\hatcurXvkisoredxxxxxB
\else
??????\fi
\fi
}
\newcommand{\hatcurhtreccenxxxxxA}{HATS579-019}                        % Original HTR name of target
\newcommand{\hatcurfieldeccenxxxxxA}{\ensuremath{string}}              % HTR field
\newcommand{\hatcurCCraeccenxxxxxA}{\ensuremath{19^{\mathrm h}17^{\mathrm m}36.24{\mathrm s}}}                       % Right Ascension
\newcommand{\hatcurCCdececcenxxxxxA}{\ensuremath{-22{\arcdeg}23{\arcmin}23.7{\arcsec}}}                      % Declination
\newcommand{\hatcurCCmageccenxxxxxA}{14.018}                           % apparent V-band magnitude
\newcommand{\hatcurCCtwomasseccenxxxxxA}{2MASS~19173618-2223236}       % 2MASS identifier
\newcommand{\hatcurCCgsceccenxxxxxA}{GSC~6308-00430}                   % GSC(1.2) identifier
\newcommand{\hatcurCCtassmveccenxxxxxA}{\ensuremath{14.018\pm0.080}}   % APASS V-band magnitude
\newcommand{\hatcurCCtassmvshorteccenxxxxxA}{\ensuremath{14.0}}        % APASS V-band magnitude
\newcommand{\hatcurCCtassmBeccenxxxxxA}{\ensuremath{14.657\pm0.020}}   % APASS B-band magnitude
\newcommand{\hatcurCCtassmBshorteccenxxxxxA}{\ensuremath{14.7}}        % APASS B-band magnitude
\newcommand{\hatcurCCtassmIeccenxxxxxA}{\ensuremath{nff\pmnff}}        % TASS I-band magnitude
\newcommand{\hatcurCCtassmIshorteccenxxxxxA}{\ensuremath{0.0}}         % TASS I-band magnitude
\newcommand{\hatcurCCtassmgeccenxxxxxA}{\ensuremath{14.268\pm0.010}}   % APASS g-band magnitude
\newcommand{\hatcurCCtassmgshorteccenxxxxxA}{\ensuremath{14.3}}        % APASS g-band magnitude
\newcommand{\hatcurCCtassmreccenxxxxxA}{\ensuremath{13.865\pm0.040}}   % APASS r-band magnitude
\newcommand{\hatcurCCtassmrshorteccenxxxxxA}{\ensuremath{13.9}}        % APASS r-band magnitude
\newcommand{\hatcurCCtassmieccenxxxxxA}{\ensuremath{13.51\pm0.23}}     % APASS i-band magnitude
\newcommand{\hatcurCCtassmishorteccenxxxxxA}{\ensuremath{13.5}}        % APASS i-band magnitude
\newcommand{\hatcurCCtwomassJmageccenxxxxxA}{\ensuremath{12.647\pm0.023}} % 2MASS ORIG MAG
\newcommand{\hatcurCCtwomassHmageccenxxxxxA}{\ensuremath{12.315\pm0.025}} % 2MASS ORIG MAG
\newcommand{\hatcurCCtwomassKmageccenxxxxxA}{\ensuremath{12.243\pm0.025}} % 2MASS ORIG MAG
\newcommand{\hatcurCCcitJmageccenxxxxxA}{\ensuremath{12.662\pm0.023}}  % 2MASS CIT MAG
\newcommand{\hatcurCCcitHmageccenxxxxxA}{\ensuremath{12.310\pm0.026}}  % 2MASS CIT MAG
\newcommand{\hatcurCCcitKmageccenxxxxxA}{\ensuremath{12.267\pm0.025}}  % 2MASS CIT MAG
\newcommand{\hatcurCCbbJmageccenxxxxxA}{\ensuremath{12.714\pm0.025}}   % 2MASS BB MAG
\newcommand{\hatcurCCbbHmageccenxxxxxA}{\ensuremath{12.331\pm0.026}}   % 2MASS BB MAG
\newcommand{\hatcurCCbbKmageccenxxxxxA}{\ensuremath{12.287\pm0.025}}   % 2MASS BB MAG
\newcommand{\hatcurCCesoJmageccenxxxxxA}{\ensuremath{12.716\pm0.027}}  % 2MASS ESO MAG
\newcommand{\hatcurCCesoHmageccenxxxxxA}{\ensuremath{12.326\pm0.027}}  % 2MASS ESO MAG
\newcommand{\hatcurCCesoKmageccenxxxxxA}{\ensuremath{12.286\pm0.026}}  % 2MASS ESO MAG
\newcommand{\hatcurCCesoJHmageccenxxxxxA}{\ensuremath{0.391\pm0.037}}  % 2MASS ESO JH COLOR
\newcommand{\hatcurCCesoJKmageccenxxxxxA}{\ensuremath{0.431\pm0.037}}  % 2MASS ESO JK COLOR
\newcommand{\hatcurCCesoHKmageccenxxxxxA}{\ensuremath{0.0400\pm0.0090}} % 2MASS ESO HK COLOR
\newcommand{\hatcurLCdipeccenxxxxxA}{\ensuremath{12.9}}                % BLS detected dip (mmag)
\newcommand{\hatcurLCrprstareccenxxxxxA}{\ensuremath{0.1083\pm0.0023}} % Rp/R*
\newcommand{\hatcurLCbsqeccenxxxxxA}{\ensuremath{0.049_{-0.034}^{+0.062}}} % impact parameter square
\newcommand{\hatcurLCimpeccenxxxxxA}{\ensuremath{0.221_{-0.097}^{+0.112}}} % impact parameter
\newcommand{\hatcurLCzetaeccenxxxxxA}{\ensuremath{12.19\pm0.23}}       % zeta/R*
\newcommand{\hatcurLCdureccenxxxxxA}{\ensuremath{0.1831\pm0.0035}}     % transit duration (days)
\newcommand{\hatcurLCdurshorteccenxxxxxA}{\ensuremath{0.1831}}         % transit duration (days)
\newcommand{\hatcurLCdurhreccenxxxxxA}{\ensuremath{4.395\pm0.083}}     % transit duration (hours)
\newcommand{\hatcurLCdurhrshorteccenxxxxxA}{\ensuremath{4.395}}        % transit duration (hours)
\newcommand{\hatcurLCqeccenxxxxxA}{\ensuremath{0.05060\pm0.00096}}     % fractional transit duration (days)
\newcommand{\hatcurLCqshorteccenxxxxxA}{\ensuremath{0.051}}            % fractional transit duration (days)
\newcommand{\hatcurLCingdureccenxxxxxA}{\ensuremath{0.0188\pm0.0012}}  % ingress/egress duration (days)
\newcommand{\hatcurLCPeccenxxxxxA}{\ensuremath{3.6191625\pm0.0000099}} % period (days)
\newcommand{\hatcurLCPprececcenxxxxxA}{\ensuremath{3.6191625}}         % period (days)
\newcommand{\hatcurLCPshorteccenxxxxxA}{\ensuremath{3.6192}}           % period (days)
\newcommand{\hatcurLCTeccenxxxxxA}{\ensuremath{2456611.1572\pm0.0012}} % epoch (BJD)
\newcommand{\hatcurLCTAeccenxxxxxA}{\ensuremath{2455272.0673\pm0.0035}} % TA (BJD)
\newcommand{\hatcurLCTBeccenxxxxxA}{\ensuremath{2456911.5477\pm0.0016}} % TB (BJD)
\newcommand{\hatcurLChatnetmAeccenxxxxxA}{\ensuremath{13.76551\pm0.00023}} % HATNet OOT level
\newcommand{\hatcurLCiblendAeccenxxxxxA}{\ensuremath{0.695\pm0.097}}   % HATNet iblend factor
\newcommand{\hatcurLChatnetmBeccenxxxxxA}{\ensuremath{13.76555\pm0.00011}} % HATNet OOT level
\newcommand{\hatcurLCiblendBeccenxxxxxA}{\ensuremath{0.818\pm0.058}}   % HATNet iblend factor
\newcommand{\hatcurSMEiteffeccenxxxxxA}{\ensuremath{6080\pm150}}       % Ini SME, stellar effective temperature
\newcommand{\hatcurSMEizfeheccenxxxxxA}{\ensuremath{-0.380\pm0.080}}   % Ini SME, stellar metallicity
\newcommand{\hatcurSMEizfehshorteccenxxxxxA}{\ensuremath{-0.38}}       % Ini SME, stellar metallicity
\newcommand{\hatcurSMEiloggeccenxxxxxA}{\ensuremath{4.19\pm0.19}}      % Ini SME, stellar surface gravity
\newcommand{\hatcurSMEivsineccenxxxxxA}{\ensuremath{3.63\pm0.60}}      % Ini SME, stellar rotational velocity
\newcommand{\hatcurSMEivmaceccenxxxxxA}{\ensuremath{0.0}}              % Ini SME, stellar macroturbulence
\newcommand{\hatcurSMEivmiceccenxxxxxA}{\ensuremath{0.0}}              % Ini SME, stellar microturbulence
\newcommand{\hatcurSMEiiteffeccenxxxxxA}{\ensuremath{6060\pm150}}      % Final SME, stellar effective temperature
\newcommand{\hatcurSMEiizfeheccenxxxxxA}{\ensuremath{-0.390\pm0.060}}  % Final SME, stellar metallicity
\newcommand{\hatcurSMEiizfehshorteccenxxxxxA}{\ensuremath{-0.39}}      % Final SME, stellar metallicity
\newcommand{\hatcurSMEiiloggeccenxxxxxA}{\ensuremath{4.140\pm0.030}}   % Final SME, stellar surface gravity
\newcommand{\hatcurSMEiivsineccenxxxxxA}{\ensuremath{3.8\pm1.0}}       % Final SME, stellar rotational velocity
\newcommand{\hatcurLBizeccenxxxxxA}{\ensuremath{0.1621}}               % Limb darkening parameters, Gamma1, z-band
\newcommand{\hatcurLBiizeccenxxxxxA}{\ensuremath{0.3369}}              % Limb darkening parameters, Gamma2, z-band
\newcommand{\hatcurLBiieccenxxxxxA}{\ensuremath{0.2115}}               % Limb darkening parameters, Gamma1, i-band
\newcommand{\hatcurLBiiieccenxxxxxA}{\ensuremath{0.3464}}              % Limb darkening parameters, Gamma2, i-band
\newcommand{\hatcurLBiIeccenxxxxxA}{\ensuremath{0.1947}}               % Limb darkening parameters, Gamma1, I-band
\newcommand{\hatcurLBiiIeccenxxxxxA}{\ensuremath{0.3436}}              % Limb darkening parameters, Gamma2, I-band
\newcommand{\hatcurLBigeccenxxxxxA}{\ensuremath{0.4393}}               % Limb darkening parameters, Gamma1, g-band
\newcommand{\hatcurLBiigeccenxxxxxA}{\ensuremath{0.3122}}              % Limb darkening parameters, Gamma2, g-band
\newcommand{\hatcurLBireccenxxxxxA}{\ensuremath{0.2778}}               % Limb darkening parameters, Gamma1, r-band
\newcommand{\hatcurLBiireccenxxxxxA}{\ensuremath{0.3619}}              % Limb darkening parameters, Gamma2, r-band
\newcommand{\hatcurLBiReccenxxxxxA}{\ensuremath{0.2590}}               % Limb darkening parameters, Gamma1, R-band
\newcommand{\hatcurLBiiReccenxxxxxA}{\ensuremath{0.3590}}              % Limb darkening parameters, Gamma2, R-band
\newcommand{\hatcurLBikepeccenxxxxxA}{\ensuremath{0.1000}}             % Limb darkening parameters, Gamma1, Kep-band
\newcommand{\hatcurLBiikepeccenxxxxxA}{\ensuremath{0.1000}}            % Limb darkening parameters, Gamma2, Kep-band
\newcommand{\hatcurISOmeccenxxxxxA}{\ensuremath{0.976\pm0.065}}        % stellar mass
\newcommand{\hatcurISOmshorteccenxxxxxA}{\ensuremath{0.98}}            % stellar mass
\newcommand{\hatcurISOmlongeccenxxxxxA}{\ensuremath{0.976\pm0.065}}    % stellar mass
\newcommand{\hatcurISOreccenxxxxxA}{\ensuremath{1.28\pm0.16}}          % stellar radius
\newcommand{\hatcurISOrshorteccenxxxxxA}{\ensuremath{1.28}}            % stellar radius
\newcommand{\hatcurISOrlongeccenxxxxxA}{\ensuremath{1.28\pm0.16}}      % stellar radius
\newcommand{\hatcurISOrhoeccenxxxxxA}{\ensuremath{0.66_{-0.19}^{+0.33}}} % stellar density (cgs)
\newcommand{\hatcurISOrholongeccenxxxxxA}{\ensuremath{0.66_{-0.19}^{+0.33}}} % stellar density (cgs)
\newcommand{\hatcurISOloggeccenxxxxxA}{\ensuremath{4.21\pm0.10}}       % stellar surface gravity from isochrones
\newcommand{\hatcurISOlumeccenxxxxxA}{\ensuremath{1.97\pm0.58}}        % stellar luminosity
\newcommand{\hatcurISOlumshorteccenxxxxxA}{\ensuremath{1.97}}          % stellar luminosity
\newcommand{\hatcurISOmveccenxxxxxA}{\ensuremath{4.08\pm0.32}}         % stellar absolute magnitude
\newcommand{\hatcurISOvieccenxxxxxA}{\ensuremath{0.612\pm0.041}}       % stellar V-I index
\newcommand{\hatcurISOageeccenxxxxxA}{\ensuremath{7.5\pm2.5}}          % stellar age
\newcommand{\hatcurISOsigmaeccenxxxxxA}{\ensuremath{0.00070\pm0.00024}} % system mass-correction sigma parameter
\newcommand{\hatcurISOMJeccenxxxxxA}{\ensuremath{3.06\pm0.29}}         % stellar absolute J magnitude
\newcommand{\hatcurISOMHeccenxxxxxA}{\ensuremath{2.76\pm0.28}}         % stellar absolute H magnitude
\newcommand{\hatcurISOMKeccenxxxxxA}{\ensuremath{2.71\pm0.28}}         % stellar absolute K magnitude
\newcommand{\hatcurISOJKeccenxxxxxA}{\ensuremath{0.360\pm0.030}}       % J-K color index from isochrones.
\newcommand{\hatcurISOspececcenxxxxxA}{G0}                             % stellar spectral type
\newcommand{\hatcurRVKeccenxxxxxA}{\ensuremath{114\pm14}}              % RV semi-amplitude [m/s]
\newcommand{\hatcurRVrkeccenxxxxxA}{\ensuremath{0.113_{-0.105}^{+0.081}}} % sqrt(e)*cos(omega)
\newcommand{\hatcurRVrheccenxxxxxA}{\ensuremath{-0.33_{-0.16}^{+0.29}}} % sqrt(e)*sin(omega)
\newcommand{\hatcurRVkeccenxxxxxA}{\ensuremath{0.043\pm0.041}}         % e*cos(omega)
\newcommand{\hatcurRVheccenxxxxxA}{\ensuremath{-0.12\pm0.11}}          % e*sin(omega)
\newcommand{\hatcurRVtroneeccenxxxxxA}{\ensuremath{0\pm0}}             % RV linear trend tr1 factor
\newcommand{\hatcurRVtrtwoeccenxxxxxA}{\ensuremath{0\pm0}}             % RV linear trend tr2 factor
\newcommand{\hatcurRVgammaAeccenxxxxxA}{\ensuremath{-58412\pm32}}      % RV gamma velocity, relative scale
\newcommand{\hatcurRVjitterAeccenxxxxxA}{\ensuremath{0\pm25}}          % RV jitter (m/s)
\newcommand{\hatcurRVfitrmsAeccenxxxxxA}{\ensuremath{0.0}}             % RVfitrms
\newcommand{\hatcurRVgammaBeccenxxxxxA}{\ensuremath{-58324\pm11}}      % RV gamma velocity, relative scale
\newcommand{\hatcurRVjitterBeccenxxxxxA}{\ensuremath{0.1\pm5.1}}       % RV jitter (m/s)
\newcommand{\hatcurRVfitrmsBeccenxxxxxA}{\ensuremath{0.0}}             % RVfitrms
\newcommand{\hatcurRVecceneccenxxxxxA}{\ensuremath{0.14\pm0.10}}       % eccentricity
\newcommand{\hatcurRVeccentwosiglimeccenxxxxxA}{\ensuremath{<0.340}}   % eccentricity
\newcommand{\hatcurRVomegaeccenxxxxxA}{\ensuremath{284\pm82}}          % argument of pericenter
\newcommand{\hatcurPPieccenxxxxxA}{\ensuremath{88.58_{-0.97}^{+0.68}}} % orbital inclination
\newcommand{\hatcurPPgeccenxxxxxA}{\ensuremath{10.9_{-2.5}^{+4.7}}}    % planetary surface gravity (m/s^2)
\newcommand{\hatcurPPloggeccenxxxxxA}{\ensuremath{3.04_{-0.11}^{+0.16}}} % planetary surface gravity (log cgs)
\newcommand{\hatcurPPareccenxxxxxA}{\ensuremath{7.69_{-0.82}^{+1.11}}} % relative orbital radius (a/R*)
\newcommand{\hatcurPPareleccenxxxxxA}{\ensuremath{0.0458\pm0.0010}}    % semimajor axis (AU)
\newcommand{\hatcurPPrhoeccenxxxxxA}{\ensuremath{0.40_{-0.13}^{+0.27}}} % planetary density (cgs)
\newcommand{\hatcurPPmeccenxxxxxA}{\ensuremath{0.83\pm0.10}}           % planetary mass (M_jup)
\newcommand{\hatcurPPmshorteccenxxxxxA}{\ensuremath{0.83}}             % planetary mass (M_jup)
\newcommand{\hatcurPPmlongeccenxxxxxA}{\ensuremath{0.83\pm0.10}}       % planetary mass (M_jup)
\newcommand{\hatcurPPmeeccenxxxxxA}{\ensuremath{265\pm32}}             % planetary mass (M_earth)
\newcommand{\hatcurPPmeshorteccenxxxxxA}{\ensuremath{264.8}}           % planetary mass (M_earth)
\newcommand{\hatcurPPmelongeccenxxxxxA}{\ensuremath{265\pm32}}         % planetary mass (M_earth)
\newcommand{\hatcurPPreccenxxxxxA}{\ensuremath{1.35\pm0.18}}           % planetary radius (R_jup)
\newcommand{\hatcurPPrshorteccenxxxxxA}{\ensuremath{1.35}}             % planetary radius (R_jup)
\newcommand{\hatcurPPrlongeccenxxxxxA}{\ensuremath{1.35\pm0.18}}       % planetary radius (R_jup)
\newcommand{\hatcurPPreeccenxxxxxA}{\ensuremath{15.1\pm2.0}}           % planetary radius (R_earth)
\newcommand{\hatcurPPreshorteccenxxxxxA}{\ensuremath{15.1}}            % planetary radius (R_earth)
\newcommand{\hatcurPPrelongeccenxxxxxA}{\ensuremath{15.1\pm2.0}}       % planetary radius (R_earth)
\newcommand{\hatcurPPmrcorreccenxxxxxA}{\ensuremath{-0.06}}            % mass/radius correlation
\newcommand{\hatcurPPteffeccenxxxxxA}{\ensuremath{1550\pm94}}          % planetary temperature (K)
\newcommand{\hatcurPPthetaeccenxxxxxA}{\ensuremath{0.0570_{-0.0098}^{+0.0138}}} % Safranov number
\newcommand{\hatcurPPfluxperieccenxxxxxA}{\ensuremath{1.78\pm0.25}}    % flux @ periastron (CGS)
\newcommand{\hatcurPPfluxperidimeccenxxxxxA}{\ensuremath{9}}           % flux @ periastron (CGS) units.
\newcommand{\hatcurPPfluxapeccenxxxxxA}{\ensuremath{10.0_{-3.9}^{+5.2}}} % flux @ apastron (CGS)
\newcommand{\hatcurPPfluxapdimeccenxxxxxA}{\ensuremath{8}}             % flux @ apastron (CGS) units.
\newcommand{\hatcurPPfluxavgeccenxxxxxA}{\ensuremath{1.30\pm0.32}}     % flux on average (CGS)
\newcommand{\hatcurPPfluxavgdimeccenxxxxxA}{\ensuremath{9}}            % flux average (CGS) units.
\newcommand{\hatcurPPfluxavglogeccenxxxxxA}{\ensuremath{9.12\pm0.11}}  % log10 flux on average (CGS)
\newcommand{\hatcurXsecphaseeccenxxxxxA}{\ensuremath{0.528\pm0.026}}   % Phase of secondary eclipse
\newcommand{\hatcurXsecondaryeccenxxxxxA}{\ensuremath{2456613.067\pm0.095}} % Secondary eclipse epoch
\newcommand{\hatcurXsecdureccenxxxxxA}{\ensuremath{0.144\pm0.031}}     % sec eclipse duration (days)
\newcommand{\hatcurXsecingdureccenxxxxxA}{\ensuremath{0.0146\pm0.0037}} % sec I/E duration (days)
\newcommand{\hatcurPPphiconjeccenxxxxxA}{\ensuremath{0.404_{-0.362}^{+0.050}}} % phase diff between conjunction and periastron
\newcommand{\hatcurPPperieccenxxxxxA}{\ensuremath{2456609.7\pm1.1}}    % time of periastron passage.
\newcommand{\hatcurPPaequiveccenxxxxxA}{\ensuremath{0.0325_{-0.0038}^{+0.0051}}} % equivalent semi-major axis
\newcommand{\hatcurPPtcirceccenxxxxxA}{\ensuremath{123_{-47}^{+68}}}   % circularization timescale
\newcommand{\hatcurPPtinfalleccenxxxxxA}{\ensuremath{1100_{-480}^{+950}}} % infall timescale
\newcommand{\hatcurXdisteccenxxxxxA}{\ensuremath{820\pm110}}           % distance (pc), no reddenning correction
\newcommand{\hatcurXAveccenxxxxxA}{\ensuremath{0.42\pm0.15}}           % Av (mag)
\newcommand{\hatcurXdistredeccenxxxxxA}{\ensuremath{800\pm100}}        % distance with Av correction (pc)
\newcommand{\hatcurXEBVeccenxxxxxA}{\ensuremath{0.134\pm0.048}}        % E(B-V) (mag)
\newcommand{\hatcurXmvisoredeccenxxxxxA}{\ensuremath{14.022\pm0.074}}  % Expected m_v with reddening (mag)
\newcommand{\hatcurXmiisoredeccenxxxxxA}{\ensuremath{13.190\pm0.031}}  % Expected m_i with reddening (mag)
\newcommand{\hatcurXmjisoredeccenxxxxxA}{\ensuremath{12.700\pm0.018}}  % Expected m_j with reddening (mag)
\newcommand{\hatcurXmhisoredeccenxxxxxA}{\ensuremath{12.356\pm0.016}}  % Expected m_h with reddening (mag)
\newcommand{\hatcurXmkisoredeccenxxxxxA}{\ensuremath{12.273\pm0.018}}  % Expected m_k with reddening (mag)
\newcommand{\hatcurXviisoredeccenxxxxxA}{\ensuremath{0.832\pm0.052}}   % Expected V-I with reddening (mag)
\newcommand{\hatcurXvkisoredeccenxxxxxA}{\ensuremath{1.749\pm0.082}}   % Expected V-K with reddening (mag)
\newcommand{\hatcurXjhisoredeccenxxxxxA}{\ensuremath{0.345\pm0.016}}   % Expected J-H with reddening (mag)
\newcommand{\hatcurXjkisoredeccenxxxxxA}{\ensuremath{0.427\pm0.018}}   % Expected J-K with reddening (mag)
\newcommand{\hatcurCCpmraeccenxxxxxA}{\ensuremath{4.5\pm1.9}}          % proper motion, in RA
\newcommand{\hatcurCCpmdececcenxxxxxA}{\ensuremath{-21.5\pm1.9}}       % proper motion, in DEC
\newcommand{\hatcurCCpmeccenxxxxxA}{\ensuremath{22.0\pm2.7}}           % proper motion
\newcommand{\hatcurhtreccenxxxxxB}{HATS579-023}                        % Original HTR name of target
\newcommand{\hatcurfieldeccenxxxxxB}{\ensuremath{string}}              % HTR field
\newcommand{\hatcurCCraeccenxxxxxB}{\ensuremath{19^{\mathrm h}16^{\mathrm m}48.72{\mathrm s}}}                       % Right Ascension
\newcommand{\hatcurCCdececcenxxxxxB}{\ensuremath{-19{\arcdeg}21{\arcmin}21.2{\arcsec}}}                      % Declination
\newcommand{\hatcurCCmageccenxxxxxB}{12.767}                           % apparent V-band magnitude
\newcommand{\hatcurCCtwomasseccenxxxxxB}{2MASS~19164857-1921212}       % 2MASS identifier
\newcommand{\hatcurCCgsceccenxxxxxB}{GSC~6304-00396}                   % GSC(1.2) identifier
\newcommand{\hatcurCCtassmveccenxxxxxB}{\ensuremath{12.767\pm0.020}}   % APASS V-band magnitude
\newcommand{\hatcurCCtassmvshorteccenxxxxxB}{\ensuremath{12.8}}        % APASS V-band magnitude
\newcommand{\hatcurCCtassmBeccenxxxxxB}{\ensuremath{13.284\pm0.010}}   % APASS B-band magnitude
\newcommand{\hatcurCCtassmBshorteccenxxxxxB}{\ensuremath{13.3}}        % APASS B-band magnitude
\newcommand{\hatcurCCtassmIeccenxxxxxB}{\ensuremath{100\pm100}}        % TASS I-band magnitude
\newcommand{\hatcurCCtassmIshorteccenxxxxxB}{\ensuremath{100.0}}       % TASS I-band magnitude
\newcommand{\hatcurCCtassmgeccenxxxxxB}{\ensuremath{12.972\pm0.010}}   % APASS g-band magnitude
\newcommand{\hatcurCCtassmgshorteccenxxxxxB}{\ensuremath{13.0}}        % APASS g-band magnitude
\newcommand{\hatcurCCtassmreccenxxxxxB}{\ensuremath{12.654\pm0.020}}   % APASS r-band magnitude
\newcommand{\hatcurCCtassmrshorteccenxxxxxB}{\ensuremath{12.7}}        % APASS r-band magnitude
\newcommand{\hatcurCCtassmieccenxxxxxB}{\ensuremath{12.595\pm0.020}}   % APASS i-band magnitude
\newcommand{\hatcurCCtassmishorteccenxxxxxB}{\ensuremath{12.6}}        % APASS i-band magnitude
\newcommand{\hatcurCCtwomassJmageccenxxxxxB}{\ensuremath{11.725\pm0.023}} % 2MASS ORIG MAG
\newcommand{\hatcurCCtwomassHmageccenxxxxxB}{\ensuremath{11.507\pm0.024}} % 2MASS ORIG MAG
\newcommand{\hatcurCCtwomassKmageccenxxxxxB}{\ensuremath{11.391\pm0.023}} % 2MASS ORIG MAG
\newcommand{\hatcurCCcitJmageccenxxxxxB}{\ensuremath{11.744\pm0.024}}  % 2MASS CIT MAG
\newcommand{\hatcurCCcitHmageccenxxxxxB}{\ensuremath{11.501\pm0.024}}  % 2MASS CIT MAG
\newcommand{\hatcurCCcitKmageccenxxxxxB}{\ensuremath{11.415\pm0.023}}  % 2MASS CIT MAG
\newcommand{\hatcurCCbbJmageccenxxxxxB}{\ensuremath{11.790\pm0.024}}   % 2MASS BB MAG
\newcommand{\hatcurCCbbHmageccenxxxxxB}{\ensuremath{11.523\pm0.025}}   % 2MASS BB MAG
\newcommand{\hatcurCCbbKmageccenxxxxxB}{\ensuremath{11.435\pm0.023}}   % 2MASS BB MAG
\newcommand{\hatcurCCesoJmageccenxxxxxB}{\ensuremath{11.792\pm0.025}}  % 2MASS ESO MAG
\newcommand{\hatcurCCesoHmageccenxxxxxB}{\ensuremath{11.520\pm0.030}}  % 2MASS ESO MAG
\newcommand{\hatcurCCesoKmageccenxxxxxB}{\ensuremath{11.434\pm0.024}}  % 2MASS ESO MAG
\newcommand{\hatcurCCesoJHmageccenxxxxxB}{\ensuremath{0.272\pm0.038}}  % 2MASS ESO JH COLOR
\newcommand{\hatcurCCesoJKmageccenxxxxxB}{\ensuremath{0.3580\pm0.0090}} % 2MASS ESO JK COLOR
\newcommand{\hatcurCCesoHKmageccenxxxxxB}{\ensuremath{0.086\pm0.038}}  % 2MASS ESO HK COLOR
\newcommand{\hatcurLCdipeccenxxxxxB}{\ensuremath{5.1}}                 % BLS detected dip (mmag)
\newcommand{\hatcurLCrprstareccenxxxxxB}{\ensuremath{0.0625\pm0.0022}} % Rp/R*
\newcommand{\hatcurLCbsqeccenxxxxxB}{\ensuremath{0.31_{-0.11}^{+0.11}}} % impact parameter square
\newcommand{\hatcurLCimpeccenxxxxxB}{\ensuremath{0.556_{-0.109}^{+0.095}}} % impact parameter
\newcommand{\hatcurLCzetaeccenxxxxxB}{\ensuremath{11.52_{-0.11}^{+0.16}}} % zeta/R*
\newcommand{\hatcurLCdureccenxxxxxB}{\ensuremath{0.1893\pm0.0032}}     % transit duration (days)
\newcommand{\hatcurLCdurshorteccenxxxxxB}{\ensuremath{0.1893}}         % transit duration (days)
\newcommand{\hatcurLCdurhreccenxxxxxB}{\ensuremath{4.544\pm0.076}}     % transit duration (hours)
\newcommand{\hatcurLCdurhrshorteccenxxxxxB}{\ensuremath{4.544}}        % transit duration (hours)
\newcommand{\hatcurLCqeccenxxxxxB}{\ensuremath{0.0602\pm0.0010}}       % fractional transit duration (days)
\newcommand{\hatcurLCqshorteccenxxxxxB}{\ensuremath{0.060}}            % fractional transit duration (days)
\newcommand{\hatcurLCingdureccenxxxxxB}{\ensuremath{0.0157\pm0.0031}}  % ingress/egress duration (days)
\newcommand{\hatcurLCPeccenxxxxxB}{\ensuremath{3.142835\pm0.000010}}   % period (days)
\newcommand{\hatcurLCPprececcenxxxxxB}{\ensuremath{3.1428348}}         % period (days)
\newcommand{\hatcurLCPshorteccenxxxxxB}{\ensuremath{3.1428}}           % period (days)
\newcommand{\hatcurLCTeccenxxxxxB}{\ensuremath{2456852.3839\pm0.0012}} % epoch (BJD)
\newcommand{\hatcurLCTAeccenxxxxxB}{\ensuremath{2455271.5380\pm0.0045}} % TA (BJD)
\newcommand{\hatcurLCTBeccenxxxxxB}{\ensuremath{2457169.8103\pm0.0021}} % TB (BJD)
\newcommand{\hatcurLChatnetmeccenxxxxxB}{\ensuremath{12.687830\pm0.000070}} % HATNet OOT level
\newcommand{\hatcurLCiblendeccenxxxxxB}{\ensuremath{0.760\pm0.090}}    % HATNet iblend factor
\newcommand{\hatcurSMEiteffeccenxxxxxB}{\ensuremath{6315\pm67}}        % Ini SME, stellar effective temperature
\newcommand{\hatcurSMEizfeheccenxxxxxB}{\ensuremath{-0.290\pm0.020}}   % Ini SME, stellar metallicity
\newcommand{\hatcurSMEizfehshorteccenxxxxxB}{\ensuremath{-0.29}}       % Ini SME, stellar metallicity
\newcommand{\hatcurSMEiloggeccenxxxxxB}{\ensuremath{3.64\pm0.12}}      % Ini SME, stellar surface gravity
\newcommand{\hatcurSMEivsineccenxxxxxB}{\ensuremath{5.89\pm0.50}}      % Ini SME, stellar rotational velocity
\newcommand{\hatcurSMEivmaceccenxxxxxB}{\ensuremath{0.0}}              % Ini SME, stellar macroturbulence
\newcommand{\hatcurSMEivmiceccenxxxxxB}{\ensuremath{0.0}}              % Ini SME, stellar microturbulence
\newcommand{\hatcurSMEiiteffeccenxxxxxB}{\ensuremath{6408\pm75}}       % Final SME, stellar effective temperature
\newcommand{\hatcurSMEiizfeheccenxxxxxB}{\ensuremath{-0.100\pm0.040}}  % Final SME, stellar metallicity
\newcommand{\hatcurSMEiizfehshorteccenxxxxxB}{\ensuremath{-0.1}}       % Final SME, stellar metallicity
\newcommand{\hatcurSMEiiloggeccenxxxxxB}{\ensuremath{4\pm0}}           % Final SME, stellar surface gravity
\newcommand{\hatcurSMEiivsineccenxxxxxB}{\ensuremath{6.57\pm0.50}}     % Final SME, stellar rotational velocity
\newcommand{\hatcurLBizeccenxxxxxB}{\ensuremath{0.1276}}               % Limb darkening parameters, Gamma1, z-band
\newcommand{\hatcurLBiizeccenxxxxxB}{\ensuremath{0.3601}}              % Limb darkening parameters, Gamma2, z-band
\newcommand{\hatcurLBiieccenxxxxxB}{\ensuremath{0.1762}}               % Limb darkening parameters, Gamma1, i-band
\newcommand{\hatcurLBiiieccenxxxxxB}{\ensuremath{0.3706}}              % Limb darkening parameters, Gamma2, i-band
\newcommand{\hatcurLBiIeccenxxxxxB}{\ensuremath{0.1580}}               % Limb darkening parameters, Gamma1, I-band
\newcommand{\hatcurLBiiIeccenxxxxxB}{\ensuremath{0.3685}}              % Limb darkening parameters, Gamma2, I-band
\newcommand{\hatcurLBigeccenxxxxxB}{\ensuremath{0.4153}}               % Limb darkening parameters, Gamma1, g-band
\newcommand{\hatcurLBiigeccenxxxxxB}{\ensuremath{0.3276}}              % Limb darkening parameters, Gamma2, g-band
\newcommand{\hatcurLBireccenxxxxxB}{\ensuremath{0.2509}}               % Limb darkening parameters, Gamma1, r-band
\newcommand{\hatcurLBiireccenxxxxxB}{\ensuremath{0.3789}}              % Limb darkening parameters, Gamma2, r-band
\newcommand{\hatcurLBiReccenxxxxxB}{\ensuremath{0.2295}}               % Limb darkening parameters, Gamma1, R-band
\newcommand{\hatcurLBiiReccenxxxxxB}{\ensuremath{0.3783}}              % Limb darkening parameters, Gamma2, R-band
\newcommand{\hatcurLBikepeccenxxxxxB}{\ensuremath{0.1000}}             % Limb darkening parameters, Gamma1, Kep-band
\newcommand{\hatcurLBiikepeccenxxxxxB}{\ensuremath{0.1000}}            % Limb darkening parameters, Gamma2, Kep-band
\newcommand{\hatcurISOmeccenxxxxxB}{\ensuremath{1.449\pm0.065}}        % stellar mass
\newcommand{\hatcurISOmshorteccenxxxxxB}{\ensuremath{1.45}}            % stellar mass
\newcommand{\hatcurISOmlongeccenxxxxxB}{\ensuremath{1.449\pm0.065}}    % stellar mass
\newcommand{\hatcurISOreccenxxxxxB}{\ensuremath{2.07\pm0.19}}          % stellar radius
\newcommand{\hatcurISOrshorteccenxxxxxB}{\ensuremath{2.07}}            % stellar radius
\newcommand{\hatcurISOrlongeccenxxxxxB}{\ensuremath{2.07\pm0.19}}      % stellar radius
\newcommand{\hatcurISOrhoeccenxxxxxB}{\ensuremath{0.231\pm0.056}}      % stellar density (cgs)
\newcommand{\hatcurISOrholongeccenxxxxxB}{\ensuremath{0.231\pm0.056}}  % stellar density (cgs)
\newcommand{\hatcurISOloggeccenxxxxxB}{\ensuremath{3.966\pm0.062}}     % stellar surface gravity from isochrones
\newcommand{\hatcurISOlumeccenxxxxxB}{\ensuremath{6.5\pm1.3}}          % stellar luminosity
\newcommand{\hatcurISOlumshorteccenxxxxxB}{\ensuremath{6.49}}          % stellar luminosity
\newcommand{\hatcurISOmveccenxxxxxB}{\ensuremath{2.72\pm0.21}}         % stellar absolute magnitude
\newcommand{\hatcurISOvieccenxxxxxB}{\ensuremath{0.519\pm0.019}}       % stellar V-I index
\newcommand{\hatcurISOageeccenxxxxxB}{\ensuremath{2.48\pm0.30}}        % stellar age
\newcommand{\hatcurISOsigmaeccenxxxxxB}{\ensuremath{0.00070\pm0.00014}} % system mass-correction sigma parameter
\newcommand{\hatcurISOMJeccenxxxxxB}{\ensuremath{1.88\pm0.20}}         % stellar absolute J magnitude
\newcommand{\hatcurISOMHeccenxxxxxB}{\ensuremath{1.65\pm0.20}}         % stellar absolute H magnitude
\newcommand{\hatcurISOMKeccenxxxxxB}{\ensuremath{1.61\pm0.20}}         % stellar absolute K magnitude
\newcommand{\hatcurISOJKeccenxxxxxB}{\ensuremath{0.280\pm0.020}}       % J-K color index from isochrones.
\newcommand{\hatcurISOspececcenxxxxxB}{F}                              % stellar spectral type
\newcommand{\hatcurRVKeccenxxxxxB}{\ensuremath{258.8\pm7.9}}           % RV semi-amplitude [m/s]
\newcommand{\hatcurRVrkeccenxxxxxB}{\ensuremath{0.039\pm0.089}}        % sqrt(e)*cos(omega)
\newcommand{\hatcurRVrheccenxxxxxB}{\ensuremath{-0.177_{-0.062}^{+0.105}}} % sqrt(e)*sin(omega)
\newcommand{\hatcurRVkeccenxxxxxB}{\ensuremath{0.007\pm0.019}}         % e*cos(omega)
\newcommand{\hatcurRVheccenxxxxxB}{\ensuremath{-0.034\pm0.027}}        % e*sin(omega)
\newcommand{\hatcurRVtroneeccenxxxxxB}{\ensuremath{0\pm0}}             % RV linear trend tr1 factor
\newcommand{\hatcurRVtrtwoeccenxxxxxB}{\ensuremath{0\pm0}}             % RV linear trend tr2 factor
\newcommand{\hatcurRVgammaAeccenxxxxxB}{\ensuremath{-21666\pm11}}      % RV gamma velocity, relative scale
\newcommand{\hatcurRVjitterAeccenxxxxxB}{\ensuremath{30\pm14}}         % RV jitter (m/s)
\newcommand{\hatcurRVfitrmsAeccenxxxxxB}{\ensuremath{0.0}}             % RVfitrms
\newcommand{\hatcurRVgammaBeccenxxxxxB}{\ensuremath{67.3\pm7.5}}       % RV gamma velocity, relative scale
\newcommand{\hatcurRVjitterBeccenxxxxxB}{\ensuremath{0.00\pm0.92}}     % RV jitter (m/s)
\newcommand{\hatcurRVfitrmsBeccenxxxxxB}{\ensuremath{0.0}}             % RVfitrms
\newcommand{\hatcurRVecceneccenxxxxxB}{\ensuremath{0.041\pm0.022}}     % eccentricity
\newcommand{\hatcurRVeccentwosiglimeccenxxxxxB}{\ensuremath{<0.085}}   % eccentricity
\newcommand{\hatcurRVomegaeccenxxxxxB}{\ensuremath{277\pm69}}          % argument of pericenter
\newcommand{\hatcurPPieccenxxxxxB}{\ensuremath{83.7\pm1.7}}            % orbital inclination
\newcommand{\hatcurPPgeccenxxxxxB}{\ensuremath{37.8\pm7.8}}            % planetary surface gravity (m/s^2)
\newcommand{\hatcurPPloggeccenxxxxxB}{\ensuremath{3.578\pm0.090}}      % planetary surface gravity (log cgs)
\newcommand{\hatcurPPareccenxxxxxB}{\ensuremath{4.94\pm0.38}}          % relative orbital radius (a/R*)
\newcommand{\hatcurPPareleccenxxxxxB}{\ensuremath{0.04752\pm0.00071}}  % semimajor axis (AU)
\newcommand{\hatcurPPrhoeccenxxxxxB}{\ensuremath{1.50\pm0.50}}         % planetary density (cgs)
\newcommand{\hatcurPPmeccenxxxxxB}{\ensuremath{2.40\pm0.11}}           % planetary mass (M_jup)
\newcommand{\hatcurPPmshorteccenxxxxxB}{\ensuremath{2.40}}             % planetary mass (M_jup)
\newcommand{\hatcurPPmlongeccenxxxxxB}{\ensuremath{2.40\pm0.11}}       % planetary mass (M_jup)
\newcommand{\hatcurPPmeeccenxxxxxB}{\ensuremath{761\pm34}}             % planetary mass (M_earth)
\newcommand{\hatcurPPmeshorteccenxxxxxB}{\ensuremath{761.3}}           % planetary mass (M_earth)
\newcommand{\hatcurPPmelongeccenxxxxxB}{\ensuremath{761\pm34}}         % planetary mass (M_earth)
\newcommand{\hatcurPPreccenxxxxxB}{\ensuremath{1.26\pm0.15}}           % planetary radius (R_jup)
\newcommand{\hatcurPPrshorteccenxxxxxB}{\ensuremath{1.26}}             % planetary radius (R_jup)
\newcommand{\hatcurPPrlongeccenxxxxxB}{\ensuremath{1.26\pm0.15}}       % planetary radius (R_jup)
\newcommand{\hatcurPPreeccenxxxxxB}{\ensuremath{14.1\pm1.7}}           % planetary radius (R_earth)
\newcommand{\hatcurPPreshorteccenxxxxxB}{\ensuremath{14.1}}            % planetary radius (R_earth)
\newcommand{\hatcurPPrelongeccenxxxxxB}{\ensuremath{14.1\pm1.7}}       % planetary radius (R_earth)
\newcommand{\hatcurPPmrcorreccenxxxxxB}{\ensuremath{0.67}}             % mass/radius correlation
\newcommand{\hatcurPPteffeccenxxxxxB}{\ensuremath{2043\pm83}}          % planetary temperature (K)
\newcommand{\hatcurPPthetaeccenxxxxxB}{\ensuremath{0.123\pm0.015}}     % Safranov number
\newcommand{\hatcurPPfluxperieccenxxxxxB}{\ensuremath{4.29\pm0.71}}    % flux @ periastron (CGS)
\newcommand{\hatcurPPfluxperidimeccenxxxxxB}{\ensuremath{9}}           % flux @ periastron (CGS) units.
\newcommand{\hatcurPPfluxapeccenxxxxxB}{\ensuremath{3.62\pm0.64}}      % flux @ apastron (CGS)
\newcommand{\hatcurPPfluxapdimeccenxxxxxB}{\ensuremath{9}}             % flux @ apastron (CGS) units.
\newcommand{\hatcurPPfluxavgeccenxxxxxB}{\ensuremath{3.93\pm0.65}}     % flux on average (CGS)
\newcommand{\hatcurPPfluxavgdimeccenxxxxxB}{\ensuremath{9}}            % flux average (CGS) units.
\newcommand{\hatcurPPfluxavglogeccenxxxxxB}{\ensuremath{9.595\pm0.071}} % log10 flux on average (CGS)
\newcommand{\hatcurXsecphaseeccenxxxxxB}{\ensuremath{0.504\pm0.012}}   % Phase of secondary eclipse
\newcommand{\hatcurXsecondaryeccenxxxxxB}{\ensuremath{2456853.969\pm0.039}} % Secondary eclipse epoch
\newcommand{\hatcurXsecdureccenxxxxxB}{\ensuremath{0.1824\pm0.0075}}   % sec eclipse duration (days)
\newcommand{\hatcurXsecingdureccenxxxxxB}{\ensuremath{0.0144\pm0.0026}} % sec I/E duration (days)
\newcommand{\hatcurPPphiconjeccenxxxxxB}{\ensuremath{0.31_{-0.78}^{+0.15}}} % phase diff between conjunction and periastron
\newcommand{\hatcurPPperieccenxxxxxB}{\ensuremath{2456851.4\pm1.5}}    % time of periastron passage.
\newcommand{\hatcurPPaequiveccenxxxxxB}{\ensuremath{0.0186\pm0.0015}}  % equivalent semi-major axis
\newcommand{\hatcurPPtcirceccenxxxxxB}{\ensuremath{410_{-170}^{+270}}} % circularization timescale
\newcommand{\hatcurPPtinfalleccenxxxxxB}{\ensuremath{53_{-17}^{+23}}}  % infall timescale
\newcommand{\hatcurXdisteccenxxxxxB}{\ensuremath{924\pm85}}            % distance (pc), no reddenning correction
\newcommand{\hatcurXAveccenxxxxxB}{\ensuremath{0.226\pm0.068}}         % Av (mag)
\newcommand{\hatcurXdistredeccenxxxxxB}{\ensuremath{922\pm84}}         % distance with Av correction (pc)
\newcommand{\hatcurXEBVeccenxxxxxB}{\ensuremath{0.073\pm0.022}}        % E(B-V) (mag)
\newcommand{\hatcurXmvisoredeccenxxxxxB}{\ensuremath{12.770\pm0.020}}  % Expected m_v with reddening (mag)
\newcommand{\hatcurXmiisoredeccenxxxxxB}{\ensuremath{12.133\pm0.017}}  % Expected m_i with reddening (mag)
\newcommand{\hatcurXmjisoredeccenxxxxxB}{\ensuremath{11.769\pm0.014}}  % Expected m_j with reddening (mag)
\newcommand{\hatcurXmhisoredeccenxxxxxB}{\ensuremath{11.515\pm0.016}}  % Expected m_h with reddening (mag)
\newcommand{\hatcurXmkisoredeccenxxxxxB}{\ensuremath{11.454\pm0.017}}  % Expected m_k with reddening (mag)
\newcommand{\hatcurXviisoredeccenxxxxxB}{\ensuremath{0.637\pm0.021}}   % Expected V-I with reddening (mag)
\newcommand{\hatcurXvkisoredeccenxxxxxB}{\ensuremath{1.316\pm0.027}}   % Expected V-K with reddening (mag)
\newcommand{\hatcurXjhisoredeccenxxxxxB}{\ensuremath{0.2550\pm0.0092}} % Expected J-H with reddening (mag)
\newcommand{\hatcurXjkisoredeccenxxxxxB}{\ensuremath{0.3150\pm0.0078}} % Expected J-K with reddening (mag)
\newcommand{\hatcurCCpmraeccenxxxxxB}{\ensuremath{6.2\pm4.0}}          % proper motion, in RA
\newcommand{\hatcurCCpmdececcenxxxxxB}{\ensuremath{5.2\pm2.7}}         % proper motion, in DEC
\newcommand{\hatcurCCpmeccenxxxxxB}{\ensuremath{8.1\pm4.8}}            % proper motion

\newcommand{\hatcurCCbbHmageccen}[1]{\ifnum#1=11 %
\hatcurCCbbHmageccenxxxxxA
\else
\ifnum#1=12 %
\hatcurCCbbHmageccenxxxxxB
\else
??????\fi
\fi
}
\newcommand{\hatcurCCbbJmageccen}[1]{\ifnum#1=11 %
\hatcurCCbbJmageccenxxxxxA
\else
\ifnum#1=12 %
\hatcurCCbbJmageccenxxxxxB
\else
??????\fi
\fi
}
\newcommand{\hatcurCCbbKmageccen}[1]{\ifnum#1=11 %
\hatcurCCbbKmageccenxxxxxA
\else
\ifnum#1=12 %
\hatcurCCbbKmageccenxxxxxB
\else
??????\fi
\fi
}
\newcommand{\hatcurCCcitHmageccen}[1]{\ifnum#1=11 %
\hatcurCCcitHmageccenxxxxxA
\else
\ifnum#1=12 %
\hatcurCCcitHmageccenxxxxxB
\else
??????\fi
\fi
}
\newcommand{\hatcurCCcitJmageccen}[1]{\ifnum#1=11 %
\hatcurCCcitJmageccenxxxxxA
\else
\ifnum#1=12 %
\hatcurCCcitJmageccenxxxxxB
\else
??????\fi
\fi
}
\newcommand{\hatcurCCcitKmageccen}[1]{\ifnum#1=11 %
\hatcurCCcitKmageccenxxxxxA
\else
\ifnum#1=12 %
\hatcurCCcitKmageccenxxxxxB
\else
??????\fi
\fi
}
\newcommand{\hatcurCCdececcen}[1]{\ifnum#1=11 %
\hatcurCCdececcenxxxxxA
\else
\ifnum#1=12 %
\hatcurCCdececcenxxxxxB
\else
??????\fi
\fi
}
\newcommand{\hatcurCCesoHKmageccen}[1]{\ifnum#1=11 %
\hatcurCCesoHKmageccenxxxxxA
\else
\ifnum#1=12 %
\hatcurCCesoHKmageccenxxxxxB
\else
??????\fi
\fi
}
\newcommand{\hatcurCCesoHmageccen}[1]{\ifnum#1=11 %
\hatcurCCesoHmageccenxxxxxA
\else
\ifnum#1=12 %
\hatcurCCesoHmageccenxxxxxB
\else
??????\fi
\fi
}
\newcommand{\hatcurCCesoJHmageccen}[1]{\ifnum#1=11 %
\hatcurCCesoJHmageccenxxxxxA
\else
\ifnum#1=12 %
\hatcurCCesoJHmageccenxxxxxB
\else
??????\fi
\fi
}
\newcommand{\hatcurCCesoJKmageccen}[1]{\ifnum#1=11 %
\hatcurCCesoJKmageccenxxxxxA
\else
\ifnum#1=12 %
\hatcurCCesoJKmageccenxxxxxB
\else
??????\fi
\fi
}
\newcommand{\hatcurCCesoJmageccen}[1]{\ifnum#1=11 %
\hatcurCCesoJmageccenxxxxxA
\else
\ifnum#1=12 %
\hatcurCCesoJmageccenxxxxxB
\else
??????\fi
\fi
}
\newcommand{\hatcurCCesoKmageccen}[1]{\ifnum#1=11 %
\hatcurCCesoKmageccenxxxxxA
\else
\ifnum#1=12 %
\hatcurCCesoKmageccenxxxxxB
\else
??????\fi
\fi
}
\newcommand{\hatcurCCgsceccen}[1]{\ifnum#1=11 %
\hatcurCCgsceccenxxxxxA
\else
\ifnum#1=12 %
\hatcurCCgsceccenxxxxxB
\else
??????\fi
\fi
}
\newcommand{\hatcurCCmageccen}[1]{\ifnum#1=11 %
\hatcurCCmageccenxxxxxA
\else
\ifnum#1=12 %
\hatcurCCmageccenxxxxxB
\else
??????\fi
\fi
}
\newcommand{\hatcurCCpmdececcen}[1]{\ifnum#1=11 %
\hatcurCCpmdececcenxxxxxA
\else
\ifnum#1=12 %
\hatcurCCpmdececcenxxxxxB
\else
??????\fi
\fi
}
\newcommand{\hatcurCCpmeccen}[1]{\ifnum#1=11 %
\hatcurCCpmeccenxxxxxA
\else
\ifnum#1=12 %
\hatcurCCpmeccenxxxxxB
\else
??????\fi
\fi
}
\newcommand{\hatcurCCpmraeccen}[1]{\ifnum#1=11 %
\hatcurCCpmraeccenxxxxxA
\else
\ifnum#1=12 %
\hatcurCCpmraeccenxxxxxB
\else
??????\fi
\fi
}
\newcommand{\hatcurCCraeccen}[1]{\ifnum#1=11 %
\hatcurCCraeccenxxxxxA
\else
\ifnum#1=12 %
\hatcurCCraeccenxxxxxB
\else
??????\fi
\fi
}
\newcommand{\hatcurCCtassmBeccen}[1]{\ifnum#1=11 %
\hatcurCCtassmBeccenxxxxxA
\else
\ifnum#1=12 %
\hatcurCCtassmBeccenxxxxxB
\else
??????\fi
\fi
}
\newcommand{\hatcurCCtassmBshorteccen}[1]{\ifnum#1=11 %
\hatcurCCtassmBshorteccenxxxxxA
\else
\ifnum#1=12 %
\hatcurCCtassmBshorteccenxxxxxB
\else
??????\fi
\fi
}
\newcommand{\hatcurCCtassmgeccen}[1]{\ifnum#1=11 %
\hatcurCCtassmgeccenxxxxxA
\else
\ifnum#1=12 %
\hatcurCCtassmgeccenxxxxxB
\else
??????\fi
\fi
}
\newcommand{\hatcurCCtassmgshorteccen}[1]{\ifnum#1=11 %
\hatcurCCtassmgshorteccenxxxxxA
\else
\ifnum#1=12 %
\hatcurCCtassmgshorteccenxxxxxB
\else
??????\fi
\fi
}
\newcommand{\hatcurCCtassmieccen}[1]{\ifnum#1=11 %
\hatcurCCtassmieccenxxxxxA
\else
\ifnum#1=12 %
\hatcurCCtassmieccenxxxxxB
\else
??????\fi
\fi
}
\newcommand{\hatcurCCtassmIeccen}[1]{\ifnum#1=11 %
\hatcurCCtassmIeccenxxxxxA
\else
\ifnum#1=12 %
\hatcurCCtassmIeccenxxxxxB
\else
??????\fi
\fi
}
\newcommand{\hatcurCCtassmishorteccen}[1]{\ifnum#1=11 %
\hatcurCCtassmishorteccenxxxxxA
\else
\ifnum#1=12 %
\hatcurCCtassmishorteccenxxxxxB
\else
??????\fi
\fi
}
\newcommand{\hatcurCCtassmIshorteccen}[1]{\ifnum#1=11 %
\hatcurCCtassmIshorteccenxxxxxA
\else
\ifnum#1=12 %
\hatcurCCtassmIshorteccenxxxxxB
\else
??????\fi
\fi
}
\newcommand{\hatcurCCtassmreccen}[1]{\ifnum#1=11 %
\hatcurCCtassmreccenxxxxxA
\else
\ifnum#1=12 %
\hatcurCCtassmreccenxxxxxB
\else
??????\fi
\fi
}
\newcommand{\hatcurCCtassmrshorteccen}[1]{\ifnum#1=11 %
\hatcurCCtassmrshorteccenxxxxxA
\else
\ifnum#1=12 %
\hatcurCCtassmrshorteccenxxxxxB
\else
??????\fi
\fi
}
\newcommand{\hatcurCCtassmveccen}[1]{\ifnum#1=11 %
\hatcurCCtassmveccenxxxxxA
\else
\ifnum#1=12 %
\hatcurCCtassmveccenxxxxxB
\else
??????\fi
\fi
}
\newcommand{\hatcurCCtassmvshorteccen}[1]{\ifnum#1=11 %
\hatcurCCtassmvshorteccenxxxxxA
\else
\ifnum#1=12 %
\hatcurCCtassmvshorteccenxxxxxB
\else
??????\fi
\fi
}
\newcommand{\hatcurCCtwomasseccen}[1]{\ifnum#1=11 %
\hatcurCCtwomasseccenxxxxxA
\else
\ifnum#1=12 %
\hatcurCCtwomasseccenxxxxxB
\else
??????\fi
\fi
}
\newcommand{\hatcurCCtwomassHmageccen}[1]{\ifnum#1=11 %
\hatcurCCtwomassHmageccenxxxxxA
\else
\ifnum#1=12 %
\hatcurCCtwomassHmageccenxxxxxB
\else
??????\fi
\fi
}
\newcommand{\hatcurCCtwomassJmageccen}[1]{\ifnum#1=11 %
\hatcurCCtwomassJmageccenxxxxxA
\else
\ifnum#1=12 %
\hatcurCCtwomassJmageccenxxxxxB
\else
??????\fi
\fi
}
\newcommand{\hatcurCCtwomassKmageccen}[1]{\ifnum#1=11 %
\hatcurCCtwomassKmageccenxxxxxA
\else
\ifnum#1=12 %
\hatcurCCtwomassKmageccenxxxxxB
\else
??????\fi
\fi
}
\newcommand{\hatcurfieldeccen}[1]{\ifnum#1=11 %
\hatcurfieldeccenxxxxxA
\else
\ifnum#1=12 %
\hatcurfieldeccenxxxxxB
\else
??????\fi
\fi
}
\newcommand{\hatcurhtreccen}[1]{\ifnum#1=11 %
\hatcurhtreccenxxxxxA
\else
\ifnum#1=12 %
\hatcurhtreccenxxxxxB
\else
??????\fi
\fi
}
\newcommand{\hatcurISOageeccen}[1]{\ifnum#1=11 %
\hatcurISOageeccenxxxxxA
\else
\ifnum#1=12 %
\hatcurISOageeccenxxxxxB
\else
??????\fi
\fi
}
\newcommand{\hatcurISOJKeccen}[1]{\ifnum#1=11 %
\hatcurISOJKeccenxxxxxA
\else
\ifnum#1=12 %
\hatcurISOJKeccenxxxxxB
\else
??????\fi
\fi
}
\newcommand{\hatcurISOloggeccen}[1]{\ifnum#1=11 %
\hatcurISOloggeccenxxxxxA
\else
\ifnum#1=12 %
\hatcurISOloggeccenxxxxxB
\else
??????\fi
\fi
}
\newcommand{\hatcurISOlumeccen}[1]{\ifnum#1=11 %
\hatcurISOlumeccenxxxxxA
\else
\ifnum#1=12 %
\hatcurISOlumeccenxxxxxB
\else
??????\fi
\fi
}
\newcommand{\hatcurISOlumshorteccen}[1]{\ifnum#1=11 %
\hatcurISOlumshorteccenxxxxxA
\else
\ifnum#1=12 %
\hatcurISOlumshorteccenxxxxxB
\else
??????\fi
\fi
}
\newcommand{\hatcurISOmeccen}[1]{\ifnum#1=11 %
\hatcurISOmeccenxxxxxA
\else
\ifnum#1=12 %
\hatcurISOmeccenxxxxxB
\else
??????\fi
\fi
}
\newcommand{\hatcurISOMHeccen}[1]{\ifnum#1=11 %
\hatcurISOMHeccenxxxxxA
\else
\ifnum#1=12 %
\hatcurISOMHeccenxxxxxB
\else
??????\fi
\fi
}
\newcommand{\hatcurISOMJeccen}[1]{\ifnum#1=11 %
\hatcurISOMJeccenxxxxxA
\else
\ifnum#1=12 %
\hatcurISOMJeccenxxxxxB
\else
??????\fi
\fi
}
\newcommand{\hatcurISOMKeccen}[1]{\ifnum#1=11 %
\hatcurISOMKeccenxxxxxA
\else
\ifnum#1=12 %
\hatcurISOMKeccenxxxxxB
\else
??????\fi
\fi
}
\newcommand{\hatcurISOmlongeccen}[1]{\ifnum#1=11 %
\hatcurISOmlongeccenxxxxxA
\else
\ifnum#1=12 %
\hatcurISOmlongeccenxxxxxB
\else
??????\fi
\fi
}
\newcommand{\hatcurISOmshorteccen}[1]{\ifnum#1=11 %
\hatcurISOmshorteccenxxxxxA
\else
\ifnum#1=12 %
\hatcurISOmshorteccenxxxxxB
\else
??????\fi
\fi
}
\newcommand{\hatcurISOmveccen}[1]{\ifnum#1=11 %
\hatcurISOmveccenxxxxxA
\else
\ifnum#1=12 %
\hatcurISOmveccenxxxxxB
\else
??????\fi
\fi
}
\newcommand{\hatcurISOreccen}[1]{\ifnum#1=11 %
\hatcurISOreccenxxxxxA
\else
\ifnum#1=12 %
\hatcurISOreccenxxxxxB
\else
??????\fi
\fi
}
\newcommand{\hatcurISOrhoeccen}[1]{\ifnum#1=11 %
\hatcurISOrhoeccenxxxxxA
\else
\ifnum#1=12 %
\hatcurISOrhoeccenxxxxxB
\else
??????\fi
\fi
}
\newcommand{\hatcurISOrholongeccen}[1]{\ifnum#1=11 %
\hatcurISOrholongeccenxxxxxA
\else
\ifnum#1=12 %
\hatcurISOrholongeccenxxxxxB
\else
??????\fi
\fi
}
\newcommand{\hatcurISOrlongeccen}[1]{\ifnum#1=11 %
\hatcurISOrlongeccenxxxxxA
\else
\ifnum#1=12 %
\hatcurISOrlongeccenxxxxxB
\else
??????\fi
\fi
}
\newcommand{\hatcurISOrshorteccen}[1]{\ifnum#1=11 %
\hatcurISOrshorteccenxxxxxA
\else
\ifnum#1=12 %
\hatcurISOrshorteccenxxxxxB
\else
??????\fi
\fi
}
\newcommand{\hatcurISOsigmaeccen}[1]{\ifnum#1=11 %
\hatcurISOsigmaeccenxxxxxA
\else
\ifnum#1=12 %
\hatcurISOsigmaeccenxxxxxB
\else
??????\fi
\fi
}
\newcommand{\hatcurISOspececcen}[1]{\ifnum#1=11 %
\hatcurISOspececcenxxxxxA
\else
\ifnum#1=12 %
\hatcurISOspececcenxxxxxB
\else
??????\fi
\fi
}
\newcommand{\hatcurISOvieccen}[1]{\ifnum#1=11 %
\hatcurISOvieccenxxxxxA
\else
\ifnum#1=12 %
\hatcurISOvieccenxxxxxB
\else
??????\fi
\fi
}
\newcommand{\hatcurLBigeccen}[1]{\ifnum#1=11 %
\hatcurLBigeccenxxxxxA
\else
\ifnum#1=12 %
\hatcurLBigeccenxxxxxB
\else
??????\fi
\fi
}
\newcommand{\hatcurLBiieccen}[1]{\ifnum#1=11 %
\hatcurLBiieccenxxxxxA
\else
\ifnum#1=12 %
\hatcurLBiieccenxxxxxB
\else
??????\fi
\fi
}
\newcommand{\hatcurLBiIeccen}[1]{\ifnum#1=11 %
\hatcurLBiIeccenxxxxxA
\else
\ifnum#1=12 %
\hatcurLBiIeccenxxxxxB
\else
??????\fi
\fi
}
\newcommand{\hatcurLBiigeccen}[1]{\ifnum#1=11 %
\hatcurLBiigeccenxxxxxA
\else
\ifnum#1=12 %
\hatcurLBiigeccenxxxxxB
\else
??????\fi
\fi
}
\newcommand{\hatcurLBiiieccen}[1]{\ifnum#1=11 %
\hatcurLBiiieccenxxxxxA
\else
\ifnum#1=12 %
\hatcurLBiiieccenxxxxxB
\else
??????\fi
\fi
}
\newcommand{\hatcurLBiiIeccen}[1]{\ifnum#1=11 %
\hatcurLBiiIeccenxxxxxA
\else
\ifnum#1=12 %
\hatcurLBiiIeccenxxxxxB
\else
??????\fi
\fi
}
\newcommand{\hatcurLBiikepeccen}[1]{\ifnum#1=11 %
\hatcurLBiikepeccenxxxxxA
\else
\ifnum#1=12 %
\hatcurLBiikepeccenxxxxxB
\else
??????\fi
\fi
}
\newcommand{\hatcurLBiireccen}[1]{\ifnum#1=11 %
\hatcurLBiireccenxxxxxA
\else
\ifnum#1=12 %
\hatcurLBiireccenxxxxxB
\else
??????\fi
\fi
}
\newcommand{\hatcurLBiiReccen}[1]{\ifnum#1=11 %
\hatcurLBiiReccenxxxxxA
\else
\ifnum#1=12 %
\hatcurLBiiReccenxxxxxB
\else
??????\fi
\fi
}
\newcommand{\hatcurLBiizeccen}[1]{\ifnum#1=11 %
\hatcurLBiizeccenxxxxxA
\else
\ifnum#1=12 %
\hatcurLBiizeccenxxxxxB
\else
??????\fi
\fi
}
\newcommand{\hatcurLBikepeccen}[1]{\ifnum#1=11 %
\hatcurLBikepeccenxxxxxA
\else
\ifnum#1=12 %
\hatcurLBikepeccenxxxxxB
\else
??????\fi
\fi
}
\newcommand{\hatcurLBireccen}[1]{\ifnum#1=11 %
\hatcurLBireccenxxxxxA
\else
\ifnum#1=12 %
\hatcurLBireccenxxxxxB
\else
??????\fi
\fi
}
\newcommand{\hatcurLBiReccen}[1]{\ifnum#1=11 %
\hatcurLBiReccenxxxxxA
\else
\ifnum#1=12 %
\hatcurLBiReccenxxxxxB
\else
??????\fi
\fi
}
\newcommand{\hatcurLBizeccen}[1]{\ifnum#1=11 %
\hatcurLBizeccenxxxxxA
\else
\ifnum#1=12 %
\hatcurLBizeccenxxxxxB
\else
??????\fi
\fi
}
\newcommand{\hatcurLCbsqeccen}[1]{\ifnum#1=11 %
\hatcurLCbsqeccenxxxxxA
\else
\ifnum#1=12 %
\hatcurLCbsqeccenxxxxxB
\else
??????\fi
\fi
}
\newcommand{\hatcurLCdipeccen}[1]{\ifnum#1=11 %
\hatcurLCdipeccenxxxxxA
\else
\ifnum#1=12 %
\hatcurLCdipeccenxxxxxB
\else
??????\fi
\fi
}
\newcommand{\hatcurLCdureccen}[1]{\ifnum#1=11 %
\hatcurLCdureccenxxxxxA
\else
\ifnum#1=12 %
\hatcurLCdureccenxxxxxB
\else
??????\fi
\fi
}
\newcommand{\hatcurLCdurhreccen}[1]{\ifnum#1=11 %
\hatcurLCdurhreccenxxxxxA
\else
\ifnum#1=12 %
\hatcurLCdurhreccenxxxxxB
\else
??????\fi
\fi
}
\newcommand{\hatcurLCdurhrshorteccen}[1]{\ifnum#1=11 %
\hatcurLCdurhrshorteccenxxxxxA
\else
\ifnum#1=12 %
\hatcurLCdurhrshorteccenxxxxxB
\else
??????\fi
\fi
}
\newcommand{\hatcurLCdurshorteccen}[1]{\ifnum#1=11 %
\hatcurLCdurshorteccenxxxxxA
\else
\ifnum#1=12 %
\hatcurLCdurshorteccenxxxxxB
\else
??????\fi
\fi
}
\newcommand{\hatcurLChatnetmAeccen}[1]{\ifnum#1=11 %
\hatcurLChatnetmAeccenxxxxxA
\else
??????\fi
}
\newcommand{\hatcurLChatnetmBeccen}[1]{\ifnum#1=11 %
\hatcurLChatnetmBeccenxxxxxA
\else
??????\fi
}
\newcommand{\hatcurLChatnetmeccen}[1]{\ifnum#1=12 %
\hatcurLChatnetmeccenxxxxxB
\else
??????\fi
}
\newcommand{\hatcurLCiblendAeccen}[1]{\ifnum#1=11 %
\hatcurLCiblendAeccenxxxxxA
\else
??????\fi
}
\newcommand{\hatcurLCiblendBeccen}[1]{\ifnum#1=11 %
\hatcurLCiblendBeccenxxxxxA
\else
??????\fi
}
\newcommand{\hatcurLCiblendeccen}[1]{\ifnum#1=12 %
\hatcurLCiblendeccenxxxxxB
\else
??????\fi
}
\newcommand{\hatcurLCimpeccen}[1]{\ifnum#1=11 %
\hatcurLCimpeccenxxxxxA
\else
\ifnum#1=12 %
\hatcurLCimpeccenxxxxxB
\else
??????\fi
\fi
}
\newcommand{\hatcurLCingdureccen}[1]{\ifnum#1=11 %
\hatcurLCingdureccenxxxxxA
\else
\ifnum#1=12 %
\hatcurLCingdureccenxxxxxB
\else
??????\fi
\fi
}
\newcommand{\hatcurLCPeccen}[1]{\ifnum#1=11 %
\hatcurLCPeccenxxxxxA
\else
\ifnum#1=12 %
\hatcurLCPeccenxxxxxB
\else
??????\fi
\fi
}
\newcommand{\hatcurLCPprececcen}[1]{\ifnum#1=11 %
\hatcurLCPprececcenxxxxxA
\else
\ifnum#1=12 %
\hatcurLCPprececcenxxxxxB
\else
??????\fi
\fi
}
\newcommand{\hatcurLCPshorteccen}[1]{\ifnum#1=11 %
\hatcurLCPshorteccenxxxxxA
\else
\ifnum#1=12 %
\hatcurLCPshorteccenxxxxxB
\else
??????\fi
\fi
}
\newcommand{\hatcurLCqeccen}[1]{\ifnum#1=11 %
\hatcurLCqeccenxxxxxA
\else
\ifnum#1=12 %
\hatcurLCqeccenxxxxxB
\else
??????\fi
\fi
}
\newcommand{\hatcurLCqshorteccen}[1]{\ifnum#1=11 %
\hatcurLCqshorteccenxxxxxA
\else
\ifnum#1=12 %
\hatcurLCqshorteccenxxxxxB
\else
??????\fi
\fi
}
\newcommand{\hatcurLCrprstareccen}[1]{\ifnum#1=11 %
\hatcurLCrprstareccenxxxxxA
\else
\ifnum#1=12 %
\hatcurLCrprstareccenxxxxxB
\else
??????\fi
\fi
}
\newcommand{\hatcurLCTAeccen}[1]{\ifnum#1=11 %
\hatcurLCTAeccenxxxxxA
\else
\ifnum#1=12 %
\hatcurLCTAeccenxxxxxB
\else
??????\fi
\fi
}
\newcommand{\hatcurLCTBeccen}[1]{\ifnum#1=11 %
\hatcurLCTBeccenxxxxxA
\else
\ifnum#1=12 %
\hatcurLCTBeccenxxxxxB
\else
??????\fi
\fi
}
\newcommand{\hatcurLCTeccen}[1]{\ifnum#1=11 %
\hatcurLCTeccenxxxxxA
\else
\ifnum#1=12 %
\hatcurLCTeccenxxxxxB
\else
??????\fi
\fi
}
\newcommand{\hatcurLCzetaeccen}[1]{\ifnum#1=11 %
\hatcurLCzetaeccenxxxxxA
\else
\ifnum#1=12 %
\hatcurLCzetaeccenxxxxxB
\else
??????\fi
\fi
}
\newcommand{\hatcurPPaequiveccen}[1]{\ifnum#1=11 %
\hatcurPPaequiveccenxxxxxA
\else
\ifnum#1=12 %
\hatcurPPaequiveccenxxxxxB
\else
??????\fi
\fi
}
\newcommand{\hatcurPPareccen}[1]{\ifnum#1=11 %
\hatcurPPareccenxxxxxA
\else
\ifnum#1=12 %
\hatcurPPareccenxxxxxB
\else
??????\fi
\fi
}
\newcommand{\hatcurPPareleccen}[1]{\ifnum#1=11 %
\hatcurPPareleccenxxxxxA
\else
\ifnum#1=12 %
\hatcurPPareleccenxxxxxB
\else
??????\fi
\fi
}
\newcommand{\hatcurPPfluxapdimeccen}[1]{\ifnum#1=11 %
\hatcurPPfluxapdimeccenxxxxxA
\else
\ifnum#1=12 %
\hatcurPPfluxapdimeccenxxxxxB
\else
??????\fi
\fi
}
\newcommand{\hatcurPPfluxapeccen}[1]{\ifnum#1=11 %
\hatcurPPfluxapeccenxxxxxA
\else
\ifnum#1=12 %
\hatcurPPfluxapeccenxxxxxB
\else
??????\fi
\fi
}
\newcommand{\hatcurPPfluxavgdimeccen}[1]{\ifnum#1=11 %
\hatcurPPfluxavgdimeccenxxxxxA
\else
\ifnum#1=12 %
\hatcurPPfluxavgdimeccenxxxxxB
\else
??????\fi
\fi
}
\newcommand{\hatcurPPfluxavgeccen}[1]{\ifnum#1=11 %
\hatcurPPfluxavgeccenxxxxxA
\else
\ifnum#1=12 %
\hatcurPPfluxavgeccenxxxxxB
\else
??????\fi
\fi
}
\newcommand{\hatcurPPfluxavglogeccen}[1]{\ifnum#1=11 %
\hatcurPPfluxavglogeccenxxxxxA
\else
\ifnum#1=12 %
\hatcurPPfluxavglogeccenxxxxxB
\else
??????\fi
\fi
}
\newcommand{\hatcurPPfluxperidimeccen}[1]{\ifnum#1=11 %
\hatcurPPfluxperidimeccenxxxxxA
\else
\ifnum#1=12 %
\hatcurPPfluxperidimeccenxxxxxB
\else
??????\fi
\fi
}
\newcommand{\hatcurPPfluxperieccen}[1]{\ifnum#1=11 %
\hatcurPPfluxperieccenxxxxxA
\else
\ifnum#1=12 %
\hatcurPPfluxperieccenxxxxxB
\else
??????\fi
\fi
}
\newcommand{\hatcurPPgeccen}[1]{\ifnum#1=11 %
\hatcurPPgeccenxxxxxA
\else
\ifnum#1=12 %
\hatcurPPgeccenxxxxxB
\else
??????\fi
\fi
}
\newcommand{\hatcurPPieccen}[1]{\ifnum#1=11 %
\hatcurPPieccenxxxxxA
\else
\ifnum#1=12 %
\hatcurPPieccenxxxxxB
\else
??????\fi
\fi
}
\newcommand{\hatcurPPloggeccen}[1]{\ifnum#1=11 %
\hatcurPPloggeccenxxxxxA
\else
\ifnum#1=12 %
\hatcurPPloggeccenxxxxxB
\else
??????\fi
\fi
}
\newcommand{\hatcurPPmeccen}[1]{\ifnum#1=11 %
\hatcurPPmeccenxxxxxA
\else
\ifnum#1=12 %
\hatcurPPmeccenxxxxxB
\else
??????\fi
\fi
}
\newcommand{\hatcurPPmeeccen}[1]{\ifnum#1=11 %
\hatcurPPmeeccenxxxxxA
\else
\ifnum#1=12 %
\hatcurPPmeeccenxxxxxB
\else
??????\fi
\fi
}
\newcommand{\hatcurPPmelongeccen}[1]{\ifnum#1=11 %
\hatcurPPmelongeccenxxxxxA
\else
\ifnum#1=12 %
\hatcurPPmelongeccenxxxxxB
\else
??????\fi
\fi
}
\newcommand{\hatcurPPmeshorteccen}[1]{\ifnum#1=11 %
\hatcurPPmeshorteccenxxxxxA
\else
\ifnum#1=12 %
\hatcurPPmeshorteccenxxxxxB
\else
??????\fi
\fi
}
\newcommand{\hatcurPPmlongeccen}[1]{\ifnum#1=11 %
\hatcurPPmlongeccenxxxxxA
\else
\ifnum#1=12 %
\hatcurPPmlongeccenxxxxxB
\else
??????\fi
\fi
}
\newcommand{\hatcurPPmrcorreccen}[1]{\ifnum#1=11 %
\hatcurPPmrcorreccenxxxxxA
\else
\ifnum#1=12 %
\hatcurPPmrcorreccenxxxxxB
\else
??????\fi
\fi
}
\newcommand{\hatcurPPmshorteccen}[1]{\ifnum#1=11 %
\hatcurPPmshorteccenxxxxxA
\else
\ifnum#1=12 %
\hatcurPPmshorteccenxxxxxB
\else
??????\fi
\fi
}
\newcommand{\hatcurPPperieccen}[1]{\ifnum#1=11 %
\hatcurPPperieccenxxxxxA
\else
\ifnum#1=12 %
\hatcurPPperieccenxxxxxB
\else
??????\fi
\fi
}
\newcommand{\hatcurPPphiconjeccen}[1]{\ifnum#1=11 %
\hatcurPPphiconjeccenxxxxxA
\else
\ifnum#1=12 %
\hatcurPPphiconjeccenxxxxxB
\else
??????\fi
\fi
}
\newcommand{\hatcurPPreccen}[1]{\ifnum#1=11 %
\hatcurPPreccenxxxxxA
\else
\ifnum#1=12 %
\hatcurPPreccenxxxxxB
\else
??????\fi
\fi
}
\newcommand{\hatcurPPreeccen}[1]{\ifnum#1=11 %
\hatcurPPreeccenxxxxxA
\else
\ifnum#1=12 %
\hatcurPPreeccenxxxxxB
\else
??????\fi
\fi
}
\newcommand{\hatcurPPrelongeccen}[1]{\ifnum#1=11 %
\hatcurPPrelongeccenxxxxxA
\else
\ifnum#1=12 %
\hatcurPPrelongeccenxxxxxB
\else
??????\fi
\fi
}
\newcommand{\hatcurPPreshorteccen}[1]{\ifnum#1=11 %
\hatcurPPreshorteccenxxxxxA
\else
\ifnum#1=12 %
\hatcurPPreshorteccenxxxxxB
\else
??????\fi
\fi
}
\newcommand{\hatcurPPrhoeccen}[1]{\ifnum#1=11 %
\hatcurPPrhoeccenxxxxxA
\else
\ifnum#1=12 %
\hatcurPPrhoeccenxxxxxB
\else
??????\fi
\fi
}
\newcommand{\hatcurPPrlongeccen}[1]{\ifnum#1=11 %
\hatcurPPrlongeccenxxxxxA
\else
\ifnum#1=12 %
\hatcurPPrlongeccenxxxxxB
\else
??????\fi
\fi
}
\newcommand{\hatcurPPrshorteccen}[1]{\ifnum#1=11 %
\hatcurPPrshorteccenxxxxxA
\else
\ifnum#1=12 %
\hatcurPPrshorteccenxxxxxB
\else
??????\fi
\fi
}
\newcommand{\hatcurPPtcirceccen}[1]{\ifnum#1=11 %
\hatcurPPtcirceccenxxxxxA
\else
\ifnum#1=12 %
\hatcurPPtcirceccenxxxxxB
\else
??????\fi
\fi
}
\newcommand{\hatcurPPteffeccen}[1]{\ifnum#1=11 %
\hatcurPPteffeccenxxxxxA
\else
\ifnum#1=12 %
\hatcurPPteffeccenxxxxxB
\else
??????\fi
\fi
}
\newcommand{\hatcurPPthetaeccen}[1]{\ifnum#1=11 %
\hatcurPPthetaeccenxxxxxA
\else
\ifnum#1=12 %
\hatcurPPthetaeccenxxxxxB
\else
??????\fi
\fi
}
\newcommand{\hatcurPPtinfalleccen}[1]{\ifnum#1=11 %
\hatcurPPtinfalleccenxxxxxA
\else
\ifnum#1=12 %
\hatcurPPtinfalleccenxxxxxB
\else
??????\fi
\fi
}
\newcommand{\hatcurRVecceneccen}[1]{\ifnum#1=11 %
\hatcurRVecceneccenxxxxxA
\else
\ifnum#1=12 %
\hatcurRVecceneccenxxxxxB
\else
??????\fi
\fi
}
\newcommand{\hatcurRVeccentwosiglimeccen}[1]{\ifnum#1=11 %
\hatcurRVeccentwosiglimeccenxxxxxA
\else
\ifnum#1=12 %
\hatcurRVeccentwosiglimeccenxxxxxB
\else
??????\fi
\fi
}
\newcommand{\hatcurRVfitrmsAeccen}[1]{\ifnum#1=11 %
\hatcurRVfitrmsAeccenxxxxxA
\else
\ifnum#1=12 %
\hatcurRVfitrmsAeccenxxxxxB
\else
??????\fi
\fi
}
\newcommand{\hatcurRVfitrmsBeccen}[1]{\ifnum#1=11 %
\hatcurRVfitrmsBeccenxxxxxA
\else
\ifnum#1=12 %
\hatcurRVfitrmsBeccenxxxxxB
\else
??????\fi
\fi
}
\newcommand{\hatcurRVgammaAeccen}[1]{\ifnum#1=11 %
\hatcurRVgammaAeccenxxxxxA
\else
\ifnum#1=12 %
\hatcurRVgammaAeccenxxxxxB
\else
??????\fi
\fi
}
\newcommand{\hatcurRVgammaBeccen}[1]{\ifnum#1=11 %
\hatcurRVgammaBeccenxxxxxA
\else
\ifnum#1=12 %
\hatcurRVgammaBeccenxxxxxB
\else
??????\fi
\fi
}
\newcommand{\hatcurRVheccen}[1]{\ifnum#1=11 %
\hatcurRVheccenxxxxxA
\else
\ifnum#1=12 %
\hatcurRVheccenxxxxxB
\else
??????\fi
\fi
}
\newcommand{\hatcurRVjitterAeccen}[1]{\ifnum#1=11 %
\hatcurRVjitterAeccenxxxxxA
\else
\ifnum#1=12 %
\hatcurRVjitterAeccenxxxxxB
\else
??????\fi
\fi
}
\newcommand{\hatcurRVjitterBeccen}[1]{\ifnum#1=11 %
\hatcurRVjitterBeccenxxxxxA
\else
\ifnum#1=12 %
\hatcurRVjitterBeccenxxxxxB
\else
??????\fi
\fi
}
\newcommand{\hatcurRVkeccen}[1]{\ifnum#1=11 %
\hatcurRVkeccenxxxxxA
\else
\ifnum#1=12 %
\hatcurRVkeccenxxxxxB
\else
??????\fi
\fi
}
\newcommand{\hatcurRVKeccen}[1]{\ifnum#1=11 %
\hatcurRVKeccenxxxxxA
\else
\ifnum#1=12 %
\hatcurRVKeccenxxxxxB
\else
??????\fi
\fi
}
\newcommand{\hatcurRVomegaeccen}[1]{\ifnum#1=11 %
\hatcurRVomegaeccenxxxxxA
\else
\ifnum#1=12 %
\hatcurRVomegaeccenxxxxxB
\else
??????\fi
\fi
}
\newcommand{\hatcurRVrheccen}[1]{\ifnum#1=11 %
\hatcurRVrheccenxxxxxA
\else
\ifnum#1=12 %
\hatcurRVrheccenxxxxxB
\else
??????\fi
\fi
}
\newcommand{\hatcurRVrkeccen}[1]{\ifnum#1=11 %
\hatcurRVrkeccenxxxxxA
\else
\ifnum#1=12 %
\hatcurRVrkeccenxxxxxB
\else
??????\fi
\fi
}
\newcommand{\hatcurRVtroneeccen}[1]{\ifnum#1=11 %
\hatcurRVtroneeccenxxxxxA
\else
\ifnum#1=12 %
\hatcurRVtroneeccenxxxxxB
\else
??????\fi
\fi
}
\newcommand{\hatcurRVtrtwoeccen}[1]{\ifnum#1=11 %
\hatcurRVtrtwoeccenxxxxxA
\else
\ifnum#1=12 %
\hatcurRVtrtwoeccenxxxxxB
\else
??????\fi
\fi
}
\newcommand{\hatcurSMEiiloggeccen}[1]{\ifnum#1=11 %
\hatcurSMEiiloggeccenxxxxxA
\else
\ifnum#1=12 %
\hatcurSMEiiloggeccenxxxxxB
\else
??????\fi
\fi
}
\newcommand{\hatcurSMEiiteffeccen}[1]{\ifnum#1=11 %
\hatcurSMEiiteffeccenxxxxxA
\else
\ifnum#1=12 %
\hatcurSMEiiteffeccenxxxxxB
\else
??????\fi
\fi
}
\newcommand{\hatcurSMEiivsineccen}[1]{\ifnum#1=11 %
\hatcurSMEiivsineccenxxxxxA
\else
\ifnum#1=12 %
\hatcurSMEiivsineccenxxxxxB
\else
??????\fi
\fi
}
\newcommand{\hatcurSMEiizfeheccen}[1]{\ifnum#1=11 %
\hatcurSMEiizfeheccenxxxxxA
\else
\ifnum#1=12 %
\hatcurSMEiizfeheccenxxxxxB
\else
??????\fi
\fi
}
\newcommand{\hatcurSMEiizfehshorteccen}[1]{\ifnum#1=11 %
\hatcurSMEiizfehshorteccenxxxxxA
\else
\ifnum#1=12 %
\hatcurSMEiizfehshorteccenxxxxxB
\else
??????\fi
\fi
}
\newcommand{\hatcurSMEiloggeccen}[1]{\ifnum#1=11 %
\hatcurSMEiloggeccenxxxxxA
\else
\ifnum#1=12 %
\hatcurSMEiloggeccenxxxxxB
\else
??????\fi
\fi
}
\newcommand{\hatcurSMEiteffeccen}[1]{\ifnum#1=11 %
\hatcurSMEiteffeccenxxxxxA
\else
\ifnum#1=12 %
\hatcurSMEiteffeccenxxxxxB
\else
??????\fi
\fi
}
\newcommand{\hatcurSMEivmaceccen}[1]{\ifnum#1=11 %
\hatcurSMEivmaceccenxxxxxA
\else
\ifnum#1=12 %
\hatcurSMEivmaceccenxxxxxB
\else
??????\fi
\fi
}
\newcommand{\hatcurSMEivmiceccen}[1]{\ifnum#1=11 %
\hatcurSMEivmiceccenxxxxxA
\else
\ifnum#1=12 %
\hatcurSMEivmiceccenxxxxxB
\else
??????\fi
\fi
}
\newcommand{\hatcurSMEivsineccen}[1]{\ifnum#1=11 %
\hatcurSMEivsineccenxxxxxA
\else
\ifnum#1=12 %
\hatcurSMEivsineccenxxxxxB
\else
??????\fi
\fi
}
\newcommand{\hatcurSMEizfeheccen}[1]{\ifnum#1=11 %
\hatcurSMEizfeheccenxxxxxA
\else
\ifnum#1=12 %
\hatcurSMEizfeheccenxxxxxB
\else
??????\fi
\fi
}
\newcommand{\hatcurSMEizfehshorteccen}[1]{\ifnum#1=11 %
\hatcurSMEizfehshorteccenxxxxxA
\else
\ifnum#1=12 %
\hatcurSMEizfehshorteccenxxxxxB
\else
??????\fi
\fi
}
\newcommand{\hatcurXAveccen}[1]{\ifnum#1=11 %
\hatcurXAveccenxxxxxA
\else
\ifnum#1=12 %
\hatcurXAveccenxxxxxB
\else
??????\fi
\fi
}
\newcommand{\hatcurXdisteccen}[1]{\ifnum#1=11 %
\hatcurXdisteccenxxxxxA
\else
\ifnum#1=12 %
\hatcurXdisteccenxxxxxB
\else
??????\fi
\fi
}
\newcommand{\hatcurXdistredeccen}[1]{\ifnum#1=11 %
\hatcurXdistredeccenxxxxxA
\else
\ifnum#1=12 %
\hatcurXdistredeccenxxxxxB
\else
??????\fi
\fi
}
\newcommand{\hatcurXEBVeccen}[1]{\ifnum#1=11 %
\hatcurXEBVeccenxxxxxA
\else
\ifnum#1=12 %
\hatcurXEBVeccenxxxxxB
\else
??????\fi
\fi
}
\newcommand{\hatcurXjhisoredeccen}[1]{\ifnum#1=11 %
\hatcurXjhisoredeccenxxxxxA
\else
\ifnum#1=12 %
\hatcurXjhisoredeccenxxxxxB
\else
??????\fi
\fi
}
\newcommand{\hatcurXjkisoredeccen}[1]{\ifnum#1=11 %
\hatcurXjkisoredeccenxxxxxA
\else
\ifnum#1=12 %
\hatcurXjkisoredeccenxxxxxB
\else
??????\fi
\fi
}
\newcommand{\hatcurXmhisoredeccen}[1]{\ifnum#1=11 %
\hatcurXmhisoredeccenxxxxxA
\else
\ifnum#1=12 %
\hatcurXmhisoredeccenxxxxxB
\else
??????\fi
\fi
}
\newcommand{\hatcurXmiisoredeccen}[1]{\ifnum#1=11 %
\hatcurXmiisoredeccenxxxxxA
\else
\ifnum#1=12 %
\hatcurXmiisoredeccenxxxxxB
\else
??????\fi
\fi
}
\newcommand{\hatcurXmjisoredeccen}[1]{\ifnum#1=11 %
\hatcurXmjisoredeccenxxxxxA
\else
\ifnum#1=12 %
\hatcurXmjisoredeccenxxxxxB
\else
??????\fi
\fi
}
\newcommand{\hatcurXmkisoredeccen}[1]{\ifnum#1=11 %
\hatcurXmkisoredeccenxxxxxA
\else
\ifnum#1=12 %
\hatcurXmkisoredeccenxxxxxB
\else
??????\fi
\fi
}
\newcommand{\hatcurXmvisoredeccen}[1]{\ifnum#1=11 %
\hatcurXmvisoredeccenxxxxxA
\else
\ifnum#1=12 %
\hatcurXmvisoredeccenxxxxxB
\else
??????\fi
\fi
}
\newcommand{\hatcurXsecdureccen}[1]{\ifnum#1=11 %
\hatcurXsecdureccenxxxxxA
\else
\ifnum#1=12 %
\hatcurXsecdureccenxxxxxB
\else
??????\fi
\fi
}
\newcommand{\hatcurXsecingdureccen}[1]{\ifnum#1=11 %
\hatcurXsecingdureccenxxxxxA
\else
\ifnum#1=12 %
\hatcurXsecingdureccenxxxxxB
\else
??????\fi
\fi
}
\newcommand{\hatcurXsecondaryeccen}[1]{\ifnum#1=11 %
\hatcurXsecondaryeccenxxxxxA
\else
\ifnum#1=12 %
\hatcurXsecondaryeccenxxxxxB
\else
??????\fi
\fi
}
\newcommand{\hatcurXsecphaseeccen}[1]{\ifnum#1=11 %
\hatcurXsecphaseeccenxxxxxA
\else
\ifnum#1=12 %
\hatcurXsecphaseeccenxxxxxB
\else
??????\fi
\fi
}
\newcommand{\hatcurXviisoredeccen}[1]{\ifnum#1=11 %
\hatcurXviisoredeccenxxxxxA
\else
\ifnum#1=12 %
\hatcurXviisoredeccenxxxxxB
\else
??????\fi
\fi
}
\newcommand{\hatcurXvkisoredeccen}[1]{\ifnum#1=11 %
\hatcurXvkisoredeccenxxxxxA
\else
\ifnum#1=12 %
\hatcurXvkisoredeccenxxxxxB
\else
??????\fi
\fi
}

\newcommand{\hatcurxxxxxA}{HATS-11}
\newcommand{\hatcurbxxxxxA}{HATS-11b}
\newcommand{\hatcurcxxxxxA}{HATS-11c}

\newcommand{\hatcurplanetnumxxxxxA}{11}

\newcommand{\hatcurRVgammaabsxxxxxA}{$-58.324 \pm 0.012$}                           % Absolute Gamma velocity

\newcommand{\hatcurRVgammarelxxxxxA}{***TBD***}                           % Relative Gamma velocity. Typically that of the Keck RVs.

\newcommand{\hatcurCCtassvixxxxxA}{***TBD***}                  % TASS V-I

\newcommand{\hatcurSMEversionxxxxxA}{ii}                                       % Final SME version:i or ii?

\newcommand{\hatcurisoshortxxxxxA}{YY}
\newcommand{\hatcurisofullxxxxxA}{Yonsei-Yale (YY)}
\newcommand{\hatcurisocitexxxxxA}{yi:2001}
\newcommand{\hatcurlumindxxxxxA}{\arstar}
\newcommand{\hatcurjhkfilsetxxxxxA}{ESO}
\newcommand{\hatcurSMEteffxxxxxA}{\ifthenelse{\equal{\hatcurSMEversionxxxxxA}{i}}{\hatcurSMEiteff{\hatcurplanetnumxxxxxA}}{\hatcurSMEiiteff{\hatcurplanetnumxxxxxA}}}
\newcommand{\hatcurSMEzfehxxxxxA}{\ifthenelse{\equal{\hatcurSMEversionxxxxxA}{i}}{\hatcurSMEizfeh{\hatcurplanetnumxxxxxA}}{\hatcurSMEiizfeh{\hatcurplanetnumxxxxxA}}}
\newcommand{\hatcurSMEzfehshortxxxxxA}{\ifthenelse{\equal{\hatcurSMEversionxxxxxA}{i}}{\hatcurSMEizfehshort{\hatcurplanetnumxxxxxA}}{\hatcurSMEiizfehshort{\hatcurplanetnumxxxxxA}}}
\newcommand{\hatcurSMEloggxxxxxA}{\ifthenelse{\equal{\hatcurSMEversionxxxxxA}{i}}{\hatcurSMEilogg{\hatcurplanetnumxxxxxA}}{\hatcurSMEiilogg{\hatcurplanetnumxxxxxA}}}
\newcommand{\hatcurSMEvsinxxxxxA}{\ifthenelse{\equal{\hatcurSMEversionxxxxxA}{i}}{\hatcurSMEivsin{\hatcurplanetnumxxxxxA}}{\hatcurSMEiivsin{\hatcurplanetnumxxxxxA}}}
\newcommand{\hatcurSMEvmacxxxxxA}{\ifthenelse{\equal{\hatcurSMEversionxxxxxA}{i}}{\hatcurSMEivmac{\hatcurplanetnumxxxxxA}}{\hatcurSMEiivmac{\hatcurplanetnumxxxxxA}}}
\newcommand{\hatcurSMEvmicxxxxxA}{\ifthenelse{\equal{\hatcurSMEversionxxxxxA}{i}}{\hatcurSMEivmic{\hatcurplanetnumxxxxxA}}{\hatcurSMEiivmic{\hatcurplanetnumxxxxxA}}}

\newcommand{\hatcurxxxxxB}{HATS-12}
\newcommand{\hatcurbxxxxxB}{HATS-12b}
\newcommand{\hatcurcxxxxxB}{HATS-12c}

\newcommand{\hatcurplanetnumxxxxxB}{12}

\newcommand{\hatcurRVgammaabsxxxxxB}{$-21.661 \pm 0.012$}                           % Absolute Gamma velocity

\newcommand{\hatcurRVgammarelxxxxxB}{***TBD***}                           % Relative Gamma velocity. Typically that of the Keck RVs.

\newcommand{\hatcurCCtassvixxxxxB}{***TBD***}                  % TASS V-I

\newcommand{\hatcurSMEversionxxxxxB}{ii}                                       % Final SME version:i or ii?

\newcommand{\hatcurisoshortxxxxxB}{YY}
\newcommand{\hatcurisofullxxxxxB}{Yonsei-Yale (YY)}
\newcommand{\hatcurisocitexxxxxB}{yi:2001}
\newcommand{\hatcurlumindxxxxxB}{\arstar}

\newcommand{\hatcurjhkfilsetxxxxxB}{ESO}

\newcommand{\hatcurSMEteffxxxxxB}{\ifthenelse{\equal{\hatcurSMEversionxxxxxB}{i}}{\hatcurSMEiteff{\hatcurplanetnumxxxxxB}}{\hatcurSMEiiteff{\hatcurplanetnumxxxxxB}}}
\newcommand{\hatcurSMEzfehxxxxxB}{\ifthenelse{\equal{\hatcurSMEversionxxxxxB}{i}}{\hatcurSMEizfeh{\hatcurplanetnumxxxxxB}}{\hatcurSMEiizfeh{\hatcurplanetnumxxxxxB}}}
\newcommand{\hatcurSMEzfehshortxxxxxB}{\ifthenelse{\equal{\hatcurSMEversionxxxxxB}{i}}{\hatcurSMEizfehshort{\hatcurplanetnumxxxxxB}}{\hatcurSMEiizfehshort{\hatcurplanetnumxxxxxB}}}
\newcommand{\hatcurSMEloggxxxxxB}{\ifthenelse{\equal{\hatcurSMEversionxxxxxB}{i}}{\hatcurSMEilogg{\hatcurplanetnumxxxxxB}}{\hatcurSMEiilogg{\hatcurplanetnumxxxxxB}}}
\newcommand{\hatcurSMEvsinxxxxxB}{\ifthenelse{\equal{\hatcurSMEversionxxxxxB}{i}}{\hatcurSMEivsin{\hatcurplanetnumxxxxxB}}{\hatcurSMEiivsin{\hatcurplanetnumxxxxxB}}}
\newcommand{\hatcurSMEvmacxxxxxB}{\ifthenelse{\equal{\hatcurSMEversionxxxxxB}{i}}{\hatcurSMEivmac{\hatcurplanetnumxxxxxB}}{\hatcurSMEiivmac{\hatcurplanetnumxxxxxB}}}
\newcommand{\hatcurSMEvmicxxxxxB}{\ifthenelse{\equal{\hatcurSMEversionxxxxxB}{i}}{\hatcurSMEivmic{\hatcurplanetnumxxxxxB}}{\hatcurSMEiivmic{\hatcurplanetnumxxxxxB}}}

\newcommand{\hatcur}[1]{\ifnum#1=11 %
\hatcurxxxxxA
\else
\ifnum#1=12 %
\hatcurxxxxxB
\else
??????\fi
\fi
}
\newcommand{\hatcurb}[1]{\ifnum#1=11 %
\hatcurbxxxxxA
\else
\ifnum#1=12 %
\hatcurbxxxxxB
\else
??????\fi
\fi
}
\newcommand{\hatcurc}[1]{\ifnum#1=11 %
\hatcurcxxxxxA
\else
\ifnum#1=12 %
\hatcurcxxxxxB
\else
??????\fi
\fi
}
\newcommand{\hatcurCCtassvi}[1]{\ifnum#1=11 %
\hatcurCCtassvixxxxxA
\else
\ifnum#1=12 %
\hatcurCCtassvixxxxxB
\else
??????\fi
\fi
}
\newcommand{\hatcurisocite}[1]{\ifnum#1=11 %
\hatcurisocitexxxxxA
\else
\ifnum#1=12 %
\hatcurisocitexxxxxB
\else
??????\fi
\fi
}
\newcommand{\hatcurisofull}[1]{\ifnum#1=11 %
\hatcurisofullxxxxxA
\else
\ifnum#1=12 %
\hatcurisofullxxxxxB
\else
??????\fi
\fi
}
\newcommand{\hatcurisoshort}[1]{\ifnum#1=11 %
\hatcurisoshortxxxxxA
\else
\ifnum#1=12 %
\hatcurisoshortxxxxxB
\else
??????\fi
\fi
}
\newcommand{\hatcurjhkfilset}[1]{\ifnum#1=11 %
\hatcurjhkfilsetxxxxxA
\else
\ifnum#1=12 %
\hatcurjhkfilsetxxxxxB
\else
??????\fi
\fi
}
\newcommand{\hatcurlumind}[1]{\ifnum#1=11 %
\hatcurlumindxxxxxA
\else
\ifnum#1=12 %
\hatcurlumindxxxxxB
\else
??????\fi
\fi
}
\newcommand{\hatcurplanetnum}[1]{\ifnum#1=11 %
\hatcurplanetnumxxxxxA
\else
\ifnum#1=12 %
\hatcurplanetnumxxxxxB
\else
??????\fi
\fi
}
\newcommand{\hatcurRVgammaabs}[1]{\ifnum#1=11 %
\hatcurRVgammaabsxxxxxA
\else
\ifnum#1=12 %
\hatcurRVgammaabsxxxxxB
\else
??????\fi
\fi
}
\newcommand{\hatcurRVgammarel}[1]{\ifnum#1=11 %
\hatcurRVgammarelxxxxxA
\else
\ifnum#1=12 %
\hatcurRVgammarelxxxxxB
\else
??????\fi
\fi
}
\newcommand{\hatcurSMElogg}[1]{\ifnum#1=11 %
\hatcurSMEloggxxxxxA
\else
\ifnum#1=12 %
\hatcurSMEloggxxxxxB
\else
??????\fi
\fi
}
\newcommand{\hatcurSMEteff}[1]{\ifnum#1=11 %
\hatcurSMEteffxxxxxA
\else
\ifnum#1=12 %
\hatcurSMEteffxxxxxB
\else
??????\fi
\fi
}
\newcommand{\hatcurSMEversion}[1]{\ifnum#1=11 %
\hatcurSMEversionxxxxxA
\else
\ifnum#1=12 %
\hatcurSMEversionxxxxxB
\else
??????\fi
\fi
}
\newcommand{\hatcurSMEvmac}[1]{\ifnum#1=11 %
\hatcurSMEvmacxxxxxA
\else
\ifnum#1=12 %
\hatcurSMEvmacxxxxxB
\else
??????\fi
\fi
}
\newcommand{\hatcurSMEvmic}[1]{\ifnum#1=11 %
\hatcurSMEvmicxxxxxA
\else
\ifnum#1=12 %
\hatcurSMEvmicxxxxxB
\else
??????\fi
\fi
}
\newcommand{\hatcurSMEvsin}[1]{\ifnum#1=11 %
\hatcurSMEvsinxxxxxA
\else
\ifnum#1=12 %
\hatcurSMEvsinxxxxxB
\else
??????\fi
\fi
}
\newcommand{\hatcurSMEzfeh}[1]{\ifnum#1=11 %
\hatcurSMEzfehxxxxxA
\else
\ifnum#1=12 %
\hatcurSMEzfehxxxxxB
\else
??????\fi
\fi
}
\newcommand{\hatcurSMEzfehshort}[1]{\ifnum#1=11 %
\hatcurSMEzfehshortxxxxxA
\else
\ifnum#1=12 %
\hatcurSMEzfehshortxxxxxB
\else
??????\fi
\fi
}

\newcounter{planetcounter}

\newboolean{emulateapj}
\setboolean{emulateapj}{true}

\newboolean{rvtablelong}

\setboolean{rvtablelong}{true}

\newboolean{astroph}
\setboolean{astroph}{true}

\shortauthors{Rabus et~al.}
\shorttitle{\hatcur{11}\lowercase{b} and \hatcur{12}\lowercase{b}}
\ifthenelse{\boolean{emulateapj}}{
    \newcommand{\titledag}{$\dagger$}
}{
    \newcommand{\titledag}{\dagger}
}

\begin{document}

%% Titlepage
\title{%%
\hatcur{11}\lowercase{b} and \hatcur{12}\lowercase{b}: Two transiting Hot Jupiters orbiting sub-solar metallicity stars selected for the K2 Campaign 7\altaffilmark{\titledag}
}

%% Authors
\author{M. Rabus\altaffilmark{1,2}}
\author{A. Jord\'an\altaffilmark{1}}
\author{J.~D. Hartman\altaffilmark{3}}
\author{G.~\'A. Bakos\altaffilmark{3}}
\author{N. Espinoza\altaffilmark{1}} 
\author{R. Brahm\altaffilmark{1}}
\author{K. Penev\altaffilmark{3}}
\author{S. Ciceri\altaffilmark{2}}
\author{G. Zhou\altaffilmark{4}}
\author{D. Bayliss\altaffilmark{4,5}}
\author{L. Mancini\altaffilmark{2}}
\author{W. Bhatti\altaffilmark{3}}
\author{M. de Val-Borro\altaffilmark{3}}
\author{Z. Csbury\altaffilmark{3}}
\author{B. Sato\altaffilmark{6}}
\author{T.-G. Tan\altaffilmark{7}}
\author{T. Henning\altaffilmark{2}}
\author{B. Schmidt\altaffilmark{4}}
\author{J. Bento\altaffilmark{4}}
\author{V. Suc\altaffilmark{1}}
\author{R. Noyes\altaffilmark{8}}
\author{J. L\'az\'ar\altaffilmark{9}}
\author{I. Papp\altaffilmark{9}}
\author{P. S\'ari\altaffilmark{9}}

    \altaffiltext{1}{Instituto de Astrof\'isica, Facultad de F\'isica,
  Pontificia Universidad Cat\'olica de Chile, Av. Vicu\~na Mackenna
  4860, 7820436 Macul, Santiago, Chile; mrabus@astro.puc.cl}
  \altaffiltext{2}{Max
 Planck Institute for Astronomy, Heidelberg, Germany}
  \altaffiltext{3}{Department of Astrophysical Sciences, Princeton University, NJ 08544, USA}
  \altaffiltext{4}{Research School of Astronomy and Astrophysics, Australian National University, Canberra, ACT 2611, Australia
}
  \altaffiltext{5}{Observatoire Astronomique de l'Universit\'e de Gen\`eve, 51ch.
des Maillettes, 1290 Versoix, Switzerland}

  \altaffiltext{6}{Department of Earth and Planetary Sciences, Tokyo Institute of Technology, 2-12-1 Ookayama, Meguro-ku, Tokyo 152-8551, Japan}

  \altaffiltext{7}{Perth Exoplanet Survey Telescope, Perth, Australia}
  
  \altaffiltext{8}{Harvard-Smithsonian Center for Astrophysics, Cambridge, MA 02138, USA}  

  \altaffiltext{9}{Hungarian Astronomical Association, Budapest, Hungary}

\altaffiltext{$\dagger$}{%%
 The HATSouth network is operated by a collaboration consisting of
Princeton University (PU), the Max Planck Institute f\"ur Astronomie
(MPIA), the Australian National University (ANU), and the Pontificia
Universidad Cat\'olica de Chile (PUC).  The station at Las Campanas
Observatory (LCO) of the Carnegie Institute is operated by PU in
conjunction with PUC, the station at the High Energy Spectroscopic
Survey (H.E.S.S.) site is operated in conjunction with MPIA, and the
station at Siding Spring Observatory (SSO) is operated jointly with
ANU.
 Based in part on data collected at Subaru Telescope, which is
 operated by the National Astronomical Observatory of Japan. Based in
 part on observations made with the MPG~2.2\,m Telescope at the ESO
 Observatory in La Silla.
}

%% EOF authors

% #####################################################################
%% abstract
\begin{abstract}
%++++++++++++++++++++++++++++++++++++++++++++++++++++++++++++++++++++++

We report the discovery of two transiting extrasolar planets from the HATSouth survey. \hatcur{11}, a V=14.1 \hatcurISOspec{11}-star shows a periodic \hatcurLCdip{11}\,mmag dip in its \lc\ every 3.6192 days and a radial velocity variation consistent with a Keplerian orbit. \hatcur{11} has a mass of \hatcurISOm{11}\,\msun, a radius of \hatcurISOr{11}\,\rsun\ and an effective temperature of \hatcurSMEteff{11}\,K, while its  companion is a \hatcurPPmlong{11}\,\mjup, \hatcurPPrlong{11}\,\rjup\  planet in a circular orbit.
\hatcur{12} shows a periodic 5.1 mmag flux decrease every 3.1428 days and Keplerian RV variations around a V=12.8 \hatcurISOspec{12}-star. \hatcur{12} has a mass of \hatcurISOm{12}\,\msun, a radius of \hatcurISOr{12}\,\rsun\, and an effective temperature of \hatcurSMEteff{12}\,K.
For \hatcurb{12}, our measurements indicate that this is a \hatcurPPmlong{12}\,\mjup, \hatcurPPrlong{12}\,\rjup\ planet in a circular orbit.
Both host stars show sub-solar metallicity of \hatcurSMEzfeh{11} dex and \hatcurSMEzfeh{12} dex, respectively and are (slightly) evolved stars. In fact, \hatcur{11} is amongst the most metal-poor and, \hatcur{12} is amongst the most evolved stars hosting a hot Jupiter planet. Importantly, \hatcur{11} and \hatcur{12} have been observed in long cadence by Kepler as part of K2 campaign 7 (EPIC216414930 and EPIC218131080 respectively).

%++++++++++++++++++++++++++++++++++++++++++++++++++++++++++++++++++++++

\setcounter{footnote}{12}
\setcounter{footnote}{0}
\end{abstract}

% #####################################################################
%% keywords
\keywords{
    planetary systems ---
    stars: individual (
\setcounter{planetcounter}{1}
\hatcur{11},
\hatcurCCgsc{11}\loopcommanoperiod
\setcounter{planetcounter}{2}
\hatcur{12},
\hatcurCCgsc{12}\loopcommanoperiod
\setcounter{planetcounter}{3}
) 
    techniques: spectroscopic, photometric
}

%% EOF keywords
%% EOF titlepage

% #####################################################################
%% Introduction
\section{Introduction}
\label{sec:introduction}
%++++++++++++++++++++++++++++++++++++++++++++++++++++++++++++++++++++++
%++++++++++++++++++++++++++++++++++++++++++++++++++++++++++++++++++++++
%% EOF introduction

Transiting Extrasolar Planets, hereafter TEPs, allow us to measure many of their physical properties that are not accessible for non-transiting systems and thus occupy a prominent place among the nearly two thousands of exoplanets currently known. 

One of the most important features of TEPs is that their radii can be measured from the transit shape if the radii of the stellar hosts are known. The radius, coupled with the measurement of the planetary mass from radial velocity (RV) observations, allows the computation of the bulk density of the planet and the possibility of inferring properties about its internal structure and composition. 

The detection of close-in extrasolar giant planets has led to several theoretical challenges regarding their structure, formation, and subsequent evolution. One of the earliest realized challenges, and one which is yet to be solved, is the fact that close-in giant exoplanets often show radii that are far larger than those predicted by theories of giant planet evolution \citep[e.g. HAT-P-32b][]{hartman:2011:hat3233}. Moving forward in understanding the inflation mechanism will benefit from having larger samples of close-in, inflated gas giants, especially systems lying at some extreme of either planetary or stellar properties.

One of the most basic properties one can measure of a star is its metallicity, and much attention in the past has been given to the dependence of planet occurrence rate on the host star metallicity. Several studies have found that gas giants appear more frequently around metal-rich stars, both from RV programs \citep{santos:2004:feh, valenti:2005, johnson:2010} and from  {\em Kepler} data \citep{schlaufen:2011, buchhave:2012, everett:2013, wang:2015}. Discoveries of gas giants around metal-poor stars then add systems to a region of parameter space that is intrinsically less populated.

In this work we present two new inflated hot Jupiters orbiting metal poor stars discovered by the HATSouth survey: \hatcur{11}b and \hatcur{12}b. The structure of the paper is as follows. In \refsecl{obs} we summarize the detection of the photometric transit
signal and the subsequent spectroscopic and photometric observations
of each star to confirm the planets. In \refsecl{analysis} we analyze
the data to rule out false positive scenarios and characterize the star and planet. In  \refsecl{discussion} we put our new systems in the context of the sample of the well characterized TEPs known to date.

% #####################################################################
\section{Observations}
\label{sec:obs}
%++++++++++++++++++++++++++++++++++++++++++++++++++++++++++++++++++++++
%++++++++++++++++++++++++++++++++++++++++++++++++++++++++++++++++++++++

% =====================================================================
%% Photometric detection
\subsection{Photometric detection}
\label{sec:detection}
%++++++++++++++++++++++++++++++++++++++++++++++++++++++++++++++++++++++

The initial images of \hatcur{11} and \hatcur{12} were obtained with the
HATSouth wide-field \tel\, network consisting of 24 Takahashi E180 astrographs with an \aper\, of 18cm. The photons were detected with Apogee \ccdsize{4k} U16M ALTA CCDs. Details on the time span and number of images are
shown in \reftabl{photobs}. These images were processed following \citet{penev:2013:hats1}. The \lcs\ were trend-filtered following \citet{kovacs:2005:TFA} and searched for
periodic box-shaped signals using the Box Least-Squares \citep{kovacs:2002:BLS}
method. For \hatcur{11} (\hatcurCCtwomass{11}; $\alpha = \hatcurCCra{11}$, $\delta = \hatcurCCdec{11}$; J2000; V=\hatcurCCtassmv{11}) the discovery \lc\, showed a photometric precision between 10 and 14 mmag per point and 
the BLS algorithm detected a dip of \hatcurLCdip{11}\,mmag every \hatcurLCPshort{11}\,days. For the brighter \hatcur{12} star (\hatcurCCtwomass{12}; $\alpha = \hatcurCCra{12}$, $\delta = \hatcurCCdec{12}$; J2000; V=\hatcurCCtassmv{12}) the precision
was between 6 and 7 mmag per point and a periodic flux decrease of \hatcurLCdip{12}\,mmag with $P=$\hatcurLCPshort{12}\,days was found, making this the third shallowest ground-based discovery to date (after HAT-P-11 \citep{bakos:2010:hat11} and WASP-73 \citep{delrez:2014}). Both \lcs\, are shown in \reffigl{hatsouth} and the numerical data is available in \reftabl{phfu}. These initial detections triggered further spectroscopic and photometric follow-up observations in order to confirm the transit and the planetary nature as explained in the following sections.

%
%
%% ----------------
\ifthenelse{\boolean{emulateapj}}{
    \begin{figure*}[!ht]
}{
    \begin{figure}[!ht]
}
\plottwo{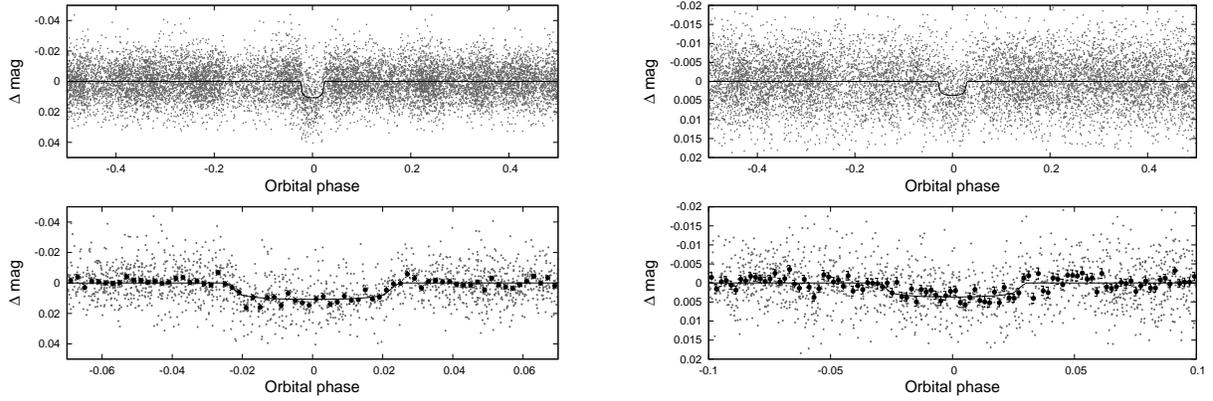}{\hatcurhtr{12}-hs.eps}
\caption[]{
    Phase-folded unbinned HATSouth light curves for \hatcur{11} (left) and \hatcur{12} (right). In each case we show two panels. The
    top panel shows the full light curve, while the bottom panel shows
    the light curve zoomed-in on the transit. The solid lines show the
    model fits to the light curves. The dark filled circles in the
    bottom panels show the light curves binned in phase with a bin
    size of 0.002.
\label{fig:hatsouth}}
\ifthenelse{\boolean{emulateapj}}{
    \end{figure*}
}{
    \end{figure}
}
%% ----------------

%% --------------------------------------------------------------------
%% Table summarizing photometric observations
%%
\ifthenelse{\boolean{emulateapj}}{
    \begin{deluxetable*}{llrrrr}
}{
    \begin{deluxetable}{llrrrr}
}
\tablewidth{0pc}
\tabletypesize{\scriptsize}
\tablecaption{
    Summary of photometric observations
    \label{tab:photobs}
}
\tablehead{
    \multicolumn{1}{c}{Instrument/Field\tablenotemark{a}} &
    \multicolumn{1}{c}{Date(s)} &
    \multicolumn{1}{c}{\# Images} &
    \multicolumn{1}{c}{Cadence\tablenotemark{b}} &
    \multicolumn{1}{c}{Filter} &
    \multicolumn{1}{c}{Precision\tablenotemark{c}} \\
    \multicolumn{1}{c}{} &
    \multicolumn{1}{c}{} &
    \multicolumn{1}{c}{} &
    \multicolumn{1}{c}{(sec)} &
    \multicolumn{1}{c}{} &
    \multicolumn{1}{c}{(mmag)}
}
\startdata
\sidehead{\textbf{\hatcur{11}}}
~~~~HS-3.1/G579 & 2010 Mar--2011 Aug & 2229 & 304 & $r$ & 11.5 \\
~~~~HS-1.2/G579 & 2010 Mar--2011 Aug & 4298 & 301 & $r$ & 10.8 \\
~~~~HS-3.2/G579 & 2010 Mar--2011 Aug & 2144 & 304 & $r$ & 11.2 \\
~~~~HS-5.2/G579 & 2010 Sep--2011 Aug & 2768 & 303 & $r$ & 10.2 \\
~~~~PEST & 2013 Jun 15 & 138 & 131 & $R_{C}$ & 4.9 \\
~~~~Swope~1\,m & 2014 Jul 03 & 69 & 189 & $i$ & 1.6 \\
~~~~LCOGT~1\,m+SBIG & 2014 Sep 03 & 48 & 196 & $i$ & 3.8 \\
~~~~AAT+IRIS2 & 2014 Sep 07 & 778 & 16 & $K_{S}$ & 3.7 \\
~~~~LCOGT~1\,m+Sinistro & 2014 Sep 11 & 46 & 288 & $i$ & 2.7 \\
\sidehead{\textbf{\hatcur{12}}}
~~~~HS-1.2/G579 & 2010 Mar--2011 Aug & 4315 & 300 & $r$ & 5.9 \\
~~~~HS-3.2/G579 & 2010 Mar--2011 Aug & 2126 & 303 & $r$ & 6.9 \\
~~~~HS-5.2/G579 & 2010 Sep--2011 Aug & 2781 & 303 & $r$ & 6.1 \\
~~~~PEST & 2013 May 24 & 193 & 130 & $R_{C}$ & 3.4 \\
~~~~MPG~2.2m+GROND & 2013 Jul 13 & 279 & 90 & $g$ & 0.8 \\
~~~~MPG~2.2m+GROND & 2013 Jul 13 & 271 & 90 & $r$ & 1.9 \\
~~~~MPG~2.2m+GROND & 2013 Jul 13 & 273 & 90 & $i$ & 0.7 \\
~~~~DK~1.54m+DFOSC & 2013 Oct 03 & 98 & 145 & $R$ & 1.9 \\
~~~~SWOPE~1m & 2015 May 26 & 82 & 59 & $g$ & 2.1 \\
\enddata
\tablenotetext{a}{
    For HATSouth data we list the HATSouth unit, CCD and field name
    from which the observations are taken. HS-1 and -2 are located at
    Las Campanas Observatory in Chile, HS-3 and -4 are located at the
    H.E.S.S. site in Namibia, and HS-5 and -6 are located at Siding
    Spring Observatory in Australia. Each unit has 4 ccds. Each field
    corresponds to one of 838 fixed pointings used to cover the full
    4$\pi$ celestial sphere. All data from a given HATSouth field and
    CCD number are reduced together, while detrending through External
    Parameter Decorrelation (EPD) is done independently for each
    unique unit+CCD+field combination.
}
\tablenotetext{b}{
    The median time between consecutive images rounded to the nearest
    second. Due to factors such as weather, the day--night cycle,
    guiding and focus corrections the cadence is only approximately
    uniform over short timescales.
}
\tablenotetext{c}{
    The RMS scatter of the residuals from our best fit transit model for each light curve at the cadence indicated in the table.
}
\ifthenelse{\boolean{emulateapj}}{
    \end{deluxetable*}
}{
    \end{deluxetable}
}
%% --------------------------------------------------------------------

% =====================================================================
\subsection{Spectroscopic Observations}
\label{sec:obsspec}
% ++++++++++++++++++++++++++++++++++++++++++++++++++++++++++++++++++++++

We started observing both candidates spectroscopically, in a process which is split into two kinds of spectroscopic observations. A first step of reconnaissance spectroscopy serves to reject possible astrophysical false positives, like e.g.\ binary stars, and to obtain a first estimate of stellar parameters. Afterwards, stable and high precision spectroscopic measurements allow us to obtain high precision radial velocity (RV) and line bisector (BS) time series for the stars. From this we can estimate the orbital parameters as well as the presence and mass of the companion for systems that are confirmed as planets, and a precise set of stellar parameters. The spectroscopic observations are summarized in \reftabl{specobs}.

\hatcur{11} was observed first with the ANU 2.3m telescope using the WiFES spectrograph and the duPont 2.5 m telescope using the echelle spectrograph. Details on the observing strategy and data reduction for the ANU 2.3m can be found in \citet{bayliss:2013:hats3} and for the duPont in \citet{brahm:2015:hats9}. From these spectra we detect no RV variations greater than $\sim$2 \kms\, and could determine the host star's spectral type. We verified with the new information that the transit is still consistent with a planetary companion. In light of this, we continued to observe \hatcur{11} with spectrographs allowing for simultaneous wavelength calibration for RV precision, namely with CORALIE at the Euler 1.2m and FEROS at the MPG 2.2m telescope. For both spectrographs we made use of the reduction procedures described in \citet{jordan:2014:hats4} which gave us the RV and BS measurements of the spectrum. Similarly, \hatcur{12} was observed with the WiFES and the duPont echelle spectrograph. RVs/BSs were obtained with FEROS and additionally with the HDS spectrograph at the Subaru telescope. Details on the data reduction with HDS can be found in \citet{sato:2002,sato:2012}. The phased high-precision RV and BS measurements are shown for each system
in \reffigl{rvbis}; the data are presented in \reftabl{rvs}. Spectra, RVs and BSs are used in \refsecl{analysis} to reject some possible systems mimicking a planetary system.

%% --------------------------------------------------------------------
%% Table summarizing spectroscopy observations
%%
\ifthenelse{\boolean{emulateapj}}{
    \begin{deluxetable*}{llrrrrr}
}{
    \begin{deluxetable}{llrrrrrrrr}
}
\tablewidth{0pc}
\tabletypesize{\scriptsize}
\tablecaption{
    Summary of spectroscopy observations
    \label{tab:specobs}
}
\tablehead{
    \multicolumn{1}{c}{Instrument}          &
    \multicolumn{1}{c}{UT Date(s)}             &
    \multicolumn{1}{c}{\# Spec.}   &
    \multicolumn{1}{c}{Res.}          &
    \multicolumn{1}{c}{S/N Range\tablenotemark{a}}           &
    \multicolumn{1}{c}{$\gamma_{\rm RV}$\tablenotemark{b}} &
    \multicolumn{1}{c}{RV Precision\tablenotemark{c}} \\
    &
    &
    &
    \multicolumn{1}{c}{$\Delta \lambda$/$\lambda$/1000} &
    &
    \multicolumn{1}{c}{(\kms)}              &
    \multicolumn{1}{c}{(\ms)}
}
\startdata
\sidehead{\textbf{\hatcur{11}}}
ANU~2.3\,m/WiFeS & 2012 Sep 9 & 1 & 3 & 152 & $\cdots$ & $\cdots$ \\
ANU~2.3\,m/WiFeS & 2012 Sep--2013 Mar & 4 & 7 & 19--70 & -53.8 & 4000 \\
du~Pont~2.5\,m/Echelle & 2013 Aug 21 & 1 & 40 & 40 & -58.8 & 500 \\
Euler~1.2\,m/Coralie & 2013 Aug--2014 Mar & 10 & 60 & 11--17 & -58.41 & 130 \\
MPG~2.2\,m/FEROS & 2013 Apr--Sep & 10 & 48 & 16--61 & -58.32 & 42 \\
\sidehead{\textbf{\hatcur{12}}}
MPG~2.2\,m/FEROS & 2012 Aug--2013 May & 12 & 48 & 57--107 & -21.66 & 42 \\
ANU~2.3\,m/WiFeS & 2012 Sep 8 & 1 & 3 & 257 & $\cdots$ & $\cdots$ \\
Subaru~8\,m/HDS+I$_{2}$ & 2012 Sep 19--22 & 9 & 60 & 53--86 & $\cdots$ & 13 \\
Subaru~8\,m/HDS & 2012 Sep 20 & 3 & 60 & 112-118 & $\cdots$ & $\cdots$ \\
du~Pont~2.5\,m/Echelle & 2013 Aug 21 & 1 & 40 & 58 & -21.72 & 500 \\
\enddata 
\tablenotetext{a}{
    S/N per resolution element near 5180\,\AA.
}
\tablenotetext{b}{
    For the CORALIE and FEROS observations of \hatcur{11}, and for the FEROS observations of \hatcur{12}, this is the zero-point RV from the best-fit orbit. For the WiFeS and du~Pont Echelle it is the mean of the observations. We do not provide this quantity for HDS for which only relative RVs are measured, or for the lower resolution WiFeS observations which were only used to measure stellar atmospheric parameters.
}
\tablenotetext{c}{
    For High-precision RV observations included in the orbit
    determination this is the scatter in the RV residuals from the
    best-fit orbit (which may include astrophysical jitter), for other
    instruments this is either an estimate of the precision (not
    including jitter), or the measured standard deviation. We do not
    provide this quantity for low-resolution observations from the
    ANU~2.3\,m/WiFeS or for I$_{2}$-free observations made with HDS,
    as RVs are not measured from these data.
}
\ifthenelse{\boolean{emulateapj}}{
    \end{deluxetable*}
}{
    \end{deluxetable}
}
%% --------------------------------------------------------------------

%
\setcounter{planetcounter}{1}
%
%% --------------------------------------------------------------------
\ifthenelse{\boolean{emulateapj}}{
    \begin{figure*} [ht]
}{
    \begin{figure}[ht]
}
\plottwo{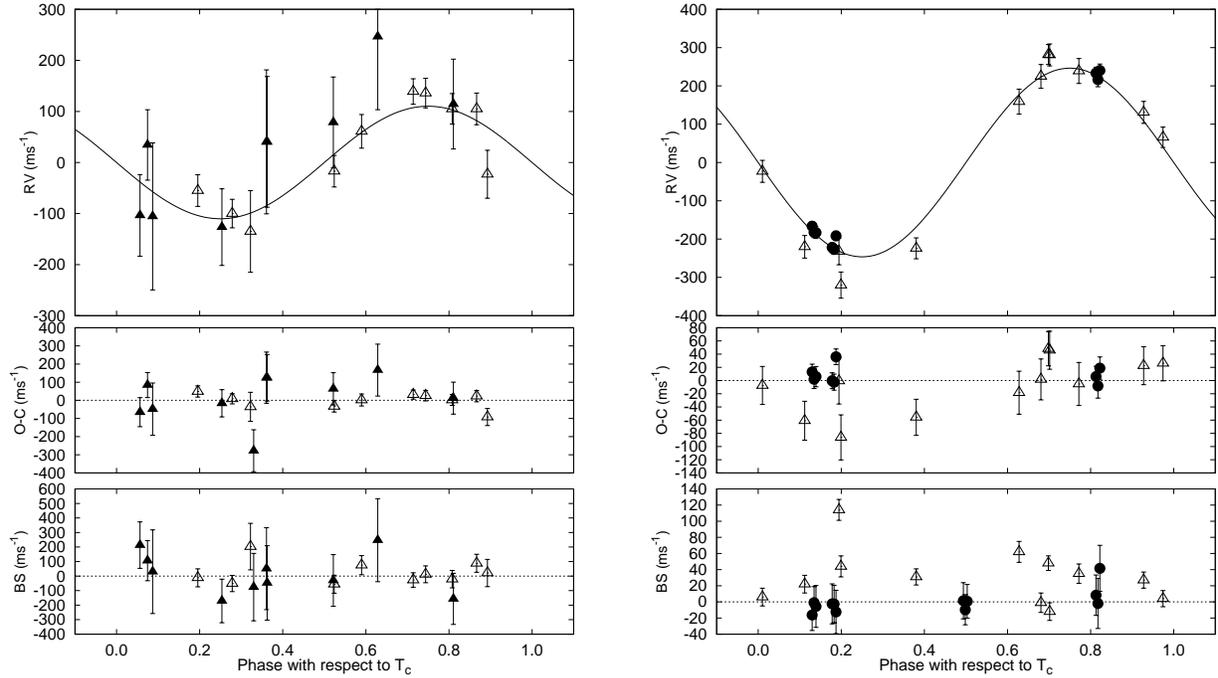}{\hatcurhtr{12}-rv.eps}
\caption{
    Phased high-precision RV measurements for \hbox{\hatcur{11}{}} (left), and \hbox{\hatcur{12}{}} (right) from FEROS (open triangles), CORALIE (filled triangles), and HDS (filled circles). In each case we show three panels. The top panel shows the phased measurements together with our best-fit model (see \reftabl{planetparam}) for each system. Solid lines show the best-fit circular orbit. Zero-phase corresponds to the time of mid-transit. The center-of-mass velocity has been subtracted. The second panel shows the velocity $O\!-\!C$ residuals from the best fit. The error bars include the jitter terms listed in \reftabl{planetparam} added in quadrature to the formal errors for each instrument. The third panel shows the bisector spans (BS), with the median value subtracted for each instrument. Note the different vertical scales of the panels.
}
\label{fig:rvbis}
\ifthenelse{\boolean{emulateapj}}{
    \end{figure*}
}{
    \end{figure}
}
%% --------------------------------------------------------------------

%% --------------------------------------------------------------------

% =====================================================================
\subsection{Photometric follow-up observations}
\label{sec:phot}
%++++++++++++++++++++++++++++++++++++++++++++++++++++++++++++++++++++++

%
\setcounter{planetcounter}{1}
%
%% --------------------------------------------------------------------
\begin{figure*}[!ht]
\plotone{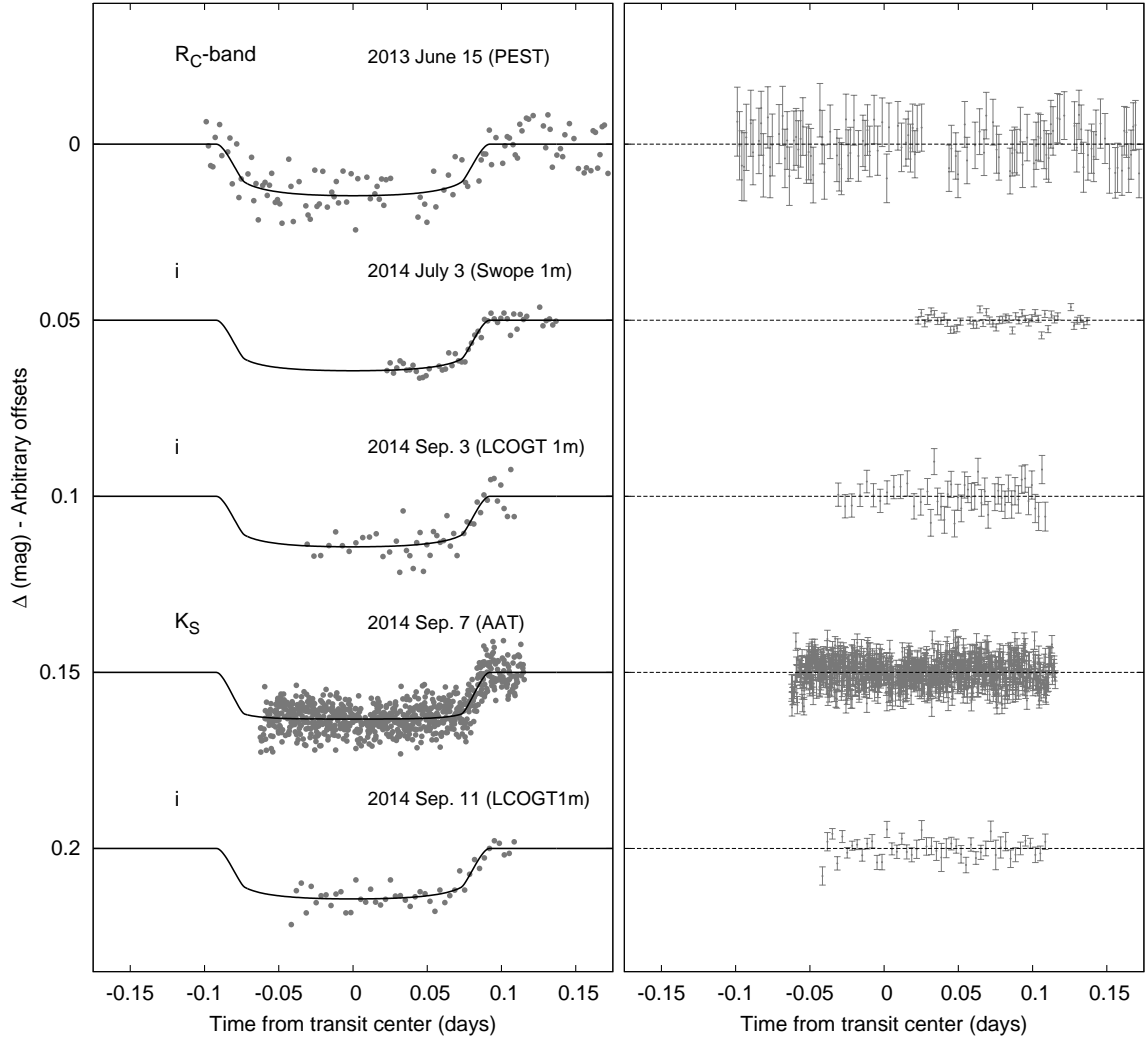}
\caption{
    Left: Unbinned transit \lcs{} for \hatcur{11}.  The light curves
    have been corrected for quadratic trends in time fitted
    simultaneously with the transit model.
    The dates of the events, filters and instruments used are
    indicated.  Light curves following the first are displaced
    vertically for clarity.  Our best fit from the global modeling
    described in \refsecl{globmod} is shown by the solid lines.
    Right: residuals from the fits are displayed in the same order as
    the left curves.  The error bars represent the photon and
    background shot noise, plus the readout noise.
}
\label{fig:lc11}
\end{figure*}
%% --------------------------------------------------------------------
\setcounter{planetcounter}{2}
%
%% --------------------------------------------------------------------
\begin{figure*}[!ht]
\plotone{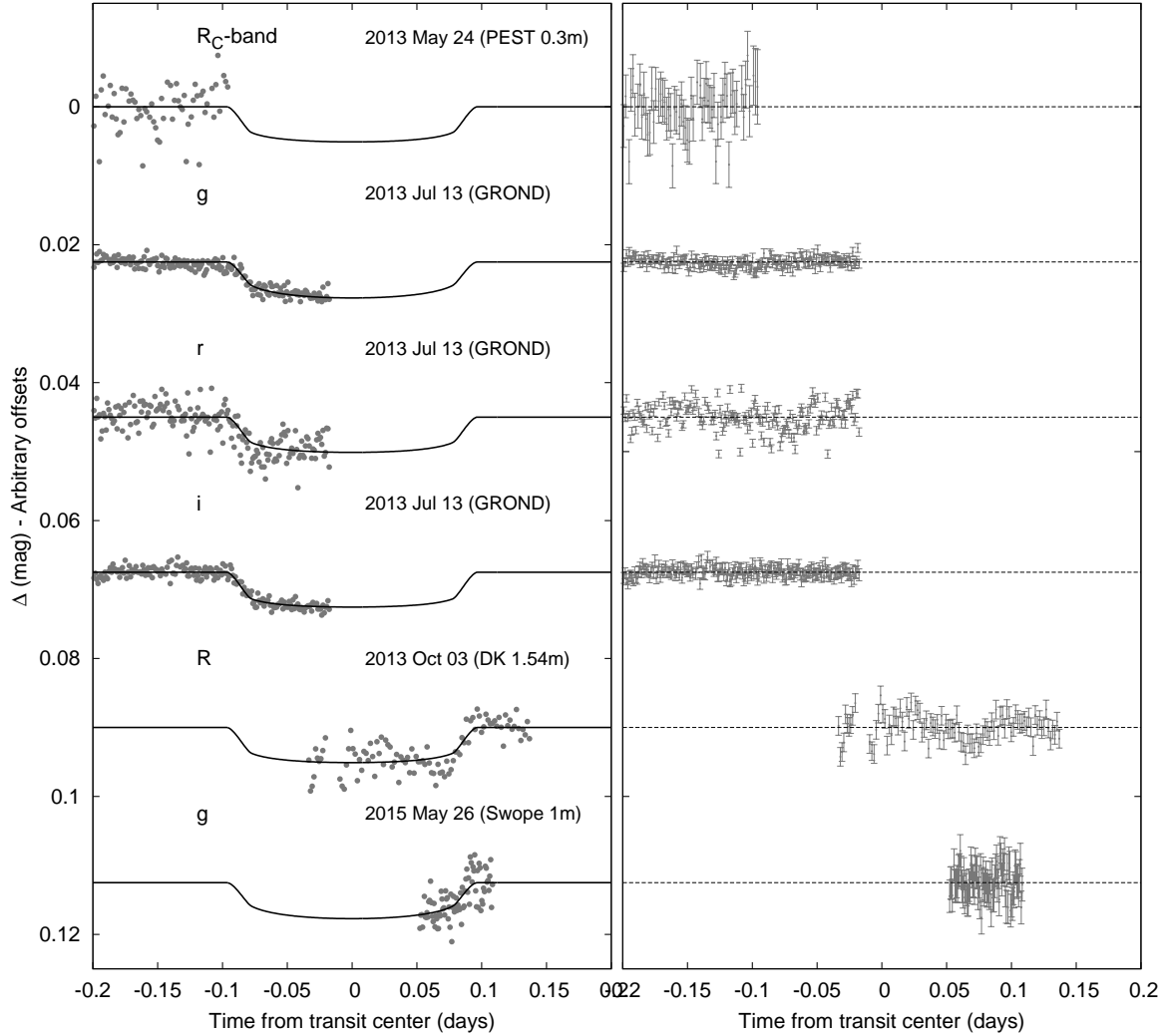}
\caption{
    Similar to \reffigl{lc11}; here we show the follow-up
    \lcs{} for \hatcur{12}.
The PEST~0.3\,m, observation shows a flat \oot\, part. However, the observation helped to refine the ephemeris and the subsequent GROND observation detected the transit, see text.
}
\label{fig:lc12}
\end{figure*}

%% Observations
%%
We also photometrically followed-up both candidates with larger aperture \tel s. This is necessary because the HATSouth survey \tel s have limited
  photometric precision whereas the \lcs\, obtained with these follow-up \tel s are of higher quality allowing us to better characterize the system. The distant in time photometric follow-up observations further help us to refine the transit ephemeris.
Photometric follow-up observations are summarized in
\reftabl{photobs}. The data are given in \reftabl{phfu}. The light
curves for \hatcur{11} are shown in \reffigl{lc11} while for \hatcur{12}
they are shown in \reffigl{lc12}.

We obtained photometric time series of \hatcur{11} with the PEST 0.3\,m, Swope 1\,m, and the LCOGT 1\,m networks. Data reduction followed the established procedures described in previous HATSouth discoveries (\citet{penev:2013:hats1}; \citet{mohlerfischer:2013:hats2}; \citet{bayliss:2013:hats3}; \citet{jordan:2014:hats4}; \citet{zhou:2014:hats5}; \citet{hartman:2015:hats6}; \citet{brahm:2015:hats9}; \citet{mancini:2015:hats13}). 

Additionally, a partial transit of \hatcur{11}b was observed on 2014-09-07, in the \Ks-band, using the IRIS2 infrared camera \citep{2004SPIE.5492..998T}
 on the \tsize{3.9} Anglo Australian Telescope, at Siding Spring Observatory, Australia. The instrument uses a Hawaii 1-RG detector, has a field of view of \fovsize{7.7}\arcmin, pixel scale of 0.4486\pxs. Exposures were 15\,s in duration, with a total of 931 exposures taken, lasting a total of 4.6 hours. The target remained above airmass 1.25, and drifted by less than 1 pixel throughout the observations. The telescope was defocused to achieve a PSF FWHM of $\sim$2.2\arcsec, in order to minimize the effect of inter- and intra-pixel variation, and prevent saturation. The observing strategy, data reduction, and light curve analysis techniques are detailed in \citet{2014MNRAS.445.2746Z}. Flat fielding is performed using a linear combination of two sets of dither frames taken before and after the observation. Aperture photometry is performed on each frame to extract the fluxes of the target and reference stars. The target \lc\, was corrected against the ensemble of reference stars, with weights to each reference star fitted for to minimize the \oot\, root mean square (RMS) scatter of the target star. 

All observations of \hatcur{11}b in different filters reproduced a similar flux decrease of \hatcurLCdip{11}\,mmag as seen in the discovery \lc, but with a higher precision, between 1.6 and 5 mmag per point. We note that mostly partial transits were observed. Each light curve allowed us to refine and improve the uncertainties of the ephemeris, which helped us to schedule subsequent transit observations. The final ephemeris is determined through the global MCMC analysis, as explained in \refsecl{globmod}, founding them to be $T_c ({\rm BJD})=\hatcurLCT{11}$ and $P=\hatcurLCP{11}\,days$, as well as the parameters $\rpl/\rstar=\hatcurLCrprstar{11}$ and $b^2=\hatcurLCbsq{11}$. 

\hatcur{12} was observed with the PEST~0.3\,m, MPG~2.2\,m, DK~1.54\,m and the Swope\,1\,m telescopes. Again, their \lcs\, were consistent with the discovery observations. For data analysis of the PEST, Swope and MPG2.2m observations we repeated our well established procedures as described in previous papers (\citet{penev:2013:hats1}; \citet{mohlerfischer:2013:hats2}; \citet{bayliss:2013:hats3}; \citet{jordan:2014:hats4}; \citet{zhou:2014:hats5}; \citet{hartman:2015:hats6}; \citet{brahm:2015:hats9}; \citet{mancini:2015:hats13}). 
For the DK\,1.54\,m observation on the night 2013-10-03, we defocused the telescope. We used DFOSC, a focal reducer type camera with a \ensuremath{\rm 2048\times\rm 4096} pixels E2V44-82 CCD. The CCD electronics was improved with a 32-bit analog-digital-converter (ADC), allowing for more levels than the usual 65536 available in 16-bit ADCs which are generally used for astronomical CCDs. The saturation obtained with this instrument is up to around 700000 \ensuremath{ADU}s with higher readout speed (\ensuremath{\rm 400\,kpix/s}) and higher gain of 0.24 \eladu, and readout noise is 9.94 \el. However, only half of the CCD is illuminated by the \tel\, and this part of the CCD is readout and generally referred to as full frame, while further windowing is generally possible. The field-of-view (FOV) of the illuminated part is \fovsize{13.7}\arcmin. This is sufficiently large to find adequate reference stars for our differential photometry. We choose a reference image and calculated the shift of all images
with respect to the reference image. From the reference image we
extracted the position of the stars. Following \citet{deeg-doyle}, the
time series photometry was generated from these observations using
optimized aperture photometry that maximizes the signal-to-noise ratio
(SNR) for each star. For all images in one night we used three fixed
apertures and choose these to be much larger than the typical point
spread function in order to minimize the impact of the time-variable
seeing.
Depending on the \tel, for this bright object we obtained a follow-up precision between 0.7 and 3.4 mmag per point. The subsequent refinement of the transit ephemeris was especially important for HATS-12, as the first photometric follow-up observation with the PEST~0.3\,m, telescope showed the flat \oot\, part. Despite that we did not detect the transit with the our initial ephemeris, this observation was still consistent with the transit having occurred after the observation, see upper most light curve in \reffigl{lc12}. We updated the ephemeris so that the following photometric observation with GROND finally revealed the transit.
Again, using the procedure described in \refsecl{globmod}, the final ephemeris was determined to be $T_c ({\rm BJD})=\hatcurLCT{12}$ and $P=\hatcurLCP{12}\,days$, as well as the parameters $\rpl/\rstar=\hatcurLCrprstar{12}$ and $b^2=\hatcurLCbsq{12}$.

%++++++++++++++++++++++++++++++++++++++++++++++++++++++++++++++++++++++

\clearpage

%
%
%% --------------------------------------------------------------------
\ifthenelse{\boolean{emulateapj}}{
    \begin{deluxetable*}{llrrrrl}
}{
    \begin{deluxetable}{llrrrrl}
}
\tablewidth{0pc}
\tablecaption{
    Light curve data for \hatcur{11} and \hatcur{12}\label{tab:phfu}.
}
\tablehead{
    \colhead{Object\tablenotemark{a}} &
    \colhead{BJD\tablenotemark{b}} & 
    \colhead{Mag\tablenotemark{c}} & 
    \colhead{\ensuremath{\sigma_{\rm Mag}}} &
    \colhead{Mag(orig)\tablenotemark{d}} & 
    \colhead{Filter} &
    \colhead{Instrument} \\
    \colhead{} &
    \colhead{\hbox{~~~~(2,400,000$+$)~~~~}} & 
    \colhead{} & 
    \colhead{} &
    \colhead{} & 
    \colhead{} &
    \colhead{}
}
\startdata

HATS-11 &  $ 55780.56087 $ & $  -0.01727 $ & $   0.00795 $ & $ \cdots $ & $ r$ &         HS\\
HATS-11 &  $ 55762.46733 $ & $  -0.01616 $ & $   0.00787 $ & $ \cdots $ & $ r$ &         HS\\
HATS-11 &  $ 55802.27850 $ & $   0.01253 $ & $   0.00731 $ & $ \cdots $ & $ r$ &         HS\\
HATS-11 &  $ 55780.56478 $ & $   0.01161 $ & $   0.00824 $ & $ \cdots $ & $ r$ &         HS\\
HATS-11 &  $ 55791.42313 $ & $  -0.01754 $ & $   0.01052 $ & $ \cdots $ & $ r$ &         HS\\
HATS-11 &  $ 55762.47083 $ & $   0.02251 $ & $   0.00825 $ & $ \cdots $ & $ r$ &         HS\\
HATS-11 &  $ 55802.28203 $ & $   0.00687 $ & $   0.00698 $ & $ \cdots $ & $ r$ &         HS\\
HATS-11 &  $ 55273.88621 $ & $   0.01511 $ & $   0.00797 $ & $ \cdots $ & $ r$ &         HS\\
HATS-11 &  $ 55791.42657 $ & $  -0.01138 $ & $   0.01137 $ & $ \cdots $ & $ r$ &         HS\\
HATS-11 &  $ 55766.09292 $ & $   0.00550 $ & $   0.00705 $ & $ \cdots $ & $ r$ &         HS\\

\enddata
\tablenotetext{a}{
    Either HATS-11, or HATS-12.
}
\tablenotetext{b}{
    Barycentric Julian Date is computed directly from the UTC time
    without correction for leap seconds.
}
\tablenotetext{c}{
    The out-of-transit level has been subtracted. For observations
    made with the HATSouth instruments (identified by ``HS'' in the
    ``Instrument'' column) these magnitudes have been corrected for
    trends using the EPD and TFA procedures applied {\em prior} to
    fitting the transit model. This procedure may lead to an
    artificial dilution in the transit depths. For \hatcur{11} the
    transit depth is 72\% and 84\% that of the true depth for the G579.1 and G579.2 observations, respectively. For \hatcur{12}
    it is 100\% and 78\% that of the true depth for the G579.4 and G580.1 observations, respectively. For observations made with
    follow-up instruments (anything other than ``HS'' in the
    ``Instrument'' column), the magnitudes have been corrected for a
    quadratic trend in time fit simultaneously with the transit.
}
\tablenotetext{d}{
    Raw magnitude values without correction for the quadratic trend in
    time. These are only reported for the follow-up observations.
}
\tablecomments{
    This table is available in a machine-readable form in the online
    journal.  A portion is shown here for guidance regarding its form
    and content.
}
\ifthenelse{\boolean{emulateapj}}{
    \end{deluxetable*}
}{
    \end{deluxetable}
}
%% --------------------------------------------------------------------

% #####################################################################
%% Analysis
\section{Analysis}
\label{sec:analysis}

% =====================================================================
\subsection{Properties of the parent star}
\label{sec:stelparam}
%++++++++++++++++++++++++++++++++++++++++++++++++++++++++++++++++++++++

%++++++++++++++++++++++++++++++++++++++++++++++++++++++++++++++++++++++

In order to obtain the physical parameters of the newly discovered planets,
we have to characterize their stellar hosts. We determined precise atmospheric
parameters for \hatcur{11} and \hatcur{12} from the median combined FEROS
spectra. The SNR of both combined spectra was $\sim$50 per resolution element.
We applied the algorithm ZASPE (Brahm et al.\ 2016, in prep) to both spectra which determines \teff, \logg,
\feh\ and \vsini\ via least squares minimization against a grid of synthetic spectra
in the most sensitive regions of the wavelength coverage to changes in the atmospheric parameters.
ZASPE obtains reliable errors and correlations between the parameters that take into account
the systematic mismatch between the data and the optimal synthetic spectra.

The \teff\ and \feh\ values from ZASPE were combined with the stellar density (\rhostar)
which was obtained through our joint light curve and RV curve analysis to determine a first
estimation of the stellar physical parameters \citep{sozzetti:2007}. In particular we search for the parameters of the
Yonsei-Yale \citep[Y2;][]{yi:2001} isochrones (stellar mass, radius and age) that produce the best
match with our estimated \teff, \feh\ and \rhostar values. Then we compute a new value
for \logg\ which is held fix in a second run of ZASPE and a subsequent comparison with the
theoretical isochrones is made. The final adopted parameters for \hatcur{11} and \hatcur{12} are given in \reftabl{stellar}.
\reffigl{iso} shows the locations of each star on a $\teffstar$--$\rhostar$ diagram.
We found after performing the  analysis just described that both stars are slightly evolved metal poor stars.
\hatcur{11} has a mass of 1.0$\pm$0.06 \msun, a radius of 1.444$\pm$0.057 \rsun\ and an age of 7.7$\pm$2 Gyr, while
the parameters for \hatcur{12} are 1.489$\pm$0.071 \msun, 2.21$\pm$0.21 \rsun\ and an age of 2.36$\pm$0.3 Gyr.
Distances are determined by comparing the measured broad-band
photometry listed in \reftabl{stellar} to the predicted magnitudes in each
filter from the isochrones. We assume a $R_{V} = 3.1$ extinction law
from \citet{cardelli:1989} to determine the extinction.

%% --------------------------------------------------------------------
\ifthenelse{\boolean{emulateapj}}{
    \begin{figure*}[!ht]
}{
    \begin{figure}[!ht]
}
\plottwo{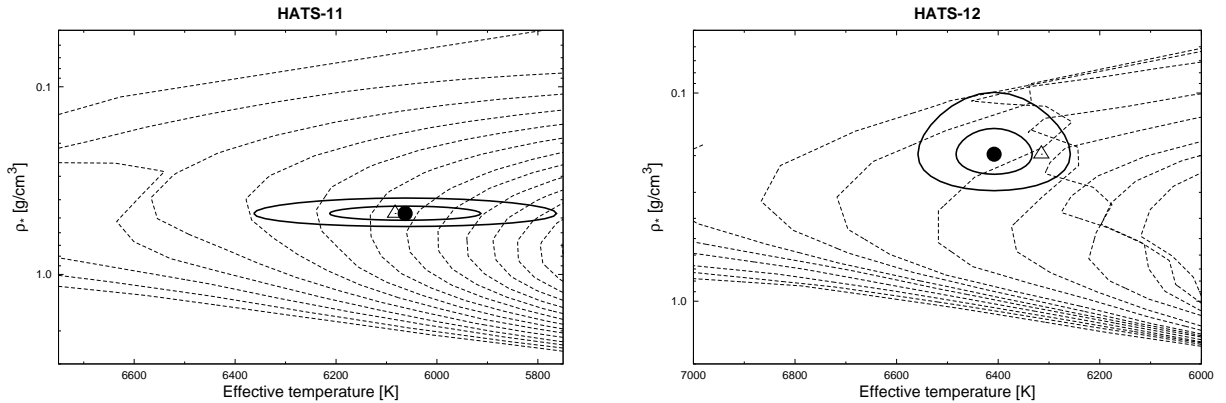}{\hatcurhtr{12}-iso-rho.eps}
\caption{
    Model isochrones from \cite{\hatcurisocite{11}} for the measured
    metallicities of \hatcur{11} (left) and \hatcur{12} (right). For \hatcur{11} we show models for ages of 0.2\,Gyr and 1.0 to 14.0\,Gyr in 1.0\,Gyr increments (ages increasing from left to right), while for \hatcur{12} we show models for ages of 0.2\,Gyr, for 0.5\,Gyr to 2.0\,Gyr in 0.25\,Gyr increments, and for 2.0\,Gyr to 5.0\,Gyr in 0.5\,Gyr increments. The
    adopted values of $\teffstar$ and \rhostar\ are shown together with
    their 1$\sigma$ and 2$\sigma$ confidence ellipsoids.  The initial
    values of \teffstar\ and \rhostar\ from the first ZASPE and \lc\
    analyses are represented with a triangle.
}
\label{fig:iso}
\ifthenelse{\boolean{emulateapj}}{
    \end{figure*}
}{
    \end{figure}
}

%
%
%% --------------------------------------------------------------------
%% Table of stellar parameters. 
%%
\ifthenelse{\boolean{emulateapj}}{
    \begin{deluxetable*}{lccl}
}{
    \begin{deluxetable}{lccl}
}
\tablewidth{0pc}
\tabletypesize{\scriptsize}
\tablecaption{
    Stellar parameters for \hatcur{11} and \hatcur{12}
    \label{tab:stellar}
}
\tablehead{
    \multicolumn{1}{c}{} &
    \multicolumn{1}{c}{\bf HATS-11} &
    \multicolumn{1}{c}{\bf HATS-12} &
    \multicolumn{1}{c}{} \\
    \multicolumn{1}{c}{~~~~~~~~Parameter~~~~~~~~} &
    \multicolumn{1}{c}{Value}                     &
    \multicolumn{1}{c}{Value}                     &
    \multicolumn{1}{c}{Source}
}
\startdata
\noalign{\vskip -3pt}
\sidehead{Astrometric properties and cross-identifications}
~~~~2MASS-ID\dotfill               & \hatcurCCtwomass{11}  & \hatcurCCtwomass{12} & \\
~~~~GSC-ID\dotfill                 & \hatcurCCgsc{11}      & \hatcurCCgsc{12}     & \\
~~~~R.A. (J2000)\dotfill            & \hatcurCCra{11}       & \hatcurCCra{12}    & 2MASS\\
~~~~Dec. (J2000)\dotfill            & \hatcurCCdec{11}      & \hatcurCCdec{12}   & 2MASS\\
~~~~$\mu_{\rm R.A.}$ (\masy)              & \hatcurCCpmra{11}     & \hatcurCCpmra{12} & UCAC4\\
~~~~$\mu_{\rm Dec.}$ (\masy)              & \hatcurCCpmdec{11}    & \hatcurCCpmdec{12} & UCAC4\\

\sidehead{Galactic space velocity components, LSR reference frame}
~~~~$u$ (\kms)\dotfill & 65.274 & 25.812 & FEROS+UCAC4 \\
~~~~$v$ (\kms)\dotfill & 88.956 & 37.784 & FEROS+UCAC4 \\
~~~~$w$ (\kms)\dotfill & 40.364 & -4.115 & FEROS+UCAC4 \\

\sidehead{Spectroscopic properties}

~~~~$\teffstar$ (K)\dotfill         &  \hatcurSMEteff{11}   & \hatcurSMEteff{12} & ZASPE\tablenotemark{a}\\
~~~~$\feh$\dotfill                  &  \hatcurSMEzfeh{11}   & \hatcurSMEzfeh{12} & ZASPE               \\
~~~~$\vsini$ (\kms)\dotfill         &  \hatcurSMEvsin{11}   & \hatcurSMEvsin{12} & ZASPE                \\
~~~~$\gamma_{\rm RV}$ (\kms)\dotfill&  \hatcurRVgammaabs{11}  & \hatcurRVgammaabs{12} & FEROS\tablenotemark{b}  \\
\sidehead{Photometric properties}

~~~~$B$ (mag)\dotfill               &  \hatcurCCtassmB{11}  & \hatcurCCtassmB{12} & APASS\tablenotemark{c} \\
~~~~$V$ (mag)\dotfill               &  \hatcurCCtassmv{11}  & \hatcurCCtassmv{12} & APASS\tablenotemark{c} \\
~~~~$g$ (mag)\dotfill               &  \hatcurCCtassmg{11}  & \hatcurCCtassmg{12} & APASS\tablenotemark{c} \\
~~~~$r$ (mag)\dotfill               &  \hatcurCCtassmr{11}  & \hatcurCCtassmr{12} & APASS\tablenotemark{c} \\
~~~~$i$ (mag)\dotfill               &  \hatcurCCtassmi{11}  & \hatcurCCtassmi{12} & APASS\tablenotemark{c} \\
~~~~$J$ (mag)\dotfill               &  \hatcurCCtwomassJmag{11} & \hatcurCCtwomassJmag{12} & 2MASS           \\
~~~~$H$ (mag)\dotfill               &  \hatcurCCtwomassHmag{11} & \hatcurCCtwomassHmag{12} & 2MASS           \\
~~~~$K_s$ (mag)\dotfill             &  \hatcurCCtwomassKmag{11} & \hatcurCCtwomassKmag{12} & 2MASS           \\
\sidehead{Derived properties}
~~~~$\mstar$ ($\msun$)\dotfill      &  \hatcurISOmlong{11}   & \hatcurISOmlong{12} & YY+$\rhostar$+ZASPE \tablenotemark{d}\\
~~~~$\rstar$ ($\rsun$)\dotfill      &  \hatcurISOrlong{11}   & \hatcurISOrlong{12} & YY+$\rhostar$+ZASPE         \\
~~~~$\loggstar$ (cgs)\dotfill       &  \hatcurISOlogg{11}    & \hatcurISOlogg{12} & YY+$\rhostar$+ZASPE         \\
~~~~$\rhostar$ (\gcmc)\dotfill       &  \hatcurISOrho{11}    & \hatcurISOrho{12} & YY+$\rhostar$+ZASPE          \\
~~~~$\lstar$ ($\lsun$)\dotfill      &  \hatcurISOlum{11}     & \hatcurISOlum{12} & YY+$\rhostar$+ZASPE         \\
~~~~$M_V$ (mag)\dotfill             &  \hatcurISOmv{11}      & \hatcurISOmv{12} & YY+$\rhostar$+ZASPE         \\
~~~~$M_K$ (mag,\hatcurjhkfilset{11})\dotfill &  \hatcurISOMK{11} & \hatcurISOMK{12} & YY+$\rhostar$+ZASPE         \\
~~~~Age (Gyr)\dotfill               &  \hatcurISOage{11}     & \hatcurISOage{12} & YY+$\rhostar$+ZASPE         \\
~~~~$A_{V}$ (mag)\dotfill               &  \hatcurXAv{11}     & \hatcurXAv{12} & YY+$\rhostar$+ZASPE         \\
~~~~Distance (pc)\dotfill           &  \hatcurXdistred{11}\phn  & \hatcurXdistred{12} & YY+$\rhostar$+ZASPE\\ [-1.5ex]
\enddata
\tablenotetext{a}{
    ZASPE = Zonal Atmospheric Stellar Parameter Estimator routine
    for the analysis of high-resolution spectra (Brahm et al.~2016, in
    preparation), applied to the FEROS spectra of \hatcur{11} and \hatcur{12}. These
    parameters rely primarily on ZASPE, but have a small dependence
    also on the iterative analysis incorporating the isochrone search
    and global modeling of the data.
}
\tablenotetext{b}{
    The error on $\gamma_{\rm RV}$ is determined from the orbital fit to
    the FEROS RV measurements, and does not include the systematic
    uncertainty in transforming the velocities from FEROS to the IAU
    standard system. 
} \tablenotetext{c}{
    From APASS DR9 \citep{henden:2014:apass}.  
}
\tablenotetext{d}{
    \hatcurisoshort{11}+\rhostar+ZASPE = Based on the \hatcurisoshort{11}
    isochrones \citep{\hatcurisocite{11}}, \rhostar\ as a luminosity
    indicator, and the ZASPE results.
}
\ifthenelse{\boolean{emulateapj}}{
    \end{deluxetable*}
}{
    \end{deluxetable}
}
%% --------------------------------------------------------------------

% =====================================================================
\subsection{Excluding blend scenarios}
\label{sec:blend}
%++++++++++++++++++++++++++++++++++++++++++++++++++++++++++++++++++++++

%++++++++++++++++++++++++++++++++++++++++++++++++++++++++++++++++++++++

In order to exclude blend scenarios we carried out an analysis following
\citet{hartman:2012:hat39hat41}. We attempt to model the available
photometric data (including light curves and catalog broad-band
photometric measurements) for each object as a blend between an
eclipsing binary star system and a third star along the line of
sight. The physical properties of the stars are constrained using the
Padova isochrones \citep{girardi:2000}, while we also require that the
brightest of the three stars in the blend have atmospheric parameters
consistent with those measured with ZASPE. 

For \hatcur{11} we find that the best-fit blend model provides a
poorer fit to the data than the best-fit planet+star model, and can be
rejected with $3\sigma$ confidence. Based on simulating composite
spectra for the blend models that we tested, we also find that any
blended eclipsing binary system that can plausibly fit the photometric
data (i.e.\ ~cannot be rejected with more than 5$\sigma$ confidence)
would show BS variations ranging from $400$\,\ms\ for the model that
provides the most marginal fit (i.e.\ ~at the 5$\sigma$ rejection
limit), to more than $1$\,\kms\ for the best-fitting blend models, as
well as having RV variations with $K > 200$\,\ms. As the measured BS
variation is 80\,\ms\ for FEROS and 140\,\ms\ for Coralie, and the RV
semiamplitude is $\hatcurRVK{11}$\,\ms, we conclude that \hatcur{11}
is not a blended stellar eclipsing binary system, and that the
observations favor a transiting planet system interpretation. 
Similarly, for \hatcur{12} we find that the best-fit blend model
provides a poorer fit to the data than the best-fit planet+star model,
and in this case can be rejected with $2\sigma$ confidence. Those
blended eclipsing binary systems that cannot be rejected with more
than 5$\sigma$ confidence based solely on the photometry would have
easily been detected as composite systems based on the FEROS and HDS
spectroscopy, and would have BS variations exceeding 1\,\kms. For
comparison, the BS RMS scatter is 34\,\ms\ and 15\,\ms\ for the
FEROS and HDS observations of \hatcur{12}, respectively. We conclude
that \hatcur{12} is also not a blended stellar eclipsing binary
system, and is instead a transiting planet system. 
However, both system could still be diluted transiting planet systems, which cannot be recognized by spectroscopic observations and only high angular resolution imaging can solve. If a blended stellar companion is present, diluting the light of the transiting system, then the true companion radius could be up to ∼60\% larger than inferred.

% =====================================================================
\subsection{Global modeling of the data}
\label{sec:globmod}
We modeled the HATSouth photometry, the follow-up photometry, and the
high-precision RV measurements following
\citet{pal:2008:hat7}, \citet{bakos:2010:hat11} and \citet{hartman:2012:hat39hat41}. We fit
\citet{mandel:2002} transit models to the light curves, allowing for a
dilution of the HATSouth transit depth as a result of blending from
neighboring stars and over-correction by the trend-filtering
method. For the follow-up light curves we include a quadratic trend in
time in our model for each event to correct for remaining systematic errors in
the photometry. We fit Keplerian orbits to the RV curves allowing the
zero-point for each instrument to vary independently in the fit, and
allowing for RV jitter which we also vary as a free parameter for
each instrument. This jitter may be astrophysical or instrumental in nature, and simply represents excess scatter in the RV observations beyond what is expected based on formal uncertainties.

We used a Differential Evolution Markov Chain Monte
Carlo procedure to explore the fitness landscape and to determine the
posterior distribution of the parameters \citep[DEMCMC;][]{terbraak:2006}. Note that we tried fitting both fixed circular orbits and free-eccentricity models to the data, and for both systems find that the data are consistent with a circular orbit. For both systems the fixed circular orbit model has a higher Bayesian evidence so we adopt the parameters obtained assuming no eccentricity for both objects. 
Furthermore, we see no structure or drift within our uncertainties that could hint to any extra component in the system. We also note that for \hatcur{11} the scatter in the CORALIE and
FEROS RV residuals is consistent with the uncertainties, so our
modeling finds jitter values of $0$ for both instruments. Similarly, we find that a jitter value of $0$ is preferred for the HDS observations of \hatcur{12}. For these instruments we list the 95\% upper limit on the jitter in the \reftabl{planetparam}. The resulting parameters for each system are listed in
\reftabl{planetparam}.
\hatcurb{11} has a mass slightly smaller than Jupiter (0.85$\pm$0.12 \mjup) and a large radius of 1.51$\pm$0.078 \rjup. It is a moderately
irradiated hot Jupiter with an equilibrium temperature of 1637$\pm$48 K. \hatcurb{12} is a rather massive hot Jupiter with 2.38$\pm$0.11 \mjup,
1.35$\pm$0.17 \rjup\, and a relatively high equilibrium temperature of 2097$\pm$ 89 K.

%
% ---------------------------------------------------------------------
\ifthenelse{\boolean{emulateapj}}{
    \begin{deluxetable*}{lcc}
}{
    \begin{deluxetable}{lcc}
}
%\tablewidth{0pc}
\tabletypesize{\scriptsize}
\tablecaption{Orbital and planetary parameters for \hatcurb{11} and \hatcurb{12}\label{tab:planetparam}}
\tablehead{
    \multicolumn{1}{c}{} &
    \multicolumn{1}{c}{\bf HATS-11b} &
    \multicolumn{1}{c}{\bf HATS-12b} \\ 
    \multicolumn{1}{c}{~~~~~~~~~~~~~~~Parameter~~~~~~~~~~~~~~~} &
    \multicolumn{1}{c}{Value} &
    \multicolumn{1}{c}{Value}
}
\startdata
\noalign{\vskip -3pt}
\sidehead{\Lc{} parameters}
~~~$P$ (days)             \dotfill    & $\hatcurLCP{11}$ & $\hatcurLCP{12}$ \\
~~~$T_c$ (${\rm BJD}$)    
      \tablenotemark{a}   \dotfill    & $\hatcurLCT{11}$ & $\hatcurLCT{12}$ \\
~~~$T_{12}$ (days)
      \tablenotemark{a}   \dotfill    & $\hatcurLCdur{11}$ & $\hatcurLCdur{12}$ \\
~~~$T_{12} = T_{34}$ (days)
      \tablenotemark{a}   \dotfill    & $\hatcurLCingdur{11}$ & $\hatcurLCingdur{12}$ \\
~~~$\arstar$              \dotfill    & $\hatcurPPar{11}$ & $\hatcurPPar{12}$ \\
~~~$\zrstar$ \tablenotemark{b}             \dotfill    & $\hatcurLCzeta{11}$\phn & $\hatcurLCzeta{12}$\phn \\
~~~$\rpl/\rstar$          \dotfill    & $\hatcurLCrprstar{11}$ & $\hatcurLCrprstar{12}$ \\
~~~$b^2$                  \dotfill    & $\hatcurLCbsq{11}$ & $\hatcurLCbsq{12}$ \\
~~~$b \equiv a \cos i/\rstar$
                          \dotfill    & $\hatcurLCimp{11}$ & $\hatcurLCimp{12}$ \\
~~~$i$ (deg)              \dotfill    & $\hatcurPPi{11}$\phn & $\hatcurPPi{12}$\phn \\
\sidehead{Limb-darkening coefficients \tablenotemark{c}}

~~~$c_1,g$ (linear term)    \dotfill    & $\cdots$ & $\hatcurLBig{12}$ \\
~~~$c_2,g$ (quadratic term) \dotfill    & $\cdots$ & $\hatcurLBiig{12}$ \\
~~~$c_1,r$                  \dotfill    & $\hatcurLBir{11}$ & $\hatcurLBir{12}$ \\
~~~$c_2,r$                  \dotfill    & $\hatcurLBiir{11}$ & $\hatcurLBiir{12}$ \\
~~~$c_1,i$                  \dotfill    & $\hatcurLBii{11}$ & $\hatcurLBii{12}$ \\
~~~$c_2,i$                  \dotfill    & $\hatcurLBiii{11}$ & $\hatcurLBiii{12}$ \\

~~~$c_1,R$                  \dotfill    & $\hatcurLBiR{11}$ & $\hatcurLBiR{12}$ \\
~~~$c_2,R$                  \dotfill    & $\hatcurLBiiR{11}$ & $\hatcurLBiiR{12}$ \\
~~~$c_1,K$                  \dotfill    & $\hatcurLBiK{11}$ & $\cdots$ \\
~~~$c_2,K$                  \dotfill    & $\hatcurLBiiK{11}$ & $\cdots$ \\

\sidehead{RV parameters}
~~~$K$ (\ms)              \dotfill    & $\hatcurRVK{11}$\phn\phn & $\hatcurRVK{12}$\phn\phn \\

~~~$e$ \tablenotemark{d}               \dotfill    & $\hatcurRVeccentwosiglimeccen{11}$ & $\hatcurRVeccentwosiglimeccen{12}$ \\
~~~RV jitter FEROS (\ms)        \dotfill    & \hatcurRVjitterBtwosiglim{11} & \hatcurRVjitterA{12} \\
~~~RV jitter Coralie (\ms)        \dotfill    & \hatcurRVjitterAtwosiglim{11} & $\cdots$ \\
~~~RV jitter HDS (\ms)        \dotfill    & $\cdots$ & \hatcurRVjitterBtwosiglim{12} \\

\sidehead{Planetary parameters}
~~~$\mpl$ ($\mjup$)       \dotfill    & $\hatcurPPmlong{11}$ & $\hatcurPPmlong{12}$ \\
~~~$\rpl$ ($\rjup$)       \dotfill    & $\hatcurPPrlong{11}$ & $\hatcurPPrlong{12}$ \\
~~~$C(\mpl,\rpl)$
    \tablenotemark{f}     \dotfill    & $\hatcurPPmrcorr{11}$ & $\hatcurPPmrcorr{12}$ \\
~~~$\rhopl$ (\gcmc)       \dotfill    & $\hatcurPPrho{11}$ & $\hatcurPPrho{12}$ \\
~~~$\log g_p$ (cgs)       \dotfill    & $\hatcurPPlogg{11}$ & $\hatcurPPlogg{12}$ \\
~~~$a$ (AU)               \dotfill    & $\hatcurPParel{11}$ & $\hatcurPParel{12}$ \\
~~~$T_{\rm eq}$ (K)        \dotfill   & $\hatcurPPteff{11}$ & $\hatcurPPteff{12}$ \\
~~~$\Theta$ \tablenotemark{g} \dotfill & $\hatcurPPtheta{11}$ & $\hatcurPPtheta{12}$ \\
~~~$\log_{10}\langle F \rangle$ (cgs) \tablenotemark{h}
                          \dotfill    & $\hatcurPPfluxavglog{11}$ & $\hatcurPPfluxavglog{12}$ \\ [-1.5ex]
\enddata
\tablenotetext{a}{
    Times are in Barycentric Julian Date calculated directly from UTC {\em without} correction for leap seconds.
    \ensuremath{T_c}: Reference epoch of
    mid transit that minimizes the correlation with the orbital
    period.
    \ensuremath{T_{12}}: total transit duration, time
    between first to last contact;
    \ensuremath{T_{12}=T_{34}}: ingress/egress time, time between first
    and second, or third and fourth contact.
}
\tablecomments{
For both objects we list the parameters assuming circular orbits. Based on Bayesian evidence, we find that such an orbit is preferred for both systems.
}
\tablenotetext{b}{
   Reciprocal of the half duration of the transit used as a jump parameter in our MCMC analysis in place of $\arstar$. It is related to $\arstar$ by the expression $\zrstar = \arstar(2\pi(1+e\sin\omega))/(P\sqrt{1-b^2}\sqrt{1-e^2})$ \citep{bakos:2010:hat11}.
}
\tablenotetext{c}{
    Values for a quadratic law, adopted from the tabulations by
    \cite{claret:2004} according to the spectroscopic (ZASPE) parameters
    listed in \reftabl{stellar}.
}
\tablenotetext{d}{
    For fixed circular orbit models we list
    the 95\% confidence upper limit on the eccentricity determined
    when $\sqrt{e}\cos\omega$ and $\sqrt{e}\sin\omega$ are allowed to
    vary in the fit.
}
\tablenotetext{e}{
 Error term, either astrophysical or instrumental in origin, added in quadrature to the formal RV errors for the listed instrument. We give the 95 \% confidence upper limit when the jitter is consistent with 0\,\ms.
}
\tablenotetext{f}{
    Correlation coefficient between the planetary mass \mpl\ and radius
    \rpl\ estimated from the posterior parameter distribution.
}
\tablenotetext{g}{
    The Safronov number is given by $\Theta = \frac{1}{2}(V_{\rm
    esc}/V_{\rm orb})^2 = (a/\rpl)(\mpl / \mstar )$
    \citep[see][]{hansen:2007}.
}
\tablenotetext{h}{
    Incoming flux per unit surface area, averaged over the orbit.
}
\ifthenelse{\boolean{emulateapj}}{
    \end{deluxetable*}
}{
    \end{deluxetable}
}

% ---------------------------------------------------------------------

%% EOF Analysis
\clearpage
% #####################################################################
%% Discussion
\section{Discussion}
\label{sec:discussion}
% ++++++++++++++++++++++++++++++++++++++++++++++++++++++++++++++++++++++
% ++++++++++++++++++++++++++++++++++++++++++++++++++++++++++++++++++++++

In this paper we have presented \hatcurb{11} and \hatcurb{12}, two inflated gas giants orbiting a metal-poor and a subsolar metallicity star with [Fe/H] of \hatcurSMEzfeh{11} and \hatcurSMEzfeh{12}, respectively. Globally fitting all observations we estimated precise system parameters as shown in \reftabls{stellar}{planetparam}.

\hatcurb{11} and \hatcurb{12} are two regular hot Jupiters that present inflated radii as compared to predictions of standard theoretical models of planetary structure. However, a distinction of these transiting systems is that both of them have low
metallicity stellar hosts which are evolved.

\begin{figure*}[h]
\epsscale{0.75}
\plotone{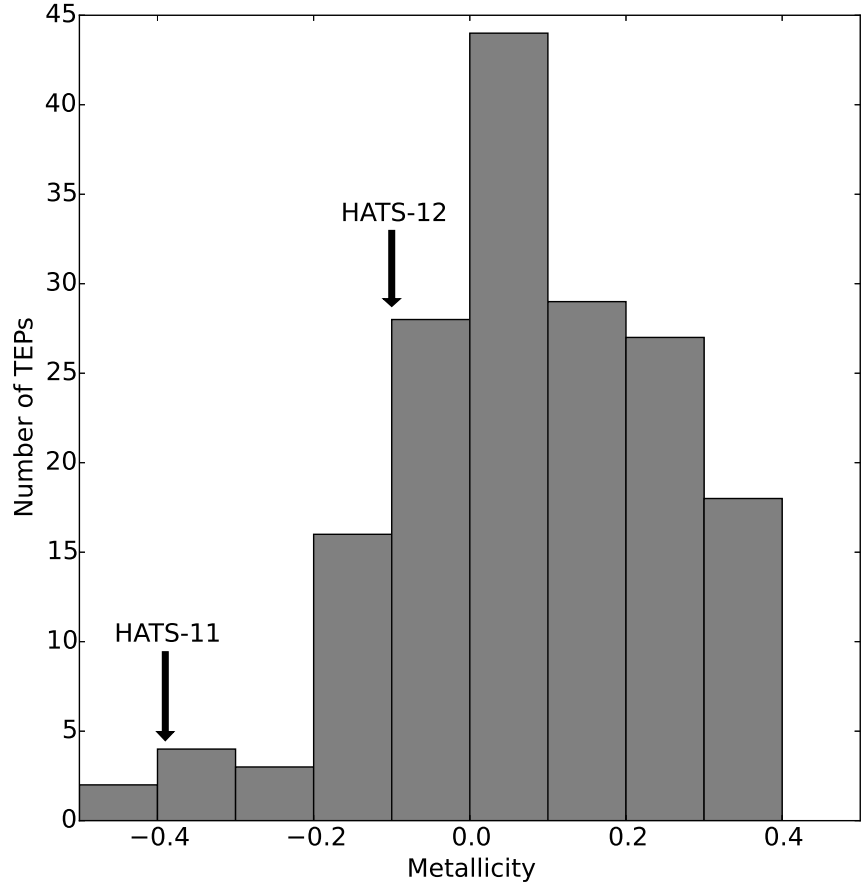}
\caption{
Histogram of stellar metallicity for 177 hot Jupiter planet hosts, see text for population parameters. \hatcur{12} shows sub-solar metallicity, whereas \hatcurb{11} transits a metal-poor star where fewer systems are well characterized.  
}
\label{fig:FeH_histogram}
\end{figure*}

\reffigl{FeH_histogram} shows a histogram of metallicity for known hot Jupiter planet hosts. The systems' parameters were taken from the Transiting Extrasolar Planet Catalogue (TEPCat)\footnotemark[1]\footnotetext[1]{available at http://www.astro.keele.ac.uk/jkt/tepcat/}.
We limited the sample to systems comparable to our discovered planets in this work by restricting it to TEPs satisfying $M_P \geq 0.47 M_J$, $0.015\leq a\leq0.5$ AU and
around stars with 5300 K $\leq \teff \leq$ 7200 K (stellar spectral type G to F). The choice of 0.47 $M_J$ for the minimum mass is made based on the findings of \citet{weiss:2013:koi94} who found a break in the radius-mass relation at that mass. We also restrict the sample to have uncertainties in \feh\ $\leq$ 0.15 dex. 
As is now well established, giant planets are found less frequently around metal-poor stars. 
We note that \hatcur{11} is amongst the most metal
poor stars detected to harbor a transiting giant planet and thus has the value of populating a sparse region of parameter space necessary to understand and validate planet formation theories. In particular, planets as massive as \hatcurb{11} around stars with such a low metallicity can help to empirically constrain limits on the metallicity of the nebulae in the context of core-accretion theory, which can give insights of the boundaries of the formation process \citep{matsuo:2007:planetformation}.

\begin{figure*}[h]
\epsscale{0.75}
\plotone{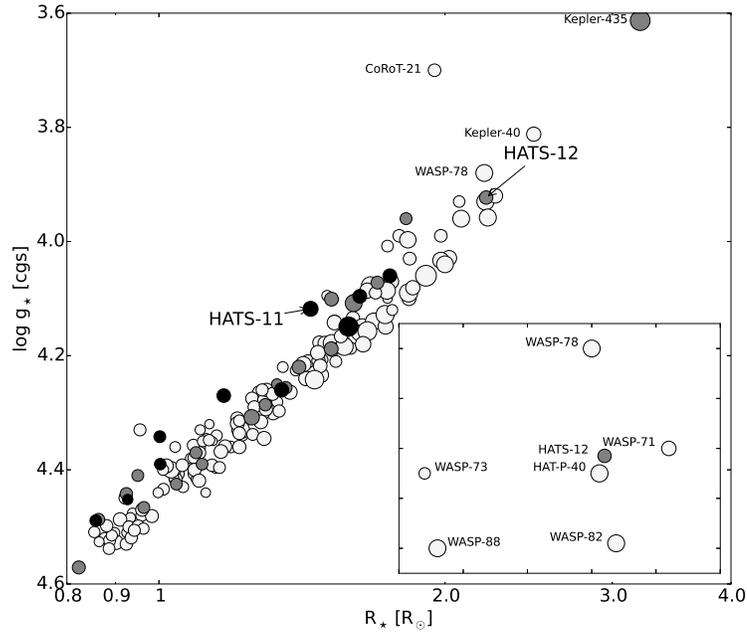}
\caption{
Stellar surface gravity versus stellar radius of hot Jupiter planet hosts. The host stars's metallicity is grey-color coded, grey circles show systems with $-0.2\leq$ \feh $\leq-0.1$ dex and black circles \feh $<-0.2$. The circle's size reflect the planet's radius. Both \hatcur{11} and \hatcur{12}, occupy a region where only few planet hosts are known. 
The inset graph in the corner shows the region around \hatcur{12}, with planet hosting stars of similar stellar radius, we note that \hatcur{12} is the only low metallicity system amongst the planet hosts with similar stellar parameters. %%
}
\label{fig:logg_Teff}
\end{figure*}

In \reffigl{logg_Teff} we show stellar surface gravity as function of stellar radius for the same sample as in \reffigl{FeH_histogram}. \hatcur{11} is a slightly evolved metal-poor star. Only two other systems resemble similar stellar parameters, namely HATS-9 \citep{brahm:2015:hats9} and WASP-48 \citep{enoch:2011:wasp48}, but none of them is metal-poor. Amongst the most evolved hot Jupiter hosts only two have a low metal content, namely \hatcur{12} and Kepler-435 \citep{almenara:2015:kepler435}. No subsolar metallicity stars with similar stellar radius and surface gravity to \hatcur{12} have been detected so far, see \reffigl{logg_Teff}. Hence, \hatcurb{11} and \hatcurb{12} add new systems to the population of low-metallicity evolved stars known to host a giant planet.

Finally, we note that \hatcur{11} and \hatcur{12} were selected as targets for Kepler two-wheeled mission (K2) Campaign 7 (EPIC216414930 and EPIC218131080 respectively) under programs GO7066 and GO7067 (PI: Bakos). These targets have now been observed, and data is expected to be released on 2016 April 30. The high-precision K2 data will allow us to improve their transit parameters, especially important for \hatcur{12}, as it shows the third shallowest ground-based transit discovery to date.
Furthermore, we will have the possibility to search intensively for additional companions through the discovery of additional transits of longer period planets, see e.g. \citet{rabus:2009:starspot}, or transit timing variations (TTVs) \citep{rabus:2009:ttv}, such as were measured in the hot Jupiter WASP-47b \citep{becker:2015:wasp47}. The discovery of \hatcurb{11} and \hatcurb{12} thus provides a strong motivation for a combination of both ground-based detection and subsequent space-based follow-up characterization that can be fruitful and efficient.

%% EOF Discussion

% #####################################################################
%% Acknowledgements
\acknowledgements 

Development of the HATSouth project was funded by NSF MRI grant
NSF/AST-0723074, operations have been supported by NASA grants NNX09AB29G and NNX12AH91H, and
follow-up observations receive partial support from grant
NSF/AST-1108686.
A.J.\ acknowledges support from FONDECYT project 1130857, BASAL CATA PFB-06, and project IC120009 ``Millennium Institute of Astrophysics (MAS)'' of the Millenium Science Initiative, Chilean Ministry of Economy. R.B.\ and N.E.\ are supported by CONICYT-PCHA/Doctorado Nacional. R.B.\ and N.E.\ acknowledge additional support from project IC120009 ``Millenium Institute of Astrophysics  (MAS)'' of the Millennium Science Initiative, Chilean Ministry of Economy.  V.S.\ acknowledges support form BASAL CATA PFB-06.  
K.P. acknowledges support from NASA grant NNX13AQ62G.
This work is based on observations made with ESO Telescopes at the La
Silla Observatory.
This paper also uses observations obtained with facilities of the Las
Cumbres Observatory Global Telescope.
Work at the Australian National University is supported by ARC Laureate
Fellowship Grant FL0992131.
We acknowledge the use of the AAVSO Photometric All-Sky Survey (APASS),
funded by the Robert Martin Ayers Sciences Fund, and the SIMBAD
database, operated at CDS, Strasbourg, France.
Operations at the MPG~2.2\,m Telescope are jointly performed by the
Max Planck Gesellschaft and the European Southern Observatory.  The
imaging system GROND has been built by the high-energy group of MPE in
collaboration with the LSW Tautenburg and ESO\@.  We thank the MPG~2.2\,m Telescope support team for their technical assistance during observations.
We are grateful to P. Sackett for her help in the early phase of the
HATSouth project.

%% EOF Acknowledgements

% #####################################################################
%% Bibliography
\clearpage
\bibliographystyle{apj}
\bibliography{hats11-12_V10_submit.bbl}

\clearpage
\LongTables

%% --------------------------------------------------------------------
%%
%
%
\tabletypesize{\scriptsize}
\ifthenelse{\boolean{emulateapj}}{
    \begin{deluxetable*}{lrrrrrl}
}{
    \begin{deluxetable}{lrrrrrl}
}
\tablewidth{0pc}
\tablecaption{
    Relative radial velocities and bisector spans for \hatcur{11} and
    \hatcur{12}.
    \label{tab:rvs}
}
\tablehead{
    \colhead{BJD} &
    \colhead{RV\tablenotemark{a}} &
    \colhead{\ensuremath{\sigma_{\rm RV}}\tablenotemark{b}} &
    \colhead{BS} &
    \colhead{\ensuremath{\sigma_{\rm BS}}\tablenotemark{c}} &
    \colhead{Phase} &
    \colhead{Instrument}\\
    \colhead{\hbox{(2,456,000$+$)}} &
    \colhead{(\ms)} &
    \colhead{(\ms)} &
    \colhead{(\ms)} &
    \colhead{(\ms)} &
    \colhead{} &
    \colhead{}
}
\startdata

\multicolumn{7}{c}{\bf HATS-11} \\
\hline\\
$ 400.76268 $ & $   105.00 $ & $    31.00 $ & $   88.0 $ & $   62.0 $ & $   0.867 $ & FEROS \\
$ 411.71442 $ & $   -23.00 $ & $    47.00 $ & $   21.0 $ & $   94.0 $ & $   0.893 $ & FEROS \\
$ 424.85662 $ & $   -17.00 $ & $    31.00 $ & $  -56.0 $ & $   62.0 $ & $   0.524 $ & FEROS \\
$ 427.74603 $ & $  -135.00 $ & $    80.00 $ & $  203.0 $ & $  160.0 $ & $   0.322 $ & FEROS \\
$ 488.81286 $ & $   -55.00 $ & $    31.00 $ & $  -12.0 $ & $   62.0 $ & $   0.195 $ & FEROS \\
$ 490.79732 $ & $   136.00 $ & $    29.00 $ & $   12.0 $ & $   58.0 $ & $   0.744 $ & FEROS \\
$ 492.73173 $ & $  -100.00 $ & $    28.00 $ & $  -51.0 $ & $   56.0 $ & $   0.278 $ & FEROS \\
$ 524.50152 $ & $  -103.59 $ & $    80.00 $ & $  213.0 $ & $  160.0 $ & $   0.057 $ & Coralie \\
$ 524.56679 $ & $    34.41 $ & $    69.00 $ & $  106.0 $ & $  138.0 $ & $   0.075 $ & Coralie \\
$ 525.49132 $ & $  -374.59 $ & $   116.00 $ & $  -76.0 $ & $  232.0 $ & $   0.330 $ & Coralie \\
$ 525.60831 $ & $    40.41 $ & $   128.00 $ & $  -47.0 $ & $  256.0 $ & $   0.362 $ & Coralie \\
$ 544.52715 $ & $    61.00 $ & $    33.00 $ & $   75.0 $ & $   66.0 $ & $   0.590 $ & FEROS \\
$ 548.59572 $ & $   139.00 $ & $    25.00 $ & $  -27.0 $ & $   50.0 $ & $   0.714 $ & FEROS \\
$ 552.55423 $ & $   105.00 $ & $    30.00 $ & $  -21.0 $ & $   60.0 $ & $   0.808 $ & FEROS \\
$ 727.88749 $ & $  -126.59 $ & $    75.00 $ & $ -171.0 $ & $  150.0 $ & $   0.254 $ & Coralie \\
$ 728.85790 $ & $    78.41 $ & $    89.00 $ & $  -30.0 $ & $  178.0 $ & $   0.522 $ & Coralie \\
$ 729.90376 $ & $   114.41 $ & $    88.00 $ & $ -157.0 $ & $  176.0 $ & $   0.811 $ & Coralie \\
$ 730.90303 $ & $  -105.59 $ & $   144.00 $ & $   30.0 $ & $  288.0 $ & $   0.087 $ & Coralie \\
$ 731.89440 $ & $    40.41 $ & $   141.00 $ & $   51.0 $ & $  282.0 $ & $   0.361 $ & Coralie \\
$ 732.86378 $ & $   246.41 $ & $   143.00 $ & $  247.0 $ & $  286.0 $ & $   0.629 $ & Coralie \\

\cutinhead{\bf HATS-12}
$ 161.57017 $ & $  -232.12 $ & $    29.00 $ & $  114.0 $ & $   13.0 $ & $   0.194 $ & FEROS \\
$ 161.58536 $ & $  -320.12 $ & $    28.00 $ & $   44.0 $ & $   13.0 $ & $   0.199 $ & FEROS \\
$ 169.67026 $ & $   238.88 $ & $    26.00 $ & $   35.0 $ & $   12.0 $ & $   0.772 $ & FEROS \\
$ 172.52620 $ & $   224.88 $ & $    24.00 $ & $   -1.0 $ & $   12.0 $ & $   0.680 $ & FEROS \\
$ 189.80413 $ & $  -221.85 $ & $    12.01 $ & $   -2.5 $ & $   24.8 $ & $   0.178 $ & HDS \\
$ 189.81885 $ & $  -227.15 $ & $    12.14 $ & $   -2.7 $ & $   23.2 $ & $   0.183 $ & HDS \\
$ 189.83357 $ & $  -191.54 $ & $    11.86 $ & $  -12.5 $ & $   26.9 $ & $   0.187 $ & HDS \\
$ 190.79653 $ & \nodata      & \nodata      & $    1.4 $ & $   22.5 $ & $   0.494 $ & HDS \\
$ 190.81126 $ & \nodata      & \nodata      & $   -9.7 $ & $   18.7 $ & $   0.498 $ & HDS \\
$ 190.82599 $ & \nodata      & \nodata      & $    0.8 $ & $   20.9 $ & $   0.503 $ & HDS \\
$ 191.79963 $ & $   233.55 $ & $    15.50 $ & $    8.3 $ & $   25.0 $ & $   0.813 $ & HDS \\
$ 191.81436 $ & $   216.36 $ & $    18.54 $ & $   -1.9 $ & $   31.0 $ & $   0.818 $ & HDS \\
$ 191.82908 $ & $   240.25 $ & $    16.91 $ & $   41.6 $ & $   28.5 $ & $   0.822 $ & HDS \\
$ 192.79563 $ & $  -166.26 $ & $    11.68 $ & $  -16.2 $ & $   19.2 $ & $   0.130 $ & HDS \\
$ 192.81036 $ & $  -182.37 $ & $    14.01 $ & $   -0.9 $ & $   19.9 $ & $   0.134 $ & HDS \\
$ 192.82509 $ & $  -183.11 $ & $    15.32 $ & $   -5.6 $ & $   26.0 $ & $   0.139 $ & HDS \\
$ 211.59628 $ & $  -220.12 $ & $    22.00 $ & $   22.0 $ & $   11.0 $ & $   0.112 $ & FEROS \\
$ 375.86732 $ & $  -224.12 $ & $    19.00 $ & $   31.0 $ & $   10.0 $ & $   0.380 $ & FEROS \\
$ 376.87669 $ & $   280.88 $ & $    21.00 $ & $  -12.0 $ & $   11.0 $ & $   0.701 $ & FEROS \\
$ 377.84700 $ & $   -23.12 $ & $    21.00 $ & $    6.0 $ & $   11.0 $ & $   0.010 $ & FEROS \\
$ 398.86710 $ & $   281.88 $ & $    17.00 $ & $   48.0 $ & $    9.0 $ & $   0.698 $ & FEROS \\
$ 402.87705 $ & $    65.88 $ & $    18.00 $ & $    4.0 $ & $   10.0 $ & $   0.974 $ & FEROS \\
$ 405.87300 $ & $   130.88 $ & $    21.00 $ & $   27.0 $ & $   10.0 $ & $   0.928 $ & FEROS \\
$ 426.93117 $ & $   158.88 $ & $    26.00 $ & $   62.0 $ & $   13.0 $ & $   0.628 $ & FEROS \\

\enddata
\tablenotetext{a}{
    The zero-point of these velocities is arbitrary. An overall offset
    $\gamma_{\rm rel}$ fitted independently to the velocities from
    each instrument has been subtracted.
}
\tablenotetext{b}{
    Internal errors excluding the component of astrophysical jitter
    considered in \refsecl{globmod}.
}
\tablenotetext{c}{
    For FEROS and Coralie we take the BS uncertainty to be twice the
    RV uncertainty. For HDS the BS uncertainty is taken to be the
    standard error on the mean of the BS values calculated for each of
    the \'{E}chelle orders.
}
\ifthenelse{\boolean{rvtablelong}}{
    \tablecomments{
        Note that for the HDS iodine-free template exposures we do not
        measure the RV but do measure the BS.  Such template exposures
        can be distinguished by the missing RV value. The HDS
        observation of \hatcur{12} without a BS measurement has too
        low S/N in the I$_{2}$-free blue spectral region to pass our
        quality threshold for calculating accurate BS values.
    }
}{
    \tablecomments{
        Note that for the HDS iodine-free template exposures we do not
        measure the RV but do measure the BS.  Such template exposures
        can be distinguished by the missing RV value. The HDS
        observation of \hatcur{12} without a BS measurement has too
        low S/N in the I$_{2}$-free blue spectral region to pass our
        quality threshold for calculating accurate BS values.
    }
} 
\ifthenelse{\boolean{emulateapj}}{
    \end{deluxetable*}
}{
    \end{deluxetable}
}

\end{document}